\documentclass[aps,pre,amsmath,amsfonts,amssymb,superscriptaddress,nofootinbib,notitlepage,a4paper]{revtex4}

\usepackage{cancel}

\usepackage{graphicx}
\usepackage{psfrag}
\usepackage[naturalnames]{hyperref}

\newcommand{\beq}{\begin{equation}} 
\newcommand{\eeq}{\end{equation}}
\newcommand{\bem}{\begin{multline}}
\newcommand{\bes}{\begin{split}} \newcommand{\ees}{\end{split}} 
\newcommand{\bea}{\begin{eqnarray}} \newcommand{\eea}{\end{eqnarray}}

\def\s{{\sigma}}
\def\t{{\tau}}
\def\us{{\underline{\sigma}}}
\def\ut{{\underline{t}}}
\def\utau{{\underline{\tau}}}
\def\uu{{\underline{u}}}
\def\ux{{\underline{x}}}
\def\di{{\partial i}}
\def\da{{\partial a}}

\def\dima{{\partial i \setminus a}}
\def\dami{{\partial a \setminus i}}

\def\ind{{\mathbb{I}}}

\def\heta{{\widehat{\eta}}}
\def\hf{{\widehat{f}}}
\def\hl{{\widehat{l}}}
\def\ho{{\widehat{o}}}

\def\hw{{\widehat{w}}}

\def\hp{{\widehat{p}}}
\def\hz{{\widehat{z}}}
\def\hrho{{\widehat{\rho}}}
\def\hO{{\widehat{O}}}
\def\hQ{{\widehat{Q}}}
\def\hP{{\widehat{P}}}

\def\dd{{\rm d}}
\def\la{{\langle}}
\def\ra{{\rangle}}
\def\Z{{\cal Z}}
\def\hZ{{\widehat{\cal Z}}}

\def\S{{\cal S}}

\def\tP{{\widetilde{P}}}

\def\tO{{\widetilde{O}}}

\def\ttip{\theta^{\rm tip}}
\def\lr{{l_{\rm r}}}
\def\ld{{l_{\rm d}}}
\def\lc{{l_{\rm c}}}
\def\lsat{{l_{\rm sat}}}
\def\lstab{{l_{\rm stab}}}
\def\lmod{{l_{\rm mod}}}
\def\lsp{{l_{\rm sp}}}
\def\llm{{l_{{\rm l},-}}}
\def\llu{{l_{\rm l,u}}}
\def\llp{{l_{{\rm l},+}}}

\def\tmin{{\theta_{\rm min}}}
\def\Ptyp{\overline{P}}
\def\Ptypt{\overline{P_t}}
\def\Ptypz{\overline{P_0}}
\def\E{\mathbb{E}}
\def\mfh{\mathfrak{h}}
\def\mfu{\mathfrak{u}}

\begin{document}
\bibliographystyle{myunsrt}

\title{The large deviations of the whitening process \\in random constraint satisfaction problems}

\author{Alfredo Braunstein}
\affiliation{Politecnico di Torino, Corso Duca degli Abruzzi 24, 10129 Torino, Italy}
\affiliation{Human Genetics Foundation, Via Nizza 52, 10126 Torino, Italy}
\affiliation{Collegio Carlo Alberto, Via Real Collegio 30, 10024 Moncalieri, Italy}
\author{Luca Dall'Asta}
\affiliation{Politecnico di Torino, Corso Duca degli Abruzzi 24, 10129 Torino, Italy}
\affiliation{Collegio Carlo Alberto, Via Real Collegio 30, 10024 Moncalieri, Italy}
\author{Guilhem Semerjian}
\email[Corresponding author : ]{guilhem@lpt.ens.fr}
\affiliation{LPTENS, Ecole Normale Sup\'erieure, PSL Research University, Sorbonne Universit\'es, UPMC Univ Paris 06, CNRS UMR 8549, 24 Rue Lhomond, 75005
 Paris, France}
\author{Lenka Zdeborov\'a}
\affiliation{Institut de Physique Th\'eorique, CEA Saclay and CNRS, Gif-sur-Yvette, France}

\begin{abstract}
Random constraint satisfaction problems undergo several phase transitions as the ratio between the number of constraints and the number of variables is varied. When this ratio exceeds the satisfiability threshold no more solutions exist; the satisfiable phase, for less constrained problems, is itself divided in an unclustered regime and a clustered one. In the latter solutions are grouped in clusters of nearby solutions separated in configuration space from solutions of other clusters. In addition the rigidity transition signals the appearance of so-called frozen variables in typical solutions: beyond this threshold most solutions belong to clusters with an extensive number of variables taking the same values in all solutions of the cluster. In this paper we refine the description of this phenomenon by estimating the location of the freezing transition, corresponding to the disappearance of all unfrozen solutions (not only typical ones). We also unveil phase transitions for the existence and uniqueness of locked solutions, in which all variables are frozen. From a technical point of view we characterize atypical solutions with a number of frozen variables different from the typical value via a large deviation study of the dynamics of a stripping process (whitening) that unveils the frozen variables of a solution, building upon recent works on atypical trajectories of the bootstrap percolation dynamics. Our results also bear some relevance from an algorithmic perspective, previous numerical studies having shown that heuristic algorithms of various kinds usually output unfrozen solutions.

\end{abstract}

\maketitle

\tableofcontents

\section{Introduction}

In Constraint Satisfaction Problems (CSP) a set of $N$ discrete variables is subject to $M$ constraints, the problem then consists in deciding whether there exists an assignment of the variables which satisfies simultaneously all the constraints. A few examples of the numerous type of constraints are $k$-SATisfiability instances, in which the variables are boolean and each constraint is the disjunction (OR) of $k$ literals (a variable or its negation), $k$-XORSATisfiability, where the disjunction is replaced by the eXclusive OR, making this problem equivalent to linear equations modulo 2, and $q$-coloring, where the variables can take $q$ values, and each constraint forbids a pair of variables to take the same value. We will use in this paper the hypergraph bicoloring problem, in which binary variables are submitted to constraints bearing on $k$-uplet of variables, that impose that both binary values are present at least once among the $k$ variables (it corresponds also to the Not All Equal version of $k$-SAT with all literals positive). The object of the theory of computational complexity~\cite{GareyJohnson79,Papadimitriou94} is to classify the worst-case difficulty of these decision problems, a seminal result in this field being the NP-completeness of the $k$-SAT problem for $k\ge 3$~\cite{Cook71,Karp72}.

Despite their proven worst-case hardness it originally turned out that generating instances of CSPs that were actually hard for heuristic algorithms was not completely obvious~\cite{Goldberg79,GoldbergPurdom82}; a simple to generate and easy to tune set of benchmark instances is provided by random CSPs, in which the $M$ constraints are drawn randomly among all possible ones. In the case of coloring this corresponds to Erd\H{o}s-R\'enyi random graphs, for problems with constraints involving $k \ge 3$ variables their hypergraph generalizations. The control parameter of these random ensembles is the density of constraints $\alpha = M/N$, which is kept constant in the large size (thermodynamic) limit $N\to \infty$. Numerical investigations of these ensembles were first presented in~\cite{CheesemanKanefsky91,MitchellSelman92} (an earlier analytical study of this random ensemble can be found in~\cite{FrancoPaull83}) and triggered an interest for these random CSPs that survived until today. Indeed these studies showed that for some values of the parameter $\alpha$ the instances generated were typically (i.e. with a probability approaching 1 in the thermodynamic limit) very difficult to solve in practice, and suggested a connection between this hardness and a threshold phenomenon (or phase transition): the probability for such an instance to be satisfiable seemed to drop abruptly from $1$ to $0$ when the parameter $\alpha$ crossed a given value $\alpha_{\rm sat}$ (its numerical value depends of course on the details of the definition of the problem). Two decades of intense and interdisciplinary efforts have produced a lot of results that we can only partially survey here. Upper bounds on the possible values of $\alpha_{\rm sat}$ can be obtained by the first moment method (with various refinements)~\cite{transition_ub}, lower bounds by the analysis of simple enough algorithms~\cite{transition_lb,Achltcs} or by the use of second moment methods~\cite{AchlioptasMoore02}, and a weak form of the existence of $\alpha_{\rm sat}$ has been established in~\cite{Friedgut99}. The arity $k$ of the constraints provides another parameter to control the ensemble of instances, in particular when it gets large these problems are somehow easier to study rigorously, the upper and lower bounds having the same leading behavior, which led to tight asymptotic expansions at large $k$ for various problems~\cite{AchlioptasPeres04}. An exact determination of $\alpha_{\rm sat}$ for small values of $k \ge 3$ was instead an open problem for a long time. 

The use of statistical physics methods originally developed for the study of spin glasses~\cite{MezardParisi87b,MezardMontanari07}, namely the replica and cavity methods, turned out to be crucial in improving the understanding of random CSPs. For what concerns in particular the determination of the threshold $\alpha_{\rm sat}$ of random $k$-SAT a formalism predicting its value for all $k$ has been presented in~\cite{MezardParisi02,MertensMezard06}, whose exactness has been recently proven rigorously in~\cite{ding2014proof} for large enough (but finite) values of $k$, see also~\cite{DiSlSu13_naeksat} for a similar result on random NAE-$k$-SAT and~\cite{Coja14} for another regularized version of $k$-SAT.

In some sense the prediction of $\alpha_{\rm sat}$ is only the tip of the iceberg of the contributions of statistical mechanics to the field of random CSPs, as this prediction relies on a detailed description of the organization and properties of the set of solutions~\cite{MonassonZecchina99b,BiroliMonasson00,MezardParisi02,KrzakalaMontanari06}, that undergo several phase transitions before the satisfiability one. Let us now briefly describe how this problem is attacked in statistical mechanics studies and which are the phase transitions thus unveiled (note that also these predictions have received a partial rigorous confirmation, see for instance~\cite{AchlioptasRicci06,AchlioptasCoja-Oghlan08,molloy_col_freezing,molloy_csp_freezing}).

A natural way to study the set of solutions of a CSP is to introduce the uniform probability measure over this set. The variables of the problem are now seen as particles, or spins, which interact one with another through the constraints that need to be satisfied. These local interactions are the building blocks of a so-called graphical model, which can be represented as a factor graph~\cite{KschischangFrey01}; in random CSPs the local structure of such a graph is typically a tree, which suggests that this uniform probability measure over solutions can be accurately described using the Belief Propagation (or sum product) message passing algorithm, that would be exact if the factor graph was globally a tree. For low enough values of $\alpha$ this is indeed the case, correlations between variables decay fast enough for the loops in the factor graph to be irrelevant, and the support of the measure is sufficiently well-connected for its marginal laws to be described accurately by the single fixed point solution of the Belief Propagation (BP) iterative equations.

When $\alpha$ increases above a threshold $\alpha_{\rm d}$ (corresponding to the dynamical transition in the language of the mean field theory of structural glasses~\cite{berthier2011theoretical}) the shape of the set of solutions undergoes a drastic modification: it becomes split in an exponential number of clusters (portions of the configuration space), solutions being well connected inside a cluster, while two different clusters are well separated. This transition can be also characterized by the divergence of some correlation length, but of a particular type: correlations between a finite number of variables are blind to the transition at $\alpha_{\rm d}$, it is only the so-called point-to-set correlation which can detect it~\cite{MontanariSemerjian06b,MezardMontanari06}. This notion of correlation is related to the reconstruction problem~\cite{Mossel04,GerschenfeldMontanari07}: the observation of the variables at some large distance from a root point of the factor graph in an uniformly chosen solution of a CSP provides a non vanishing information on the value of the root in this solution only for $\alpha>\alpha_{\rm d}$. Given an instance of a random CSP the decomposition of its solution set into ``clusters'' is somehow ambiguous, on the one hand because different authors define clusters in different and not always equivalent ways, on the other hand because a few among the exponentially many solutions should be allowed to be moved from one cluster to another without affecting crucially the statements made about this decomposition. From a physics point of view this notion of clusters mimics the definition of pure states, namely they provide a partition of the configuration space in disjoint subsets (apart from negligible intersections), such that the frontier of a cluster constitutes a free-energy barrier (the probability of the frontier of the cluster has much lower probability than its inside), while there is no such bottleneck inside one cluster. Following this intuition it is useful to think that each pure state is associated to one solution of the BP equations, hence two solutions of the CSPs are considered to belong to the same cluster if the BP iterative equations flow to the same fixed point when initialized in both of them.

In most CSPs the clusters have wildly fluctuating sizes, a phenomenon which gives rises to a further phase transition at a threshold $\alpha_{\rm c}\in [\alpha_{\rm d},\alpha_{\rm sat}]$, this transition being called the condensation one (in the mean field theory of structural glasses this phase transition is known as the Kauzmann or the ideal glass phase transition). One has indeed a so-called complexity function (or structural entropy) $\Sigma(s)$, which counts the rate of growth for the number of clusters distinguished by their size: around $e^{N\Sigma(s)}$ clusters (at the leading exponential order) contain $e^{N s}$ solutions. The rate of growth for the total number of solutions is thus expressed as $\sup_s [\Sigma(s) + s]$, and two cases can occur: this supremum is reached in a point $s_*$ with either $\Sigma(s_*) > 0$, or $\Sigma(s_*)=0$. The first case occurs for $\alpha \in [\alpha_{\rm d},\alpha_{\rm c}]$: the dominant contribution at the leading exponential order to the total number of solutions comes from an exponential number of clusters. On the other hand in the second case (i.e. for $\alpha \in [\alpha_{\rm c},\alpha_{\rm sat}]$) only a sub-exponential number of clusters supports most (i.e. a fraction which goes to 1 in the thermodynamic limit) of the solutions of the CSP. This transition is accompanied by a non-analyticity in the total entropy of solutions, and with the appearance of correlations between finite number of points. Note that the large $N$ limit allows us to define the properties of ``typical solutions'': drawing uniformly at random a solution of a CSP, the deviations from $s_*$ of the size of the cluster of the randomly chosen solution are exponentially suppressed, hence ``a typical solution lives in a cluster of size $s_*$''. The complexity $\Sigma(s)$ can be determined with the one-step of replica symmetry breaking (1RSB) cavity method, which amounts to bias the measure over solutions with a weight given to a solution depending on the size of its cluster, turning solutions which were atypical under the uniform measure into the typical ones in the biased ensemble. In particular the satisfiability threshold corresponds to the disappearance of all clusters of solutions (regardless of their sizes), and can be characterized as the vanishing of $\sup_s \Sigma(s)$.

Another concept, more directly related to the content of this article, has been introduced to describe the set of solutions of a CSP, under the name of ``frozen variables''. A variable is said to be frozen in a cluster if it takes the same value in all the solutions of this cluster. A cluster is said to be frozen if it contains an extensive number (a positive fraction) of frozen variables. A solution that belongs to a frozen cluster is called frozen, all other solutions are called unfrozen, or white. Using the cavity method it has been analyzed when typical solutions are frozen~\cite{Semerjian07,ZdeborovaKrzakala07,MontanariRicci08}, the onset of this property was called the rigidity phase transition $\alpha_{\rm r}$ in~\cite{ZdeborovaKrzakala07} and we will adopt this term also in the present paper (unfortunately in previous papers the terms ``rigidity'' and ``freezing'' have been used in sometimes confusing way, here we shall use these two terms to denote two different transitions). This notion of frozen variables has also been studied rigorously, see in particular~\cite{AchlioptasRicci06,AchlioptasCoja-Oghlan08,molloy_col_freezing,molloy_csp_freezing}. One can view the existence of frozen variables as the hallmark for a stronger notion of correlation than the one appearing at $\alpha_{\rm d}$: indeed a frozen variable can be unambiguously reconstructed from the observation of a far away set of other variables, not only inferred with a probability of success larger than a random guess (see~\cite{Semerjian07} for a discussion of this point, along with an alternative interpretation of frozen variables in terms of the minimal size of a rearrangement induced by the modification of a variable in a solution~\cite{MontanariSemerjian06}). This explains why $\alpha_{\rm d} \le \alpha_{\rm r}$, the inequality being strict in general (a notable exception being the XORSAT model~\cite{DuboisMandler02,CoccoDubois03,MezardRicci03,IbKaKrMo15,AcMo15}). As mentioned above there are various possible definitions of a cluster, hence the definition given above of a frozen variable seems to depend on the definition adopted. Fortunately one can be more precise and give a clear-cut description of which variable is frozen in a given solution of a given instance of a CSP. The answer is provided by Warning Propagation (WP)~\cite{BraunsteinMezard05,BraunsteinMezard02}, a projection of Belief Propagation onto ``sure beliefs'' (also called ``hard fields''), which is always guaranteed to converge when initialized in a solution, independently of the update rules, and which identifies the frozen variables as the one that receives sure messages in the fixed point of WP. For some type of CSPs, satisfiability and hypergraph bicoloring in particular, this identification of frozen variables by WP is equivalent to a procedure called ``whitening''~\cite{Parisi02b,BraunsteinZecchina04}, ``peeling''~\cite{ManevaMossel05} or ``coarsening''~\cite{AchlioptasRicci06,DiSlSu13_naeksat} in terms of the variables, that can be pictured more easily in configuration space. Starting from a solution, calling initially all variables frozen, some of them will be subsequently called white according to the following rule. At time $t$ consider the subcube of the configuration space which agrees on the still frozen variables, white variables being completely free. If there exists a solution at Hamming distance 1 from this subcube, i.e. if at least two values of the still frozen variable $i$ can yield solutions keeping all other frozen variables fixed and choosing arbitrarily an assignment of the currently white variables, then declare $i$ white. This dynamics stops either when all variables have been whitened, or upon discovering the smallest subcube containing the starting solution and which has no solution in its neighborhood at Hamming distance 1 (variables can be whitened one by one in a sequential fashion with arbitrary order or in parallel rounds without affecting the final outcome of the procedure). Note that the description of a cluster in terms of its frozen variables is an oversimplification: if the cluster were a subcube, then there will be no loss of information. All solutions of the cluster would agree on the frozen variables, and reciprocally all configurations that agree with the frozen variables would be solutions. However in general clusters will have very intricate shape rather distant from subcubes (see~\cite{BrKaZe11} for an application of this property to data compression), and a very thin but elongated cluster can contain only a very small fraction of solutions, yet have no frozen variables in this sense; this is another perspective explaining why $\alpha_{\rm d} < \alpha_{\rm r}$ with a strict inequality in general.

These four thresholds at which the qualitative features of the typical solutions of a random CSP change obey by definition the inequalities $\alpha_{\rm d} \le \alpha_{\rm c} \le \alpha_{\rm sat}$ and $\alpha_{\rm d} \le \alpha_{\rm r} \le \alpha_{\rm sat}$, with numerical values depending of course on the type of constraints (SAT, XORSAT, NAESAT, coloring, \dots) and on the parameters defining them (the arity $k$ for boolean variables, the number of colors $q$, \dots). For small values of $k$ or $q$ some of the inequalities above are actually equalities (for instance random $3$-SAT has $\alpha_{\rm d} =\alpha_{\rm c}$), and the ``generic'' ordering $\alpha_{\rm d} < \alpha_{\rm r} < \alpha_{\rm c} < \alpha_{\rm sat}$ is only reached for large enough values of $k,q$ (i.e. $k \ge 6$ for SAT instances). The situation gets simplified in the limit of large $k$, where at the leading order, taking the example of $k$-SAT for concreteness,
\beq
\alpha_{\rm d} \sim \alpha_{\rm r} \sim 2^k \frac{\ln k}{k} \ , \qquad
\alpha_{\rm c} \sim  \alpha_{\rm sat} \sim 2^k \ln 2 \ ,
\eeq
with subdominant corrections different for $\alpha_{\rm d}$ and $\alpha_{\rm r}$ on the one hand, and $\alpha_{\rm c}$ and $\alpha_{\rm sat}$ on the other~\cite{MertensMezard06,Semerjian07,MontanariRicci08}; see also~\cite{MoReTe11_recclus} for large $k$ asymptotics of some of these thresholds in a wide family of random CSPs. There is thus asymptotically a large regime of densities $\alpha$ between the scale of the appearance of (rigid) clusters and the satisfiability threshold. These scalings are somehow universal, for instance for hypergraph bicoloring or NAESAT their asymptotic expansions are the same modulo a global factor $1/2$.

In this paper we shall further refine the description of the set of solutions of random CSPs by quantifying the degree of freezing of atypical solutions, and in particular estimate a new threshold $\alpha_{\rm f}$ (for freezing transition) above which all solutions are frozen (at variance with $\alpha_{\rm r}$ which concerns only the typical ones). For technical reasons we have performed this study for the bicoloring of (regular) random $k$-hypergraphs (or equivalently for regular NAE-$k$-SAT instances), but we believe our conclusions are qualitatively valid for a wide family of random CSP, and even quantitatively in the large $k$ limit. We obtained results for finite values of $k$, and also an asymptotic expansion of $\alpha_{\rm f}$ at large $k$ which, once translated in terms of the $k$-SAT problem, reads at the leading order $\alpha_{\rm f} \sim 2^{k-1} \ln 2$, i.e. on the scale $\alpha_{\rm sat}/2$, much larger than $\alpha_{\rm r} \sim \alpha_{\rm sat} / k$ (forgetting the subdominant logarithmic correction): atypical unfrozen solutions survive the addition of a huge amount of constraints besides the typical rigidity threshold. Note that this asymptotic scale $\alpha_{\rm sat}/2$ appeared in~\cite{MoraMezard05} as the density above which the range of possible Hamming distances between pairs of solutions becomes gapped.
The location of this freezing transition was not computed previously by statistical physics calculations. It has been addressed numerically for the random 3-SAT problem~\cite{ArdeliusZdeborova08} where it was shown that for this case the frozen region is tiny, taking only about 0.3\% of the satisfiable phase. Two rigorous works placed an upper bound on the freezing transition. Particularly for random $k$-SAT is it known that for large $k$ all clusters are frozen when $\alpha>(4/5) \alpha_{\rm sat}$~\cite{AchlioptasRicci06}, and for random coloring there also exists a proof that there is a gap between the freezing and the satisfiability transition~\cite{coja2013upper}.  
From a technical point of view our work amounts to a characterization of the large deviation properties of the whitening dynamics introduced above (building on recent studies of rare events in the related bootstrap percolation dynamics~\cite{AlBrAsZe13,AlBrAsZe13b,GuSe15}). To this end we introduced a biased measure over the set of solutions, the bias being a function of the number of frozen variables after $T$ steps of (a parallel version of) the whitening dynamics, that induces interactions between variables at distances up to $T$ in the factor graph. The determination of $\alpha_{\rm f}$ requires to take the limit $T\to\infty$, in order to characterize the fixed points of the whitening, which we managed to do analytically. In addition we also studied the structure of the biased measure over solutions for finite values of $T$, and discovered the following phenomenon. For each value of $T$ there is a threshold $\alpha_T(k) > \alpha_{\rm r}$ on the constraint density, up to which it is possible to bias the measure over solutions, with interactions of range $T$, in such a way that the typical configurations of the biased measure are unfrozen solutions. An asymptotic expansion of $\alpha_T$ at large $k$, for $T$ fixed, yields $\alpha_T \sim \alpha_{\rm sat}/\ln^{\circ T} (k)$, where $\ln^{\circ T}$ means logarithm iterated $T$ times. Already for $T=1$ this scale $\alpha_{\rm sat}/\ln k$ is asymptotically larger than $\alpha_{\rm r}$, and increasing $T$ makes this scale closer and closer to the freezing transition, i.e. a constant fraction of the satisfiability threshold.

The results we just mentioned concern unfrozen solutions; we shall also study the other extreme case and consider the subset of ``locked'' solutions, in which all variables are frozen (this terminology is taken from~\cite{ZdeborovaMezard08} which introduced locked CSPs in which all solutions are locked in this sense), providing an estimate for the regimes of densities in which such locked solutions exist and in which all the frozen solutions are actually locked.

Our efforts to further refine the description of the many transitions undergone by the set of solutions of random CSPs is largely motivated by the relatively disappointing current state of understanding of their consequences on the performance of algorithms. The clustering transition at $\alpha_{\rm d}$ is known to imply the divergence of the equilibration time for Monte Carlo Markov chains respecting detailed balance~\cite{MontanariSemerjian06b}, hence for $\alpha > \alpha_{\rm d}$ this class of algorithms cannot sample uniformly the set of solutions in polynomial time. This result is, however, of limited interest in the algorithmic perspective, on the one hand because the main focus is on finding solutions (without requiring an uniform distribution), which can be achieved by simulated annealing~\cite{KirkpatrickGelatt83} for some range of $\alpha > \alpha_{\rm d}$ even ``off-equilibrium''~\cite{ZdeborovaKrzakala10}, and on the other hand because most efficient local search algorithms~\cite{SelmanKautz94,ArdeliusAurell06,AlavaArdelius07} do not respect detailed balance, hence $\alpha_{\rm d}$ is not particularly relevant for their analysis. Some relatively simple algorithms, that fix the state of the variables sequentially, according to some rules depending on the previous choices, have been the subject of a large literature and used to prove lower bounds on $\alpha_{\rm sat}$~\cite{transition_lb,Achltcs,Co01}. For random 3-coloring an algorithm of this class was proved to find solutions up to an average degree 4.03~\cite{AchlioptasMoore03}: this demonstrates the irrelevance of the dynamic and condensation transition for this class of algorithms, as in this case they both happen at average degree four~\cite{ZdeborovaKrzakala07}. However in the large $k$ limit the thresholds at which this type of algorithms stop to find solutions in polynomial time all scale as $\alpha_{\rm sat} / k$, with a constant prefactor that depends on the details of the algorithm, i.e. on a scale slightly lower than $\alpha_{\rm d}$. The best large $k$ asymptotics for the threshold of a proven algorithm is $2^k (\ln k)/k$~\cite{Amin_algo}, i.e. precisely the clustering threshold. On the negative side \cite{GaSu14} proved that no ``local'' algorithm (in a precise sense) can find solutions in polynomial time for densities $\alpha$ larger than $\alpha_{\rm d} \ln k$ (asymptotically at large $k$).

Statistical mechanics has contributed to the algorithmic aspects of random CSPs by using out of equilibrium physics methods to study the dynamics of known algorithms~\cite{Co01,SemerjianMonasson03}, but also suggested some new algorithms~\cite{MezardParisi02,BraunsteinMezard05,MaPaRi15,Allerton,RiSe09} based on the detailed picture of the configuration space unveiled by the static analysis. In particular the survey propagation algorithm is based on a message passing strategy, that estimates the probability for a given variable to be frozen in a cluster chosen uniformly at random, and that fixes the variables sequentially according to this information (some implementations include a backtracking mechanism~\cite{MaPaRi15}). For $k=3$ this algorithm is extremely efficient, and solves in polynomial time typical random instances with densities very close to the satisfiability threshold, certainly above the condensation threshold~\cite{BraunsteinMezard05}; less efforts have be devoted to the study of larger values of $k$. The recent study of~\cite{MaPaRi15} shows that for $k=4$ a backtracking version of SP is able to find solutions for densities slightly above the rigidity threshold. It would be desirable to have numerical studies for larger values of $k$, as the various thresholds are rather close for small values of $k$, making harder to disentangle which one is possibly at the origin of algorithmic hardness; an analysis of the performances of SP in the large $k$ limit would also be nice, but seems rather difficult because of the many implementation details that impact significantly its performances but complicates its analysis (see~\cite{Allerton,RiSe09,Coja12} for an analysis of the simpler BP guided decimation algorithm, which exhibits an asymptotic threshold on the scale $\alpha_{\rm sat}/k$); very recently \cite{Hetterich} showed that the simplest version of Survey Propagation fails for densities larger than $\alpha_{\rm d}$ (in the large $k$ limit). 

All these observations suggest a lack of current understanding of the impact of structural phase transitions on the behavior of algorithms, and point to a major challenge in the field, namely the design of an algorithm able to find solutions in polynomial time at densities strictly above $\alpha_{\rm d}$ (for large enough $k$). We believe the results of the present paper can be relevant in this perspective; indeed, it has been repeatedly reported that all efficient algorithms return unfrozen solutions (even when they are based on the existence of frozen ones as SP)~\cite{ZdeborovaKrzakala07,KrzakalaKurchan07,ManevaMossel05,BraunsteinZecchina04,MaPaRi15}. This led to the conjecture~\cite{ZdeborovaKrzakala07} that frozen solutions are truly hard to find, and that the rigidity or freezing phenomenon is more directly related to algorithmic hardness than clustering itself (an argument supporting this point of view is provided by the so called locked CSPs~\cite{ZdeborovaMezard08} where all solutions are frozen and typical instances are not solved in polynomial time by current algorithms, except XORSAT because of its linear structure). It has been shown in~\cite{ManevaMossel05} that frozen solutions must have an extensive number of frozen variables, which induce very strong correlations that seem hard to handle algorithmically, as an extensive number of variables should be set coherently to construct a frozen solution. The freezing transition would thus be an ultimate barrier for the efficient functioning of algorithms; as we saw this threshold is much higher than the one for rigidity, which shed a more optimistic light on the algorithmic challenge defined above. One could hope that as long as some (atypical) unfrozen solutions do exist some algorithms should be able to find them (this is what seems to happen in the studies of~\cite{DallAstaRamezanpour08,ZdeborovaMezard08b,MaPaRi15}). In this perspective our results on the scaling of $\alpha_T$ suggest that this hope might not be completely foolish: biasing appropriately the measure over solutions one can turn into typical some properties which were atypical in the uniform measure, up to a scale rather close to the satisfiability transition. This direction opens a wide perspective for the design of new algorithms, based on appropriately biased measures over the set of solutions; a study of this type has been recently presented in~\cite{BaInLuSaZe15,BaInLuSaZe15_long}, where the bias was based on an estimation of the local density of solutions around a configuration.

The rest of the paper is organized as follows.
In Sec.~\ref{sec_def} we first give precise definitions of the hypergraph bicoloring problem (or NAE-$k$-SAT) and briefly recall some known results on its phase transitions (\ref{sec_reminder}), then we introduce the whitening dynamical process (\ref{sec_def_whitening}) and recall its behavior on typical solutions (\ref{sec_whitening_typical}). In the next two subsections we define precisely the large deviation quantities we are interested in (\ref{sec_ld}), and state our main results (\ref{sec_main_results}); we believe a reader already familiar with the phenomenology of random CSPs and who does not want to drown him/herself in the technical details of the computation can concentrate his/her efforts on these two subsections. The justification of our results is expanded upon in Sec.~\ref{sec_statmech}, where we introduce a statistical mechanics formulation that encodes the large deviations of the whitening dynamics from its typical behavior, further technical details being deferred to Appendix~\ref{app_cavity}. Part of the discussion is split according to the value of the time horizon $T$ at which the large deviation is imposed ($T=1$ in Sec.~\ref{sec_res_T1}, $T\to\infty$ in Sec.~\ref{sec_res_Tinfty}, intermediate values of $T$ being dealt with in Sec.~\ref{sec_res_Tfinite}), some intermediate steps of the computation being relegated to Appendices~\ref{app_largek_T1}, \ref{app_Tinfty} and \ref{app_largek_Tarb}. These results have been obtained within a technical assumption known as replica symmetry; the failure of this assumption and its consequences are discussed in Sec.~\ref{sec_rsb}. We present some results of numerical experiments and algorithmic consequences of our results in Sec.~\ref{sec_numerics}, before drawing our conclusions in Sec.~\ref{sec_conclu}.

\section{Definitions and main results}
\label{sec_def}

\subsection{The hypergraph bicoloring/NAESAT problem and its known properties}
\label{sec_reminder}

\begin{figure}
\includegraphics{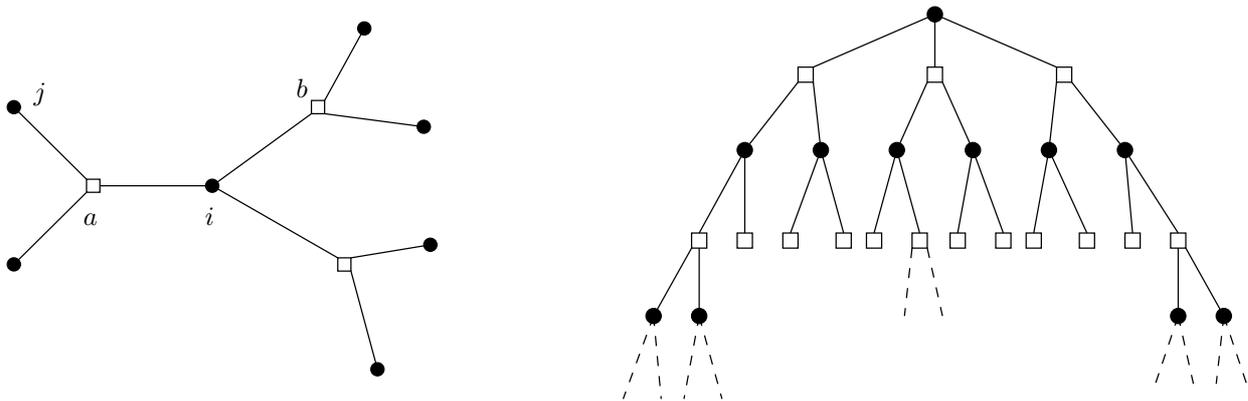}
\caption{Left panel: a factor graph representation of a hypergraph. Right panel: a portion of the tree factor graph to which random hypergraphs locally converge, here for $k=l+1=3$.}
\label{fig_fg_and_tree}
\end{figure}

An instance of the hypergraph bicoloring problem is defined as follows. $G$ denotes a hypergraph on $N$ vertices $i=1, \dots , N$, with $M$ hyperedges $a=1, \dots , M$, each of them linking a subset denoted $\da$ of $k$ distinct vertices. We also denote $\di$ the set of hyperedges in which the vertex $i$ appears. $G$ can be conveniently represented as a factor graph (see. Fig.~\ref{fig_fg_and_tree} for an illustration), i.e. a bipartite graph where empty squares (function nodes) encode the hyperedges $a$ and filled circles (variable nodes) the vertices $i$, an edge between $a$ and $i$ being drawn if and only if $a\in \di$ (or equivalently $i \in \da$). Binary variables $\s_i$ are located on each vertex of $G$; following the physics convention we represent them as Ising spins, $\s_i=\pm 1$. The global configurations of all variables are denoted $\us=(\s_1,\dots,\s_N) \in \{-1,+1\}^N$, while for a subset $S$ of the vertex indices we use $\us_S=\{\s_i \, : \, i \in S\}$ to represent the configurations of the spins in this subset. Each hyperedge imposes a constraint on the allowed configurations for the variables in its neighborhood: the $a$-th hyperedge is satisfied by $\us$ if $\us_\da$ is distinct from all $+1$ or all $-1$, in other words if around each hyperedge there is at least one spin $+1$ and at least one spin $-1$ (hence the name bicoloring). This condition can also be seen as a Not All Equal SATisfiability constraint, with all literals taken positively. In general a NAESAT constraint on the hyperedge $a$ would be defined through $k$ literal signs $\{J_i^a\}_{i\in\da}\in\{-1,1\}^k$, the constraint being that $\{J_i^a \s_i \}_{i\in\da}$ is different from all $+1$ or all $-1$. We shall denote $w_a(\us)=w_a(\us_{\da})$ the indicator function of the event ``the $a$-th hyperedge is satisfied''. A solution (also called proper bicoloring) of the instance of the hypergraph bicoloring problem on $G$ is a configuration $\us$ satisfying all hyperedges of $G$ simultaneously; $\S(G)$ will stand for the proper bicolorings of the hypergraph $G$.

We will consider the random ensemble of $l+1$-regular $k$-uniform random hypergraphs, in which $G$ is chosen uniformly at random from all hypergraphs on $N$ vertices where each constraint $a$ involves $k$ variable, and each variable $i$ appears in $l+1$ constraints; averages with respect to this ensemble will be denoted $\E_G[\bullet]$. For hypergraph bicoloring the literal signs $J_i^a$ are all equal to $+1$, in the standard definition of random NAESAT they are $\pm 1$ with equal probability. In the following we will follow the former convention for simplicity, however all our results will be obtained under an assumption of symmetry that would make them valid for any choice of the distribution of the $J_i^a$. 

For such hypergraphs the number of constraints is $M=N \frac{l+1}{k}$, hence the density of constraints $\alpha=\frac{l+1}{k}$; in the large $k$ limit Poissonian random graphs become equivalent to regular ones by concentration, hence the thresholds of the two type of ensembles have the same behavior modulo the converting factor $k$ between degree and density of constraints. In the thermodynamic (large size) limit we will be mostly interested in, these hypergraphs converge locally to regular tree structures, as represented in the right panel of Figure~\ref{fig_fg_and_tree}: for a given depth $d$, the probability that the neighborhood of a randomly chosen vertex within graph distance $d$ is a regular tree goes to 1 when $N$ diverges.

We shall now briefly sketch the cavity method derivation of the typical properties of the set of solutions $\S(G)$ and recall previously obtained results; the reader is referred to~\cite{MezardParisi01,MezardParisi03,MezardMontanari07} for more details on the cavity method, to \cite{CastellaniNapolano03,DallAstaRamezanpour08} for the original statistical mechanics works on this problem, and to \cite{CojaZdeb12,BaCoRa14,AchlioptasMoore06,CoPa12,DiSlSu13_naeksat,AchlioptasCoja-Oghlan08} for rigorous results on random hypergraph bicoloring and/or random NAE-$k$-SAT instances. To study the properties of the set $\S(G)$ it is convenient to consider the uniform measure over this set, assuming of course it is non-empty,
\beq
\mu(\us) = \frac{1}{Z(G)} \prod_{a=1}^M w_a(\us_\da) \ , 
\label{eq_mu_unbiased}
\eeq
the normalizing constant being equal to the number of solutions, $Z(G) = |\S(G)|$. In the large size limit it is then natural to define an entropy density,
$\phi(G) = \frac{1}{N} \ln Z(G)$. Such graphical models can be tackled with the Belief Propagation (BP) algorithm~\cite{KschischangFrey01}, introducing messages $\eta_{i \to a}$ and $\heta_{a \to i}$ on each directed edge of the factor graph, that corresponds to the marginal probability law of $\s_i$ in amputated (cavity) graphs where some interactions are discarded. These messages obey local consistency relations (the BP equations), of the form
\beq
\eta_{i \to a} = f(\{\heta_{b \to i}\}_{b\in \dima}) \ , \qquad
\heta_{a \to i} = \hf(\{\eta_{j \to a}\}_{j\in \dami}) \ ,
\eeq
that would be exact if the factor graph were a tree, and that are supposed to be good approximations whenever the graph is locally a tree. More explicitly the functions $f$ and $\hf$ above are defined (for a variable node of degree $l+1$ and a function node of degree $k$) through:
\bea
\eta(\s) &=& \frac{1}{z(\heta_1,\dots,\heta_l)} \, \heta_1(\s) \dots \heta_l(\s) \ , \label{eq_BP1} \\
\heta(\s) &=& \frac{1}{\hz(\eta_1,\dots,\eta_{k-1})} \sum_{\s_1,\dots,\s_{k-1}} \eta_1(\s_1) \dots \eta_{k-1}(\s_{k-1}) \, w(\s,\s_1,\dots,\s_{k-1})\ ,
\label{eq_BP2}
\eea
where $w$ is the weight function on the corresponding function node, and the functions $z$ and $\hz$ ensure the normalizations of the laws $\eta$ and $\heta$. In the present model $w(\s,\s_1,\dots,\s_{k-1})=\ind(\s,\s_1,\dots,\s_{k-1} \, \text{n.a.e.})$, with $\ind(A)$ the indicator function of the event $A$ and ``n.a.e.'' standing for ``not all equal''. 

For future convenience let us now open a brief parenthesis and discuss at this point another message passing algorithm, termed Warning Propagation (WP)~\cite{BraunsteinMezard05,BraunsteinMezard02}, that can be seen as a projection of the BP messages onto a simpler description. Let us set $\mfh_{i \to a}=1$ if $\eta_{i \to a}(\s) = \delta_{\s,1}$, $\mfh_{i \to a}=-1$ if $\eta_{i \to a}(\s) = \delta_{\s,-1}$, and $\mfh_{i \to a}=0$ otherwise. In other words we keep only the information about marginal probabilities of variables that are completely polarized in one of the two states, and wash out the non-trivial biases of intermediate cases. Denoting similarly $\mfu_{a \to i}$ the projection of $\heta_{a \to i}$, one finds the local rules relating the WP messages by projecting the BP equations (\ref{eq_BP1},\ref{eq_BP2}):
\beq
\mfh_{i \to a} = \begin{cases} 
1  & \text{if} \ \exists b \in \dima \ \text{with} \ \ \mfu_{b\to i}=1 \ \text{and} \ \nexists b \in \dima \ \text{with} \ \ \mfu_{b\to i}=-1 \\
-1 & \text{if} \ \exists b \in \dima \ \text{with} \ \ \mfu_{b\to i}=-1 \ \text{and} \ \nexists b \in \dima \ \text{with} \ \ \mfu_{b\to i}=1 \\
\frac{0}{0} & \text{if} \ \exists b \in \dima \ \text{with} \ \ \mfu_{b\to i}=1 \ \text{and} \ \exists b \in \dima \ \text{with} \ \ \mfu_{b\to i}=-1 \\
0 & \text{if} \ \forall b \in \dima \ \ \ \mfu_{b\to i}=0
\end{cases} \ .
\label{eq_WP_1}
\eeq
This rule is common to all CSPs on binary variables, and express the fact that a variable is forced as soon as one of its neighboring interactions forces it, with possible contradictions if two interactions try to impose distinct values on a single variable (the case denoted $\frac{0}{0}$ in the above equation). The next WP rule will be written here for the specific case of hypergraph bicoloring:
\beq
\mfu_{a \to i} = \begin{cases} 
1  & \text{if} \ \forall j \in \dami \ \ \ \mfh_{j \to a} = -1 \\
-1 & \text{if} \ \forall j \in \dami \ \ \ \mfh_{j \to a} = 1 \\
0 & \text{otherwise}
\end{cases} \ ,
\label{eq_WP_2}
\eeq
as indeed a variable is forced by a bicoloring constraint only when all the other variables are forced to be in the same color.

We now come back to the BP formalism. Once a solution of the BP equations for the messages $\{\eta_{i \to a},\heta_{a \to i}\}$ on all directed edges of the hypergraph is found the prediction for $\phi(G)$ corresponds to the Bethe approximation of statistical mechanics:
\beq
\phi(G) = 
-\frac{1}{N} \sum_{\la a,i \ra} \ln z_{\rm e}(\eta_{i \to a},\heta_{a \to i}) +
\frac{1}{N} \sum_{a=1}^M \ln z_{\rm c}(\{\eta_{i \to a} \}_{i \in \da}) +
\frac{1}{N} \sum_{i=1}^N \ln z_{\rm v}(\{\heta_{a \to i} \}_{a \in \di})
\eeq
with contributions coming from the edges and the two types of nodes of the factor graph,
\bea
z_{\rm e}(\eta,\heta) &=& \sum_{\s} \eta(\s) \, \heta(\s) \ , \\
z_{\rm c}(\eta_1,\dots,\eta_k)&=&\sum_{\s_1,\dots,\s_k} \eta_1(\s_1) \dots \eta_k(\s_k) \, w(\s_1,\dots,\s_k) \ , \\
z_{\rm v}(\heta_1,\dots,\heta_{l+1}) &=& \sum_{\s} \heta_1(\s) \dots \heta_{l+1}(\s) \ .
\eea
The replica symmetric (RS) version of the cavity method turns this study on a given factor graph into a prediction for the average (and also typical thanks to its self-averaging properties) value of $\phi = \E_G[\phi(G)]$ in the large size limit. In the case under study this takes a very simple form because of the locally regular structure of the hypergraph: one can search for an homogeneous (so-called factorized) solution of the BP equations with $\eta_{i \to a}=\eta$ and $\heta_{a \to i} = \heta$ on all edges. Moreover the problem being invariant under the symmetry $\us \to - \us$ one can assume these messages to be unbiased, i.e. $\eta(\s)=\heta(\s)=1/2$ independently of $\s$. Plugging this ansatz in the above equations yields the RS prediction for the typical entropy of such $k$-uniform $l+1$-regular random hypergraphs, 
\beq
s(k,l)=\lim_{N\to \infty} \E_G[ \phi(G) ] =  \ln 2 + \frac{l+1}{k} \ln\left( 1 - \frac{1}{2^{k-1}}\right) \ .
\label{eq_s_ann}
\eeq
The RS prediction for the satisfiability threshold would thus be the value where this entropy vanishes, namely
\beq
l_{s=0} = - 1 - \frac{k \ln 2}{\ln\left( 1 - \frac{1}{2^{k-1}}\right)} \ .
\eeq
Numerical values of this quantity for small values of $k$ are given in Table~\ref{table_typical}; here and in the following we give real numbers for the threshold values of $l$ which are expressed as analytic functions of $l$, even if of course the problem is only defined when $l$ is an integer. This real threshold has to be rounded to its two nearest integers to find the largest (resp. smallest) $l$ such that a property (here the positivity of the entropy) is true (resp. false) with high probability in the large size limit. In the following we shall call ``annealed entropy'' the quantity $s(k,l)$ defined in Eq.~(\ref{eq_s_ann}): this is indeed the logarithm of the average number of solutions (divided by $N$) for the NAE-$k$-SAT problem on an $l+1$-regular hypergraph, with a uniform choice for the sign of the literals, hence for this version of the problem $l_{s=0}$ is a rigorous upperbound on the satisfiability threshold thanks to the first moment method. 

The assumptions underlying the RS version of the cavity method can only be true for underconstrained problems, i.e. for sufficiently small values of $l$; in particular the negative entropy prediction for $l>l_{s=0}$ is self-contradictory. As sketched in the introduction this phenomenon, called Replica Symmetry Breaking (RSB), can be traced to the appearance of a clustered structure in the set $\S(G)$ of proper bicolorings, which for large enough values of $l$ shatters in an exponentially large number of groups of solutions called clusters or pure states, the solutions being well-connected inside each group while the groups are disconnected one from the other (in a sense that can be made more precise, see for instance~\cite{MontanariRicci08} for various alternative definitions). The long-range decorrelation hypothesis required for the BP computation to be accurate on a graph which is only locally a tree now only holds if the uniform measure $\mu$ is restricted to a given pure state. This can be exploited, in the so-called cavity method at the first level of RSB (1RSB), by considering each of the numerous solutions of the BP equation as a pure state and by introducing for each of the edges of the factor graph a probability distribution over messages, that encodes the randomness in the choice of the pure state. It turns out that the pure states are exponentially numerous and contain themselves an exponential number of solutions, different from one pure state to another. These fluctuations are encoded in the so called complexity function (or configurational entropy) $\Sigma(s)$: there are (at the leading exponential order) $\exp(N\Sigma(s))$ clusters containing $\exp(N s)$ solutions. The 1RSB cavity method allows to compute this quantity $\Sigma(s)$, or more precisely its Legendre transform $\Phi(m)=\sup_s [ \Sigma(s) + m s]$, with the conjugated parameter (called Parisi parameter) $m$. For simplicity let us give explicitly the 1RSB equations only in the factorized case, with a single distribution $P$ (resp. $\hP$) on the edges $i \to a$ (resp. $a \to i$) of a $k$-uniform $l+1$-regular hypergraph:
\bea
P(\eta) &=& \frac{1}{\Z} \int \dd \hP(\heta_1) \dots \dd \hP(\heta_l) \ \delta (\eta - f(\heta_1,\dots,\heta_l)) \ z(\heta_1,\dots,\heta_l)^m \ , 
\label{eq_1RSB_P} \\
\hP(\heta) &=& \frac{1}{\hZ} \int \dd P(\eta_1) \dots \dd P(\eta_{k-1}) \ \delta (\heta - \hf(\eta_1,\dots,\eta_{k-1})) \ \hz(\eta_1,\dots,\eta_{k-1})^m \ , 
\label{eq_1RSB_hP}
\eea
where $\Z$ and $\hZ$ are normalizing constants, and the functions $f,\hf,z$ and $\hz$ have been defined in (\ref{eq_BP1},\ref{eq_BP2}) above. From the solution of these equations on $P$ and $\hP$ the Legendre transform of the complexity is computed as
\beq
\Phi(k,l,m) = -(l+1) \ln \Z_{\rm e} + \frac{l+1}{k} \ln \Z_{\rm c} + \ln \Z_{\rm v} \ ,
\eeq
with
\bea
\Z_e &=& \int \dd P(\eta) \dd \hP(\heta) \ z_{\rm e}(\eta,\heta)^m \ , \\
\Z_c &=& \int \dd P(\eta_1) \dots \dd P(\eta_k) \ z_{\rm c}(\eta_1,\dots,\eta_k)^m  \ , \\
\Z_v &=& \int \dd \hP(\heta_1) \dots \dd \hP(\heta_{l+1}) \ z_{\rm v}(\heta_1,\dots,\heta_{l+1})^m \ .
\eea
The value $m=1$ plays a special role in this formalism, as $\Phi(m=1)=\sup_s [ \Sigma(s) + s]$ is equal to the total entropy of solutions. In Table~\ref{table_typical} we define $\ld$, the so-called dynamic or clustering threshold, as the smallest integer such that there exists a non-trivial solution of the 1RSB equations (\ref{eq_1RSB_P},\ref{eq_1RSB_hP}) at $m=1$. As long as the corresponding complexity is strictly positive the RS prediction for the total entropy is correct (technically this comes from the fact that the average $\eta$ over the 1RSB distribution $P(\eta)$ at $m=1$ obeys the RS equation, hence $\Phi(m=1)=\phi_{\rm RS}$ in general), this behavior is called a dynamic 1RSB phase. This condition is violated beyond the so-called condensation transition, as in the Legendre transform the supremum over $s$ has to be constrained to values such that $\Sigma(s)\ge 0$. The values of the condensation threshold are also given in Table~\ref{table_typical}, where $\lc$ is the smallest integer such that the complexity at $m=1$ is negative. For completeness we also reported the values $\lmod$ and $\lstab$ which corresponds to the limit of local stability of the RS factorized solution towards, respectively, a modulated antiferromagnetically ordered phase which is frustrated by the loops of the hypergraph, and a RSB solution. These two thresholds are easily found to be
\beq
\lmod = \frac{1}{k-1}\left(2^{k-1}-1\right) \ , \qquad
\lstab = \frac{1}{k-1}\left(2^{k-1}-1\right)^2 \ .
\label{eq_lstab}
\eeq
For large enough values of $k$ this last threshold is irrelevant because the continuous transition it describes is preceded by a discontinuous one.

The clustering transition, defined here as the appearance of a non-trivial solution of the 1RSB equations (\ref{eq_1RSB_P},\ref{eq_1RSB_hP}) for $m=1$, has been shown~\cite{MezardMontanari06} to be equivalent to a transition in a tree reconstruction problem, namely the possibility of inferring some information on the value of $\s_i$ in an uniformly chosen proper bicoloring of an infinite tree, given the values of $\s_j$ on the set of vertices $j$ at distance $R$ from $i$, in the limit where $R$ diverges. Another related question is the possibility, given the same information, of inferring $i$ with no possibility of error, which is possible if and only if $\s_i$ takes the same value in all the proper bicolorings allowed by the boundary condition on the $\s_j$. As discussed in~\cite{Semerjian07} this can be rephrased in terms of the 1RSB equations, this ``naive reconstruction'' being possible if and only if the solutions of (\ref{eq_1RSB_P},\ref{eq_1RSB_hP}) at $m=1$ gives a non-zero weight to ``hard fields'' (or ``hard messages'') $\eta(\s) = \delta_{\s,+1}$ and $\eta(\s) = \delta_{\s,-1}$. This condition is in turns interpreted as the existence of ``frozen variables'', i.e. of variables that take the same value in all the solutions of a cluster. It is possible to write closed equations for the weight of hard fields in the solution of (\ref{eq_1RSB_P},\ref{eq_1RSB_hP}) for $m=1$, exploiting on the one hand the fact that $\int \dd P(\eta) \, \eta$ is the (unbiased) RS solution, and on the other hand the WP equations of (\ref{eq_WP_1},\ref{eq_WP_2}) that describe the possible ways hard fields can be combined to produce another hard field. Denoting $p$ the probability of a positive hard field $\eta(\s)=\delta_{\s,1}$ (by symmetry it is also the probability of a negative hard field) under the law $P$, and similarly $\hp$ for the law $\hP$, one finds:
\beq
p = \frac{1}{2} \left(1-(1-2\hp)^l \right) \ , \qquad
\hp = \frac{1}{2} \frac{1}{1-\frac{1}{2^{k-1}}} p^{k-1} \ .
\eeq
Denoting $x=2 p$ the total probability of hard fields (of both polarities) in $P$ and regrouping these two equations yields the following self-consistent equation, 
\beq
x=1-\left(1- \frac{x^{k-1}}{2^{k-1}-1} \right)^l \ .
\label{eq_rigidity_typ}
\eeq
The rigidity threshold $\lr$, that is reported in Table~\ref{table_typical}, is defined as the smallest value of $l$ such that the equation admits a solution with $x\in ]0,1]$. 

Above the condensation threshold the RS prediction for the entropy of solutions is incorrect and has to be replaced by the entropy of the largest existing clusters, such that $\Sigma(s)=0$ (defining a static value of the Parisi parameter $m_{\rm s}<1$): the typical (most numerous) solutions are located in a sub-exponential number of clusters of this size. Increasing further the degree $l$, hence the ratio of constraints over variables, the satisfiability threshold is reached when all clusters disappear, i.e. when $\sup_s \Sigma(s) = 0$, this threshold can hence be computed by an $m=0$ 1RSB computation. Fortunately this special case allows for great simplifications in the 1RSB formalism. Indeed in the limit $m\to 0$ the reweighting factors $z^m$ in (\ref{eq_1RSB_P},\ref{eq_1RSB_hP}) becomes indicator functions of the event ``$z>0$'', hence the computation of the complexity reduces to the determination of the probability of the hard fields in the distributions $P$ and $\hP$. Using the same definition as above for $p$ and $\hp$, one finds the following relations at $m=0$,
\beq
p = \frac{(1-\hp)^l - (1-2 \hp)^l}{2 (1-\hp)^l - (1-2 \hp)^l} \ , \qquad \hp = p^{k-1} \ , 
\label{eq_sp}
\eeq
which have a non-trivial solution above the threshold $\lsp$ (standing for Survey Propagation). The associated complexity reads
\beq
\Sigma = - (l+1) \ln(1 - 2 p \hp) + \frac{l+1}{k} \ln(1-2 p^k) + \ln(2(1-\hp)^{l+1} - (1-2 \hp)^{l+1}) \ ,
\label{eq_Sigma_m0}
\eeq
which becomes negative above the satisfiability threshold $\lsat$. The values of $\lsp$ and $\lsat$ are reported in Table~\ref{table_typical}; note that this criterion for the location of the satisfiability transition has been rigorously established in~\cite{DiSlSu13_naeksat} for the regular NAESAT problem for large enough (but finite) values of $k$ (see equation (1) in~\cite{DiSlSu13_naeksat}, which is equal to our (\ref{eq_Sigma_m0}), with the change of notations $l+1 \to d$, $p \to q/2$).

It is instructive to study the scaling of these thresholds when $k$ gets large; these asymptotic expansions are naturally organized with three main levels of refinement, dominated by exponential behaviors in $k$, then polynomial contributions in $k$, and finally (iterated) logarithmic functions of $k$. Among the various thresholds defined above some of them share a leading behavior of $2^{k-1} k \ln 2$, namely
\bea
l_{s=0} &=&  2^{k-1} k \, \ln 2 - \left( 1 + k \frac{\ln 2}{2} \right) + O\left(\frac{k}{2^k}\right)  \ , \\ 
\lc &=&  2^{k-1} k \, \ln 2 - \left( 1 + k \ln 2 \right) + O\left(\frac{k^3}{2^k}\right)\ , \\
\lsat &=&  2^{k-1} k \, \ln 2 - \left( 1 + k \left( \frac{\ln 2}{2} + \frac{1}{4} \right) \right) + O\left(\frac{k^3}{2^k}\right)\ .
\label{eq_lsat_largek}
\eea
We thus see that the simple annealed bound $l_{s=0}$ gives the correct leading behavior of the satisfiability threshold in the large $k$ limit; improvingly tight asymptotically lower bounds were rigorously derived in~\cite{AchlioptasMoore06,CoPa12}.

Other thresholds are on the scale $2^{k-1} \ln k$, i.e. a factor $k \ln 2 / \ln k$ below the previous one, namely
\bea
\ld &=& 2^{k-1} \left( \ln k + \ln \ln k + O(1) \right) \ , \\
\lr &=& 2^{k-1} \left( \ln k + \ln \ln k + 1 + O\left(\frac{\ln \ln k}{\ln k} \right) \right) \ , \label{eq_lr_largek} \\
\lsp &=& 2^{k-1} \left( \ln k + \ln \ln k + 1 - \ln 2 + O\left(\frac{\ln \ln k}{\ln k} \right) \right) \ .
\eea
As $\lr$ and $\lsp$ are the limit of existence of solutions for scalar equations (\ref{eq_rigidity_typ},\ref{eq_sp}) their asymptotic expansion can be easily performed; on the contrary $\ld$ concerns a functional equation, its expansion is thus more delicate. In any case one has $\ld \le \lr$, and a lower bound of $\ld$ which is asymptotically equivalent to $\lsp$ (up to the smallest terms stated above) was proven rigorously for $q$-coloring at large $k$ in~\cite{Sly08}. Moreover a rigorous proof of ``shattering'' on this scale can be found in~\cite{AchlioptasCoja-Oghlan08}.

Finally at large $k$ the limit of local stability is irrelevant as it occurs on the scale $\frac{2^{2k-2}}{k}$ (see Eq.~\ref{eq_lstab}), much after the satisfiability transition.

One thus sees at large $k$ a clear separation of scales with three main regimes: for $l \lesssim 2^{k-1} \ln k$ the problem is satisfiable and unclustered (RS). For $2^{k-1} \ln k \lesssim l \lesssim 2^{k-1} k \ln 2$ the model is satisfiable, clustered and rigid (for the typical solutions) but non-condensed, hence the simple RS prediction for the entropy of solutions is correct, while for $2^{k-1} k \ln 2 \lesssim l$ typical instances are unsatisfiable. In this limit the regimes of clustering without frozen variables ($\ld < l < \lr $) and of condensation ($\lc < l < \lsat$) have a negligible relative width. However for small values of $k$ the various thresholds are relatively close to each other (a phenomenon amplified here by the discrete character of the parameter $l$), and their generic ordering exhibited for large enough values of $k$ is not respected. For instance when $k=3$ there is no value of $l$ with a clustered uncondensed phase, and for $k=3,4,5$ the rigidity transition computed at $m=1$ occurs after the condensation $\lc$, hence the value of $\lr$ is not relevant in this case and should be replaced by the threshold for the appearance of hard fields in the solution of the 1RSB equation at the value $m_{\rm s}$ of the Parisi parameter. To summarize this discussion, the generic behavior of the problem appears for $k \ge 6$, this is the reason why in the following we shall often use $k=6$ when discussing our results.

\begin{table}
\begin{tabular}{|c||c|c|c||c|c|c||c|c||c|c|c|c||c|c|c||}
\hline
& \multicolumn{3}{c||}{RS} & \multicolumn{3}{c||}{1RSB, $m=1$} & \multicolumn{2}{c||}{1RSB, $m=0$} & \multicolumn{4}{c||}{unfrozen} & \multicolumn{3}{c||}{locked}\\
\hline
$k$ & $l_{s=0}$ & $\lmod$ & $\lstab$ & $\ld$ & $\lr$  & $\lc$ & $\lsp$ & $\lsat$ & $l_1$  & $l_2$     & $l_5$   & $l_\infty$ & $\llm$ & $\llu$ & $\llp$ \\
\hline
3   & 6.228    & 1.5     & 4.5      & 5     & 6.673  & 5     & 5.124  & 5.742   &        &          &         &           & 3.231 & 4.636 & 5.166 \\
\hline
4   & 19.76    & 2.333   & 16.333   & 17    & 20.64  & 17    & 15.43  & 18.89   &        &          &         &           & 6.942 & 14.004 & 18.050 \\
\hline
5   & 52.70    & 3.75    & 56.25    & 47    & 51.45  & 51    & 38.78  & 51.50   & 52.656 & 52.669   & 52.679  & 52.687    & 13.88 & 35.55 & 50.40 \\
\hline
6   & 129.99   & 6.2     & 192.2    & 108   & 117.16 & 128   & 89.65  & 128.50  & 128.438 & 128.875 & 129.193 & 129.467   & 27.17 & 82.83 & 127.13 \\
\hline
7   & 307.10   & 10.5    & 661.5    &       & 255.10 &       & 198.13 & 305.34  & 297.582 & 300.079 & 301.943 & 303.563   & 52.96 & 184.08 & 303.68 \\
\hline
8   & 706.00   & 18.14   & 2304.14  &       & 541.99 &       & 426.54 & 703.99  & 668.272 & 677.633 & 684.772 & 691.040   & 103.44 & 397.91 & 702.06 \\
\hline
\end{tabular}
\caption{Various thresholds of the bicoloring problem on $l+1$-uniform $k$-regular hypergraphs. The 1RSB ones were first derived in~\cite{DallAstaRamezanpour08} (there is a discrepancy for some values of $\ld$ and $\lc$ that comes from the analytical but approximate expressions used in~\cite{DallAstaRamezanpour08}), and $\lsat$ was rigorously proven in~\cite{DiSlSu13_naeksat} to be the satisfiability threshold for the NAE-$k$-SAT problem (for large enough $k$). We stated as real numbers the thresholds that are obtained analytically as a function of $l$ even if the model studied is defined only for integer $l$, see the main text for the details of the rounding needed for the interpretation of these real values. The last seven columns constitute the main results of this paper on unfrozen and locked solutions, see Sec.~\ref{sec_main_results} for the precise definitions of these thresholds; the ones concerning unfrozen solutions are not defined for $k=3,4$ because $\lr >l_{s=0}$ in these cases.}
\label{table_typical}
\end{table}

\subsection{From Warning Propagation to the whitening dynamics}
\label{sec_def_whitening}

We would like now to define precisely the set $F(\us,G) \subset \{1,\dots,N\}$ of ``the frozen variables of the solution $\us$ of the hypergraph bicoloring problem defined on $G$''. A possible definition for $i$ to belong to this set is that $\t_i = \s_i$ in all the solutions $\utau$ of the cluster to which $\us$ belongs; but this moves the problem to the definition of the cluster containing $\us$. Taking for the latter the set described by the marginal probabilities computed from the fixed point of the BP equations reached by iterations initialized in $\us$ (provided these iterations converge), one sees that $i$ will be frozen in the cluster of $\us$ if in this BP fixed point the variable $i$ receives an ``hard message'' forcing it to $1$ or $-1$. To determine the occurrence of this property one can thus use instead the Warning Propagation equations, obtained by projecting the BP equations on the hard messages (note that these two procedures are equivalent only if ``quasi-hard fields'' do not grow under the BP equations). Let us rewrite the WP equations of Eqs.~(\ref{eq_WP_1},\ref{eq_WP_2}), adapting them to the present setting of an initialization in a solution $\us$. We shall thus consider time dependent messages $\mfh_{i \to a}^t$ and $\mfu_{a \to i}^t$ on each directed edge of the factor graph, with $t=0,1,\dots$ a discrete time index. These messages take values in $\{-1,0,1\}$, they are initialized at $t=0$ in the solution $\us \in \S(G)$ in the sense that
\beq
\mfh_{i \to a}^0 = \s_i \ ,
\eeq
then evolve at later times according to
\beq
\mfu_{a \to i}^{t+1} = \begin{cases} 
1  & \text{if} \ \forall j \in \dami \ \ \ \mfh_{j \to a}^t = -1 \\
-1 & \text{if} \ \forall j \in \dami \ \ \ \mfh_{j \to a}^t = 1 \\
0 & \text{otherwise}
\end{cases} \ , \qquad
\mfh_{i \to a}^{t+1} = \begin{cases} 
1  & \text{if} \ \exists b \in \dima \ \text{with} \ \ \mfu_{b\to i}^{t+1}=1  \\
-1 & \text{if} \ \exists b \in \dima \ \text{with} \ \ \mfu_{b\to i}^{t+1}=-1  \\
0 & \text{otherwise}
\end{cases} \ .
\label{eq_WP_3}
\eeq
One can show that this dynamics is monotonous in time (the only transitions allowed for the messages $\mfh_{i \to a}^t$ are from their initial value $\s_i=\pm 1$ to $0$), and never produce contradictions (i.e. situations with messages $\mfu_{a\to i}^t=+1$ and $\mfu_{b\to i}^t=-1$ sent to the same variable), which allowed us to slightly simplify the rules of (\ref{eq_WP_3}) with respect to (\ref{eq_WP_1}). As a consequence the dynamics converges as $t\to\infty$ (and the fixed point reached in this limit is independent from the precise update scheme, the parallel one used here can be replaced by any sequential one), we shall denote $\{\mfh_{i \to a}^*,\mfu_{a\to i}^*\}$ the limit of these messages, keeping implicit their dependency on the initial solution $\us$. A variable $i$ will then be declared frozen in $\us$ by WP if at least one of the interactions $a\in\di$ sends a message $\mfu_{a\to i}^*=\s_i$ in the fixed point. To formalize this let us translate the evolution of the messages under WP into an evolution of configurations, defining $\us^{t,{\rm WP}} \in\{-1,+1,0\}^N $ a sequence in the extended configuration space with an additional state 0, interpreted as a white, or joker, or free state, defining $\us^{0,{\rm WP}}=\us$ initially and by the following rule for $t\ge 1$:
\beq
\s_i^{t,{\rm WP}} = \begin{cases} 
1  & \text{if} \ \exists a \in \di \ \text{with} \ \ \mfu_{a\to i}^t=1  \\
-1 & \text{if} \ \exists a \in \di \ \text{with} \ \ \mfu_{a\to i}^t=-1  \\
0 & \text{otherwise}
\end{cases} \ .
\eeq
The properties of the dynamics on the messages imply that $\us^{t,{\rm WP}}$ is monotonous in the same sense as above, and converges to $\us^{*,{\rm WP}}$, we thus declare frozen the variables $i$ that have not been set to 0 along this dynamics: $F(\us,G) = \{ i : \s_i^{*,{\rm WP}}=\s_i \}$. Note that this procedure can be applied to any CSP, see~\cite{BraunsteinMezard02} for a generic construction of the WP equations.

One can wonder at this point whether an equivalent dynamics can be defined directly in the extended configuration space $\{-1,1,0\}^N$, without reference to the WP messages. Intuitively, one would start from a solution $\us=\us^0 \in \S(G)$, and iteratively ``whiten'' (i.e. set to 0) the variables $i$ that are unconstrained in the current configuration $\us^t$, thus generating the next configuration $\us^{t+1}$. A variable $i$ such that $\s_i^t \neq 0$ will be declared unconstrained if $C \cap \S(G) \neq \emptyset$, where
\beq
C = \{ \utau \in \{-1,+1\}^N \ : \ \forall \, j\neq i \ \s^t_j \neq 0 \Rightarrow \t_j=\s_j \ , \t_i = -\s_i^t  \}
\eeq
is the subcube of the original configuration space obtained by replacing the white variables of $\us^t$ by all possible values, and $\s_i^t$ by $-\s_i^t$. In other words a variable is declared unconstrained if the partial configuration obtained by flipping it and keeping the other non white variables constant can be extended to a solution by a proper replacement of the white variables. This dynamics can be seen as the growth of a subcube in the original configuration space $\{-1,1\}^N$, which starts as a single point $\us$, extends in the $i$-th direction if $\S(G)$ has at least one configuration in the current subcube flipped along that direction, and stops when all configurations at Hamming distance 1 from the current subcube are not in $\S(G)$, or when it has filled the whole configuration space. Note that this process can be applied to any subset $\S(G)$ of the configuration space, be it defined via a local graphical model or not.

Let us denote $\us^{t,{\rm whitening}}$ the sequence of extended configurations obtained by this coarsening procedure started from a solution $\us$; if the set $\S(G)$ corresponds to the proper bicolorings of a hypergraph $G$ a moment of thought reveals that the dynamics proceeds according to:
\beq
\s_i^{t,{\rm whitening}} = \begin{cases} \s_i & \text{iff} \ \exists a \in \di \ \text{such that} \ \forall j \in \dami \ \ \s_j^{t-1,{\rm whitening}}=-\s_i
\\
0 & \text{otherwise}
\end{cases}
\label{eq_def_whitening}
\eeq
Indeed an hyperedge constrains the value of one of its spin if and only if it is the only representative of its color (i.e. all its other spins are in the opposite value), and a spin is constrained as long as one of its neighboring hyperedge constrains it. This dynamics was used in~\cite{DiSlSu13_naeksat} (under the name of coarsening) as part of the proof of the satisfiability threshold for regular NAE-$k$-SAT.

It can be checked by recurrence on $t$ that for the hypergraph bicoloring problem the two procedures described above are equivalent, namely $\us^{t,{\rm WP}}=\us^{t,{\rm whitening}}$ for all $t$, in the rest of the article we will thus denote more simply $\us^t$ their common value. Let us emphasize that this equivalence is not a generic property of all CSPs. A simple counter example is provided by the XORSAT case, for which the outcomes of WP is independent of the initial configuration (because of the linear structure of its set of solutions), and equivalent to the leaf removal algorithm~\cite{CoccoDubois03,MezardRicci03}, the frozen variables forming the backbone of the hypergraph (i.e. its 2-core and the additional variables that the core logically implies). On the other hand all (non-isolated) variables would be declared frozen by the whitening algorithm: for XORSAT the Hamming distance between two solutions is strictly greater than 1 (excluding the trivial spin flips of variables of degree zero), hence the whitening is stuck in the initial configuration. It was also noted in~\cite{Parisi02b} that the equivalence does not hold for the $q$-coloring problem (the WP dynamics was called directional whitening there). We have not attempted a complete classification of the models where the two dynamics are equivalent or not; we can nevertheless state a few cases. Besides the hypergraph bicoloring (or NAE-$k$-SAT) already mentioned the equivalence also holds for $k$-SAT (in both models there can be at most one forcing message around one clause). We also considered the class of occupation models~\cite{ZdeborovaMezard08b}, where boolean variables interact via clauses involving $k$ variables, the configurations satisfying the constraint being encoded in a boolean word of $k+1$ bits, the $j$-th bit being 1 (resp. 0) if configurations with exactly $j$ out of the $k$ variables around the clause equal to 1 are satisfying (resp. unsatisfying). The equivalence between the WP and whitening dynamics holds if in this word the positions of the 1 (satisfying weights) form a contiguous interval. In so-called locked problems (no two contiguous 1 in the defining word and minimal degree of the variables at least 2) the equivalence also holds but in a rather trivial way: both the WP and the whitening dynamics are stuck in the initial configuration, all variables are declared frozen by both procedures.

\subsection{The whitening dynamics for typical solutions of the hypergraph bicoloring problem}
\label{sec_whitening_typical}

Having justified the origin of the whitening dynamics (\ref{eq_def_whitening}) we start now the study of its behavior. Thanks to the monotonicity of the dynamics a time trajectory $\s_i^t$ can be unambiguously parametrized by its initial value and the time it becomes zero (if it does so). We thus define the whitening time of variable $i$ starting from the initial solution $\us$ as $t_i(\us) = \inf \{t \, : \, \s_i^t =0 \}$; by definition $t_i(\us) \ge 1$, and if $\s_i^t =\s_i$ for all times we set $t_i(\us)=+\infty$: this is the case when the $i$-th variable is frozen in $\us$.

A compact summary of the global evolution of the whitening process is encoded in the empirical cumulative distribution of the whitening times, that for a given (satisfiable) hypergraph $G$ and one of its proper bicolorings $\us$ we define as
\beq
P_t(\us , G) = \frac{1}{N} \sum_{i=1}^N \ind (t_i(\us) \ge t+1) \ ,
\eeq
the $+1$ in the definition being for later notational convenience. By its definition $P_t$ is decreasing in $t$, has $P_0=1$, and its limit as $t\to\infty$ gives the fraction of frozen variables under the whitening starting from $\us$ (that can be interpreted as the fraction of frozen variables in the cluster of $\us$). A solution $\us \in \S(G)$ is called unfrozen if $P_t(\us,G) \to 0$ as $t \to \infty$.

One expects self-averaging properties of $P_t(\us , G)$, in the thermodynamic limit $N\to \infty$, when $G$ is chosen uniformly at random in the ensemble of $l+1$-regular $k$-uniform hypergraphs, and $\us$ is an uniformly chosen proper bicoloring of $G$ (in the uncondensed regime $l < l_{\rm c}(k)$). Let us denote $\Ptypt(k,l)$ this average value around which $P_t$ will concentrate:
\beq
\Ptypt(k,l)= \lim_{N \to \infty} \E_G [ \E_\us [  P_t(\us , G) ] ] \ ,
\eeq
the average being uniform over $\us$ and $G$ as explained above.

\begin{figure}
\includegraphics[width=8cm]{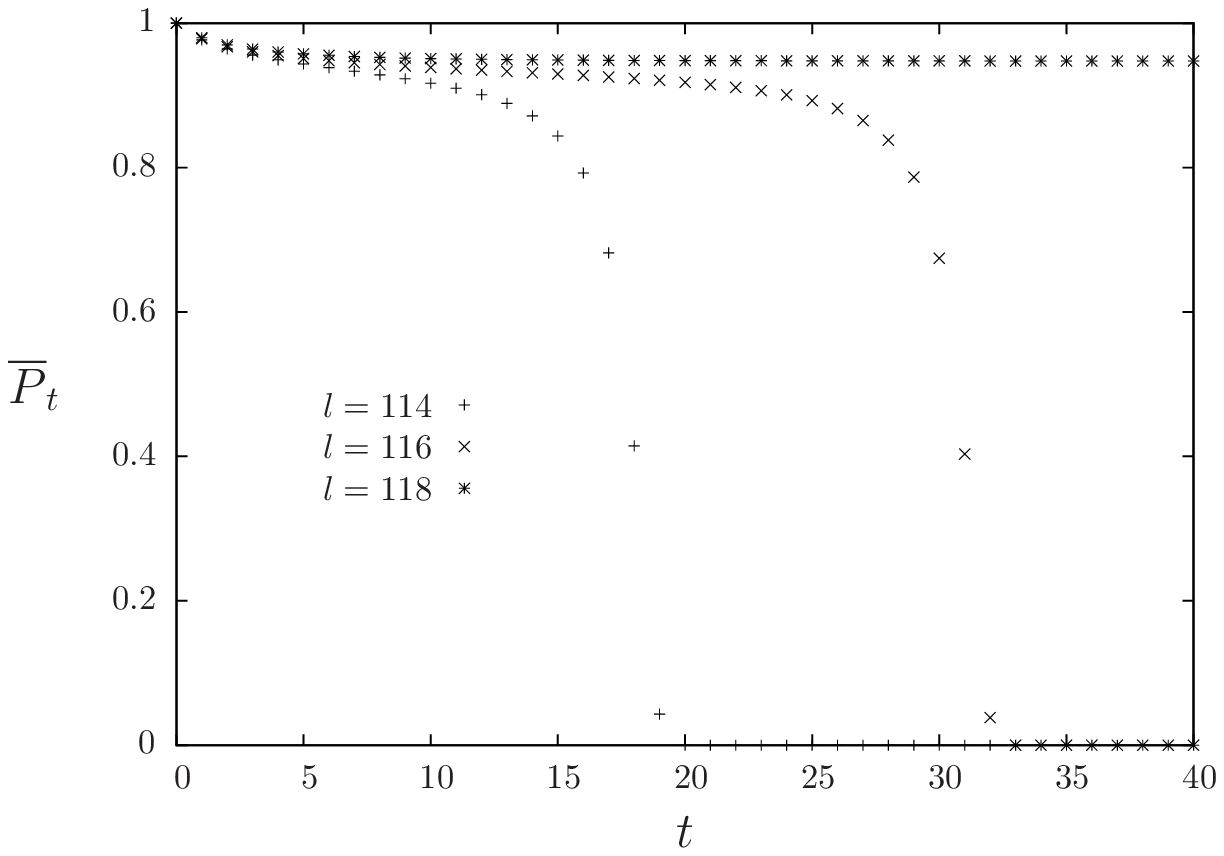}
\hspace{1cm}
\includegraphics[width=8cm]{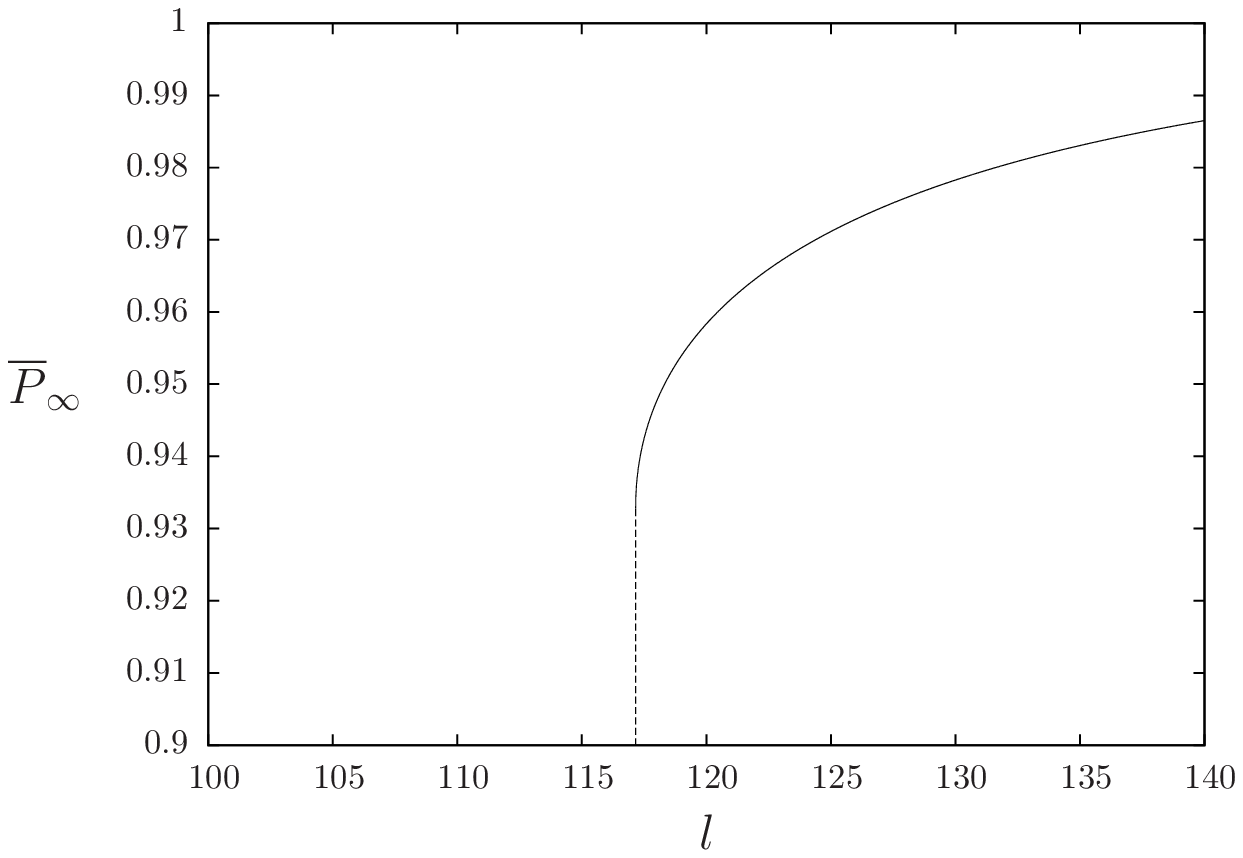}
\caption{Left panel: $\Ptyp_t$ as a function of $t$ for $k=6$, and from bottom to top, $l= 114 < \lr$, $l= 116 < \lr$ and $l= 118 > \lr$. The length of the plateau, i.e. the time to whiten most of the variables, diverges as $(\lr - l)^{-1/2}$. Right panel: the large time limit $\Ptyp_\infty = \underset{t \to \infty}{\lim} P_t$ as a function of $l$ for $k=6$. $\Ptyp_\infty$ is equal to zero for $l<\lr$, jumps discontinuously from $0$ to a strictly positive value at $\lr$, and it has a square root singularity when $l \to \lr^+$. 
}
\label{fig_Ptyp}
\end{figure}

The computation of this typical characterization of the whitening dynamics is very simple in the uncondensed regime, as we shall explain now. Note first that the state $\s_i^t$ of the variable $i$ at time $t$ depends on $\us$ only through variables at graph distance smaller than $t$ from $i$: from the evolution rule (\ref{eq_def_whitening}) $\s_i^t$ is determined by its neighbors at the previous time, and this observation can be used recursively backwards in time. In addition we have already stated the local convergence of $G$ towards a regular tree (see Fig.~\ref{fig_fg_and_tree}), hence to compute $\Ptypt$ for a fixed value of $t$, thanks to the thermodynamic limit in its definition we can safely assume that the neighborhood of $i$ that determines $\s_i^t$ is a tree. Finally the cavity method describes the marginal probability law of an uniformly drawn proper bicoloring $\us$ for a finite set of adjacent variables. The latter takes a simple form in the uncondensed regime (the RS description being valid), that is even simpler in the model under study because of its regular local structure and of the $\us \leftrightarrow -\us$ unbroken symmetry. Indeed an uniform proper bicoloring can be locally constructed as follows: the value of the root variable $\s_i$ is set to $+1$ or $-1$ with equal probability $1/2$. Then independently for each of the $l+1$ hyperedges around $i$ one chooses the value of the $k-1$ other variables uniformly at random among the $2^{k-1}-1$ configurations that do not violate the bicoloring constraint. Each of these $(l+1)(k-1)$ variables are in turn taken independently as the roots of subtrees on which this ``broadcasting generation'' continues further away from $i$. Combining the previous observations one sees that $\Ptypt$ is the probability that the root variable $\s_i$ of a regular tree has not been whitened after $t$ steps of the whitening with an initial configuration generated in this broadcasting way. Calling $\tP_t$ the same quantity on a modified tree where each variable (including the root) has $l$ descendants, one has, keeping understood the dependency on $k,l$ of $\Ptyp$ and $\tP$ for simplicity:
\beq
\tP_t = 1- \left(1 - \frac{\tP_{t-1}^{k-1}}{2^{k-1}-1} \right)^l \ , \qquad
\Ptypt = 1- \left(1 - \frac{\tP_{t-1}^{k-1}}{2^{k-1}-1} \right)^{l+1} \ ,
\qquad \tP_0=\Ptypz=1 \ . 
\label{eq_Ptyp}
\eeq
Indeed the root is frozen at time $t$ in its initial value $\s$ if at the previous time at least one of the hyperedges around it still blocks it, which can happen only if the $k-1$ other variables of this hyperedge were drawn in the initial configuration to $-\s$, and if none of them has been whitened before time $t-1$. The minor difference between $\tP$ and $\Ptyp$ compensates the difference in the number of offsprings between the root and the other variables in a regular tree. In the left panel of Fig.~\ref{fig_Ptyp} we represented the sequences $\Ptypt$ computed in this way. When $l$ increases a plateau develops in this cumulative distribution (for $k \ge 3$), which means that the distribution of whitening times becomes bimodal with a finite fraction (the height of the plateau) of variables having larger and larger whitening times. When $l$ goes above a threshold this plateau extends to infinity, the corresponding fraction of indefinitely frozen variables is plotted as a function of $l$ on the right panel of Fig.~\ref{fig_Ptyp}. This function jumps discontinuously from 0 to a strictly positive value at the threshold, and has a square root singularity above it. Comparing the equations (\ref{eq_rigidity_typ}) and (\ref{eq_Ptyp}) one easily realizes that this threshold is precisely the same as the one for the apparition of hard fields in the 1RSB equations at $m=1$, denoted $\lr$. Finally the length of the plateau in $\Ptyp$, namely the scale of the large whitening times, diverges when $l \to \lr^-$ as $(\lr - l)^{-1/2}$, such an exponent being largely universal for bifurcations in dynamical systems of the form $x_{t+1}=f(x_t)$.

\subsection{Large deviations of the whitening dynamics}
\label{sec_ld}

Even though the whitening profile $P_t(\us,G)$ concentrates in the thermodynamic limit around the typical value $\Ptypt$ we have just computed for uniformly chosen hypergraphs $G$ and solutions $\us$, there are fluctuations, arising from both choices of $G$ and $\us$, and the goal of this paper is precisely to characterize the atypical solutions $\us$ which induce large deviations of the behavior of the whitening process from the typical one. We shall completely leave aside the fluctuations arising from atypical hypergraphs $G$, in other words we will always take uniform averages over this source of randomness (for statistical mechanics studies of large deviations induced by the choice of the graph see~\cite{EngelMonasson04,Rivoire04}).

A convenient characterization of the large deviations for a given hypergraph $G$ is provided by the entropy of solutions yielding a given path $P=\{P_t\}_{t\in \mathbb{N}}$ for the empirical cumulative distribution of the whitening times,
\beq
s(P,G) = \frac{1}{N}\ln \left( 
\sum_{\us \in \S(G)} \ind( P_t(\us,G) = P_t \ \forall \, t \in \mathbb{N} ) \right) \ .
\label{eq_def_s_path}
\eeq
This entropy is formally equal to $-\infty$ for impossible paths, in particular $P_t$ should be decreasing with $t$ and verify $P_0=1$ to yield a non-trivial entropy (as we shall see in more details later on there are also more subtle constraints on the allowed whitening paths: for hypergraphs with good expansion properties the fraction of frozen variables in the final state of the whitening process is either zero or larger than a strictly positive lowerbound~\cite{ManevaMossel05}). As announced above we shall consider typical hypergraphs and define the average path entropy as
\beq
s(P,k,l) = \lim_{N\to \infty} \E_G \left[ s(P,G) \right] \ ;
\eeq
when there is no risk of confusion we will keep understood the $(k,l)$ dependency of this and other associated quantities. Note that the maximum of $s(P,k,l)$ is reached for $P=\Ptyp$, the typical cumulative distribution of whitening times, and at this maximum one has $s(\Ptyp,k,l)=s(k,l)$, the total entropy for all solutions. The difference $s(\Ptyp,k,l) - s(P,k,l)$ gives the rate of large deviation for the probability (i.e. minus the logarithm of the probability divided by $N$), with respect to an uniform choice of $\us$, to observe an atypical profile $P \neq \Ptyp$, this large deviation function being itself averaged uniformly over $G$. Obviously we do not aim here at a precise mathematical definition of a large deviation principle; the first step in this direction would be to relax the strict equality $P_t(\us,G) = P_t$ at all times in Eq.~(\ref{eq_def_s_path}) to $||P(\us,G)-P||\le \delta$ for a well-chosen norm on the space of $P$, and for a tolerance $\delta$ that would be sent to 0 only after the thermodynamic limit $N \to \infty$. 

The large deviation problem defined above is infinite dimensional, as the times at which $P_t(\us,G)$ are constrained are unbounded (of course for a finite hypergraph on $N$ vertices the whitening dynamics is stationary after at most $N$ steps, but we are ultimately interested in the thermodynamic limit). It is thus technically simpler to fix a finite time-horizon $T$ and condition the trajectory $P_t(\us,G)$ only for $t \le T$. Going even further in the simplification we shall introduce an entropy for the solutions that have a given fraction $\theta$ of still frozen variables after $T$ steps of the whitening algorithm, namely
\beq
s(T,\theta,G) = \frac{1}{N}\ln \left( 
\sum_{\us \in \S(G)} \ind(P_T(\us,G) = \theta ) \right) \ , \qquad
s(T,\theta,k,l) = \lim_{N\to \infty} \E_G \left[ s(T,\theta,G)
\right] \ ,
\label{eq_def_s_T}
\eeq
the second quantity being the typical value averaged uniformly over the graph ensemble. As explained above the maximum over $\theta$ of this entropy is reached for $\theta=\Ptyp_T$, where it equals the total entropy, the difference between the two yielding the large deviation rate for the fluctuations of $P_T$ away from its typical value. Imposing an atypical constraint $\theta \neq \Ptyp_T$ at a single time $T$ forces the system to select an atypical initial configuration $\us$ and hence an atypical trajectory $P_t \neq \Ptypt$ at all times; however for entropic reasons only the ``less atypical ones'' (i.e. the most numerous ones) fulfilling the constraint $P_t=\theta$ will contribute dominantly. We define this dominant atypical path under the constraint at time $T$ as the following average,
\beq
P_t(T,\theta,k,l) = \lim_{N\to \infty} \E_G \left[ 
\frac{\underset{\us \in \S(G)}{\sum} P_t(\us,G) \ind(P_T(\us,G) = \theta)}
{\underset{\us \in \S(G)}{\sum} \ind(P_T(\us,G) = \theta)}
\right] \ ,
\label{eq_Pt_theta}
\eeq
expecting a self-averaging phenomenon in the thermodynamic limit. By general large deviation principles it should be related to the path entropy as
\beq
P(T,\theta,k,l) = \underset{P \, : \, P_T=\theta}{\text{argmax}}\ s(P,k,l) \ .
\eeq

Let us also define the entropy of the configurations which have a fraction $\theta$ of frozen variables in their fixed point under the whitening process,
\beq
s_\infty(\theta,k,l) = \lim_{N\to \infty} \frac{1}{N} \E_G \left[ \ln \left( 
\sum_{\us \in \S(G)} \ind\left(\frac{1}{N} \sum_{i=1}^N \delta_{\s_i,\s_i^*}  = \theta \right) \right) \right] \ .
\label{eq_s_infty}
\eeq
This quantity corresponds essentially to the limit $T\to\infty$ of the entropy $s(T,\theta,k,l)$, except that the large $T$ and large $N$ limits are taken in reversed order; we will discuss the (non)-commutativity of these limits in Sec.~\ref{sec_res_Tfinite}.

In statistical mechanics jargon the entropies defined above are microcanonical ones, as we impose strictly the value of an observable (either the path $P$ at all times or its value at time $T$). It is often easier to compute them indirectly by considering the associated canonical free-energies where the strict constraints are turned into soft ones, with a thermodynamic force conjugated to the observable (here one can picture it intuitively as ``pulling'' up or down $P_t$ from its typical value), the two constructions being related in the thermodynamic limit by Legendre transforms (this is obviously a common pattern also in the more mathematical large deviation theory). We shall thus introduce the canonical potential
\beq
\phi(\varepsilon,G) = \frac{1}{N}
\ln \left( \sum_{\us \in \S(G)} e^{N \underset{t=1}{\overset{\infty}{\sum}} \epsilon_t P_t(\us,G)} \right) \ ,
\label{eq_phi_pathwise}
\eeq
where the argument is a time-dependent force, $\varepsilon=\{\epsilon_t \}_{t \in \mathbb{N}}$. This potential is the Legendre transform of the path-wise entropy of Eq.~(\ref{eq_def_s_path}):
\beq
\phi(\varepsilon,G) = \sup_{P} \left[
s(P,G) + \sum_{t=1}^\infty \epsilon_t P_t \right] \ .
\eeq
The parameters $\epsilon_t>0$ favor ``more frozen'' configurations, i.e. the ones which lead to extended configurations with fewer joker states than the typical ones, for $\epsilon_t < 0$ it favors ``less frozen'' configurations, in terms of the state of the whitening after $t$ steps. When $\varepsilon=0$ the potential $\phi$ coincides with the total entropy of all solutions.

A special case of this potential is obtained by exerting the force only at time $T$, i.e. taking $\epsilon_t=0$ for $t\neq T$ and denoting more simply $\epsilon = \epsilon_T$; in this case we shall write
\beq
\phi(T,\epsilon,G) = \frac{1}{N} 
\ln \left( \sum_{\us \in \S(G)} e^{N \epsilon P_T(\us,G)} \right)  \ ,
\label{eq_phi_T}
\eeq
which is the Legendre transform of the one-time constrained entropy of Eq.~(\ref{eq_def_s_T}):
\beq
\phi(T,\epsilon,G) =  \sup_\theta [s(T,\theta,G) + \epsilon \, \theta ] \ .
\eeq
Once this function of $\epsilon$ is known one can reconstruct $s(\theta)$ (or more precisely its concave hull) by an inverse Legendre transform, with $\epsilon$ corresponding to (minus) the slope of $s(\theta)$ at the conjugated point. If $s(\theta)$ is non-concave a first-order phase transition will occur in the canonical ensemble. 

As above we denote $\phi(\varepsilon,k,l)$ and $\phi(\epsilon,T,k,l)$ the uniform averages over $l+1$-uniform $k$-regular hypergraphs $G$ of $\phi(\varepsilon,G)$ and $\phi(\epsilon,T,G)$, respectively:
\beq
\phi(\varepsilon,k,l) = \lim_{N\to \infty} \E_G \left[ \phi(\varepsilon,G) \right] \ , \qquad
\phi(\epsilon,T,k,l) = \lim_{N\to \infty} \E_G \left[ \phi(\epsilon,T,G) \right] \ ;
\label{eq_phi_average}
\eeq
for completeness we state here the Legendre transform relationships between microcanonical and canonical graph-averaged quantities:
\beq
\phi(\varepsilon,k,l) = \sup_{P} \left[
s(P,k,l) + \sum_{t=1}^\infty \epsilon_t P_t \right] \ , \qquad
\phi(T,\epsilon,k,l) =  \sup_\theta [s(T,\theta,k,l) + \epsilon \, \theta ] \ .
\eeq
Finally we introduce the canonical equivalent of the dominant path with a constraint at time $T$ defined in (\ref{eq_Pt_theta}):
\beq
P_t(T,\epsilon,k,l) = \lim_{N\to \infty} \E_G \left[ 
\frac{\underset{\us \in \S(G)}{\sum} P_t(\us,G) e^{N \epsilon P_T(\us,G)}}
{\underset{\us \in \S(G)}{\sum} e^{N \epsilon P_T(\us,G)}}
\right] \ .
\label{eq_def_Pt_epsilon}
\eeq
Thanks to the equivalence of ensembles the microcanonical $P_t(T,\theta)$ and the canonical $P_t(T,\epsilon)$ paths should coincide at all times if $\epsilon$ and $\theta$ are conjugated parameters (i.e. if $P_t(T,\epsilon)=\theta$).

\subsection{Main results}
\label{sec_main_results}

In this Section we shall summarize the main results we have obtained before entering into the details of their derivation. We have set up statistical mechanics computations that allow to obtain quantitative predictions for the large deviations functions defined above, in particular the entropy $s(T,\theta,k,l)$ counting the number of solutions that have a fraction $\theta$ of frozen variables after $T$ steps of the whitening procedure, and also the dominant path $P_t(T,\theta,k,l)$ that gives the fraction of frozen variables after an arbitrary number $t$ of steps, conditioned on $P_T=\theta$; these quantities being defined for typical $l+1$-regular $k$-uniform hypergraphs. They are obtained numerically through the resolution of a set of equations, the number of unknowns and equations being linear in the time horizon $T$. 

\begin{figure}
\includegraphics[width=8cm]{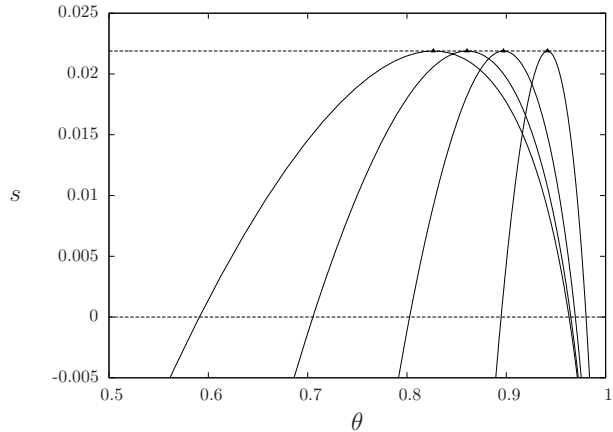}
\caption{The entropy $s(\theta)$ for $k=3$, $l=6$. From right to left $T=1,2,3,4$. The triangles mark the maximum of each curve, at $\theta=\Ptyp_T$, the value corresponding to typical solutions computed from Eq.~(\ref{eq_Ptyp}), the entropy of the maxima being independent of $T$ and equal to the entropy of all solutions (cf. Eq.~(\ref{eq_s_ann})), represented here by the horizontal dashed line.
}
\label{fig_softheta_k3l6}
\end{figure}

We first present in Fig.~\ref{fig_softheta_k3l6} the entropy curves $s(\theta)$, for a given choice of $(k,l)$, and a few values of $T$. Some obvious characteristics of these curves were anticipated above: for each $T$ the maximum of $s(T,\theta)$ is reached in $\theta=\Ptyp_T$, the outcome of the whitening process from typical configurations computed in Sec.~\ref{sec_whitening_typical}, which obviously decreases with $T$, and the value of the maximum itself is independent of $T$, as it corresponds to the entropy counting all solutions. Not all values of $\theta$ yield a positive entropy; as explained above in this RS framework negative entropies have to be interpreted as the impossibility of existence of configurations with such properties.

\begin{figure}
\includegraphics[width=8cm]{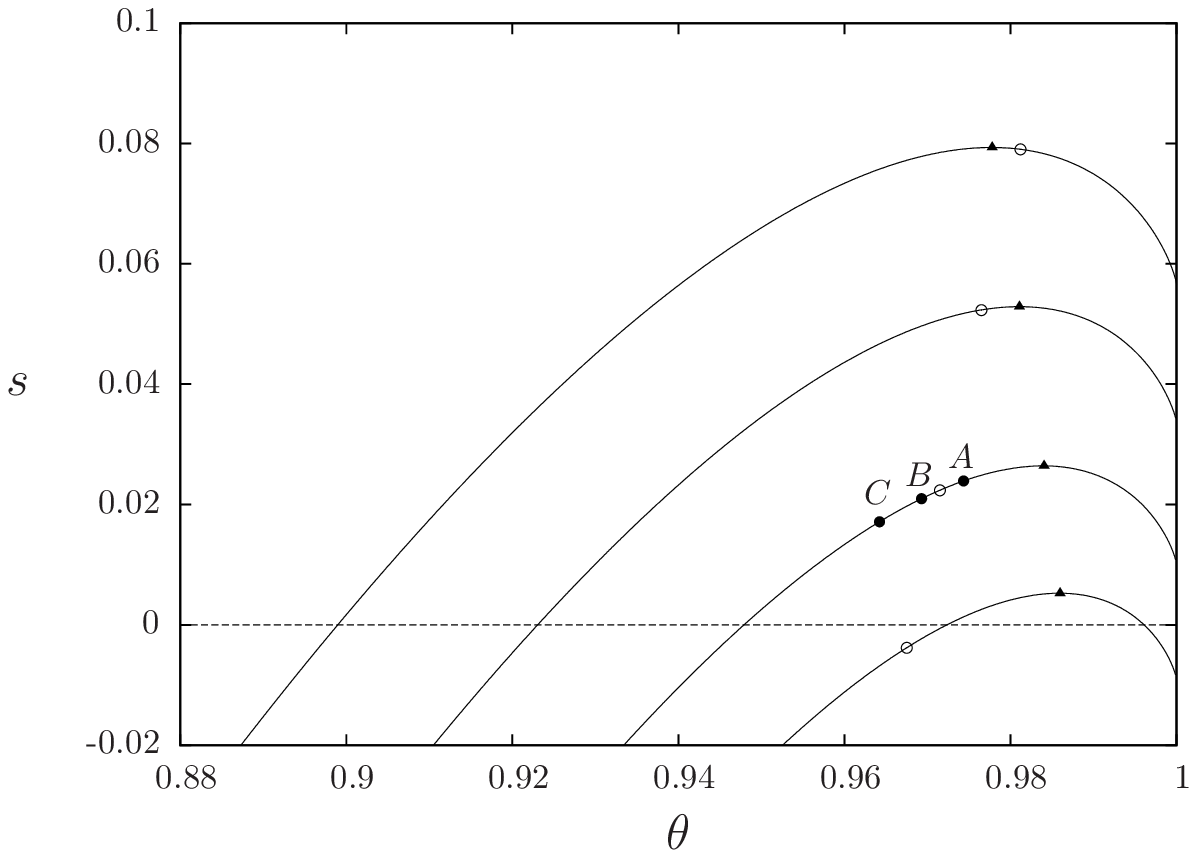}
\hspace{1cm}
\includegraphics[width=8cm]{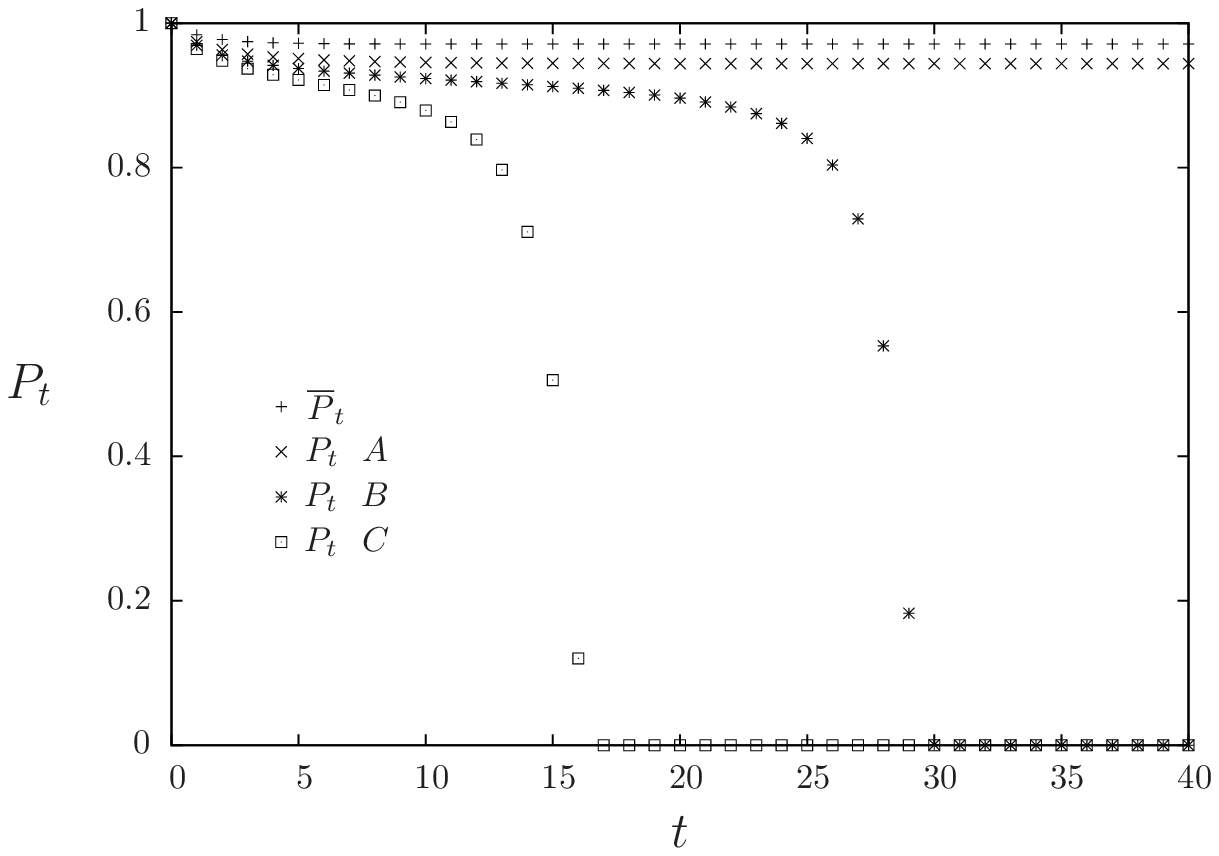}
\caption{Left panel: the entropy $s(\theta)$ for $k=6$, $T=1$, and from top to bottom $l=115$, $l=120$, $l=125$, $l=129$. On each curve the filled triangle is located at the maximum of $s(\theta)$, while the open circle denotes the tipping point (on the right of the maximum for $l<\lr$, on the left for $l>\lr$). The filled circles denote the three points for which the corresponding $P_t$ is displayed on the right panel. Right panel: $P_t$ as a function of $t$ for $k=6$, $l=125$, for the typical solutions and for three values of the bias at $T=1$ corresponding to the points $A,B$ and $C$ of the left panel. The number of steps of the whitening needed for $P_t$ to drop below the plateau diverges as $(\ttip - \theta)^{-1/2}$ when $\theta$ approaches $\ttip$ from below.}
\label{fig_softheta_T1_k6}
\end{figure}

A much finer description of the configurations selected by the constraint $P_T=\theta$ is provided by the computation of $P_t(T,\theta,k,l)$: in particular one can determine whether $P_t(T,\theta,k,l)$ tends to zero as $t\to\infty$, or if it remains strictly positive at all times $t$. In the former case the typical solutions under the constraint $P_T=\theta$ (which are of course atypical in the set of all solutions if $\theta\neq\Ptyp_T$) are unfrozen, in the latter they are frozen. We found that for a given choice of $(T,k,l)$ there is a threshold value for $\theta$, that we shall call ``tipping point'' in the following and denote $\ttip(T,k,l)$, such that for $\theta<\ttip(T,k,l)$ the constraint $P_T=\theta$ selects unfrozen solutions, while for $\theta \ge \ttip(T,k,l)$ the selected solutions are frozen. This is illustrated in Fig.~\ref{fig_softheta_T1_k6} for the simplest case $T=1$, with $k=6$. In the left panel the curves $s(\theta)$ are represented for a few values of $l$, with the location of the tipping point marked as an open circle and the maximum of the curve as a triangle. Obviously for $l<\lr$, the typical rigidity transition, the tipping point is on the right of the maximum, while it is on the left in the more interesting case $l>\lr$ (this statement is true for any value of $T$, not only $T=1$). The right panel of Fig.~\ref{fig_softheta_T1_k6} contains additional information that should help to grasp the tipping point phenomenon intuitively. For a fixed value of $l>\lr$ we plotted the whitening profile $P_t$ for typical configurations and for three values of $\theta<\Ptyp_1$. As $l>\lr$ the typical curve has a positive limit as $t\to\infty$; ``pulling down'' softly $P_{T=1}$ (i.e. exerting a force $\epsilon<0$ but not too large in absolute value in the canonical ensemble) the limit is slightly smaller but still positive (curve $A$), hence an extensive number of variables are still frozen at the end of the whitening. Pulling down stronger one reaches the tipping point at which the limit $P_{t \to \infty}$ discontinuously jumps to 0 (curves $B$ and $C$), hence for $\theta<\ttip$ the configurations selected by this bias are typically unfrozen.

As can be seen on the left panel of Fig.~\ref{fig_softheta_T1_k6} the entropy of the tipping point $s(T,\ttip(k,l),k,l)$ becomes negative when the density of constraint $l$ is increased. We thus define a threshold $l_T(k)$ from this criterion of entropy cancellation; within the RS framework this threshold is interpreted as the disappearance of unfrozen solutions in the configurations selected by a constraint imposed after $T$ steps of the whitening procedure. By definition one has $l_T(k) \in [\lr(k),l_{s=0}(k)]$, and moreover $l_T(k)$ is growing with $T$: imposing a constraint on the value of $P_T$ leaves the system the freedom to reorganize the path followed by $P_t$ for $t=1,2,\dots,T-1$, hence the later the bias is exerted the larger the entropy of the unfrozen solutions can be. We computed explicitly the values of $l_T$ for small values of $k$ and a few choices of $T$ (cf. Table~\ref{table_typical}); we also performed large $k$ asymptotic expansion of these thresholds, which yields for $T=1$
\beq
l_1(k) = \frac{ 2^{k-1}  k \ln 2}{\ln k} \left(1+ O\left(\frac{\ln \ln k}{\ln k} \right)\right) \ ,
\label{eq_l1_largek}
\eeq
and more generically for $1 \le T < \infty$,
\beq
l_T(k) = \frac{ 2^{k-1}  k \ln 2}{\ln^{\circ T}(k)} \left(1+ O\left(\frac{\ln^{\circ (T+1)}(k)}{\ln^{\circ T}(k)} \right)\right) \ ,
\label{eq_lT_largek}
\eeq
where $\ln^{\circ T}$ is the $T$-times iterated logarithmic function, i.e. $\ln^{\circ 1}(k) = \ln k$ and $\ln^{\circ (T+1)}(k) = \ln(\ln^{\circ T}(k))$.

We emphasize here that the asymptotic scale for $l_1(k)$ is $\lsat(k) / \ln k$, which is ``much'' larger than the typical rigidity $\lr(k)$ (or the asymptotically equivalent dynamic threshold $\ld(k)$ recalled in Sec.~\ref{sec_reminder}) that scales as $\lsat(k)/k$ (forgetting logarithmic corrections): even when the typical solutions are frozen there are still exponentially many unfrozen solutions, and they can be selected by a bias depending only on $P_1$. We also acknowledge here that the scaling of $l_1(k)$ was conjectured prior to our work by Amin Coja-Oghlan~\cite{Amin_private}. Larger (finite) values of $T$ push these asymptotic thresholds even closer to the satisfiability one, iterated logarithms being slower and slower functions of $k$.

We also managed to compute the limit of $l_T(k)$ as $T\to\infty$ (for fixed $k$), denoted $l_\infty(k)$, see Table~\ref{table_typical} for some numerical values at small $k$, as well as the large $k$ asymptotics of $l_\infty$:
\beq
l_\infty(k) = \frac{2^{k-1} k \ln 2}{2} \left(1+ O\left(\frac{(\ln k)^2}{k} \right)\right) \ .
\label{eq_linfty_largek}
\eeq
This threshold, that we interpret as the limit of existence of any unfrozen solution, is thus asymptotically $\lsat(k)/2$, very deep in the typically clustered and frozen regime, but a constant factor below the condensation and satisfiability threshold (in agreement with a conjecture of Dimitris Achlioptas~\cite{Dimitris_private}). We mentioned in the introduction the coincidence of this scale with a phenomenon studied in~\cite{MoraMezard05} under the name of $x$-satisfiability, i.e. the existence of pairs of solutions at Hamming distance $N x$. For density of constraints larger than $\lsat(k)/2$ it was shown in~\cite{MoraMezard05} that a range of $x$ becomes forbidden, which indicates the clustering of the solution set, in a very strong sense. Note that in general there is no direct connection between the existence of a gap in the possible Hamming distances between pairs of solutions and the existence or not of white solutions. We believe this coincidence of scale should be taken as a further evidence that the structure of clusters in the large $k$ limit simplifies a lot and is well approximated by subcubes.

An equivalent way of defining $l_\infty(k)$ is to consider the hard constraint imposing a complete whitening of all variables after $T$ steps, i.e. requiring $P_T=0$. The vanishing of $s(T,\theta=0,k,l)$ defines a sequence of thresholds $l'_T(k)$ below which such configurations do exist (according to the RS computation), the large $T$ limit of $l_T(k)$ and $l'_T(k)$ coincide, we thus denote $l_\infty(k)$ their common limit. 

In addition to this characterization of the unfrozen solutions our large deviation analysis allowed us to obtain results on the locked solutions of the bicoloring problem, namely the solutions which are fixed points of the whitening dynamics because all their variables are frozen (around each variable $i$ there is at least one clause in which $i$ is the unique representative of its color). Their entropy $s_{\rm l}(k,l)$ is by definition the common value of $s(T,\theta=1,k,l)$, independently of $T$: such solutions do not evolve under the whitening dynamics. We found $s_{\rm l}(k,l)$ to be positive on an interval of densities $[\llm,\llp]$, to be interpreted (in the RS formalism) as the domain of existence of such locked solutions. Moreover the tipping point $\ttip(T,k,l)$ defined above happens to reach 1 at another threshold $\llu \in [\llm,\llp]$ (also independent of $T$); this means that for $l<\llu$ all solutions in which an arbitrarily small but strictly positive fraction of variables are whitened after any number of steps of the whitening algorithm will ultimately whiten completely. In this case the locked solutions, if they exist (i.e. for $l > \llm$), are the unique type of frozen solutions (and coexist of course with the unfrozen ones). On the contrary for $l > \llu$ the frozen solutions can have a non-trivial fraction of frozen variables, in $]0,1[$ strictly. The numerical value of these three thresholds can be found in Table~\ref{table_typical} for small values of $k$, and admit the following asymptotic expansion at large $k$:
\bea
\llm(k) &=& 2^{k-1} \left( \ln 2 + O\left(\frac{1}{k} \right) \right) \ , \label{eq_llm_largek} \\
\llu(k) &=& 2^{k-1} \left( \ln k + \ln \ln k + O\left(\frac{\ln \ln k}{\ln k} \right) \right) \ , \label{eq_llu_largek} \\
\llp(k) &=& 2^{k-1} k \ln 2 - \left(1 + k \, \frac{1+\ln 2}{2} \right) + O \left(\frac{k^3}{2^k} \right)\ . \label{eq_llp_largek}
\eea
Note that $\llp$ is on the satisfiability scale at leading order (compare with Eq.~(\ref{eq_lsat_largek})), $\llu$ on the dynamic/rigidity scale (see Eq.~(\ref{eq_lr_largek})), while the appearance of locked solutions occurs for slightly smaller densities, $\llm \sim \lsat /k$.

Among the additional results to be discussed in the rest of this paper let us mention that we have a prediction for the entropy $s_\infty(\theta)$ (cf. Eq.~(\ref{eq_s_infty})) counting the number of solutions according to their number of frozen variables (in the fixed point of the whitening), as well as their whitening trajectories.

To conclude this section let us underline that the results we just summarized have been obtained using the replica symmetric version of the cavity method, and thus are not expected to be exact for all values of the parameters (in particular the negative entropies have to be preceded by a condensation transition for consistency reasons); a short discussion of the effects of the replica symmetry breaking will be given later on in Sec.~\ref{sec_rsb}. For this reason the threshold $l_\infty(k)$ can only be expected to be a bound on the true freezing transition, namely $l_{\rm f}(k) \le l_\infty(k)$. We expect however the large $k$ asymptotics of $l_\infty$ given in (\ref{eq_linfty_largek}) to give the correct behavior of $l_{\rm f}$, as stated in the introduction.

\section{A statistical mechanics treatment}
\label{sec_statmech}
\subsection{Factor graph representation}
\label{sec_statmech_fg}

As we have explained in Sec.~\ref{sec_ld} our goal of characterizing the solutions of the hypergraph bicoloring problem that have an atypical behavior under the whitening process will be achieved if we determine the generalized free-energy defined in Eq.~(\ref{eq_phi_pathwise}). To perform this computation let us first rewrite the summation in Eq.~(\ref{eq_phi_pathwise}) under a more convenient form:
\beq
Z(\varepsilon,G)=\sum_{\us \in \S(G)} e^{N \underset{t=1}{\overset{\infty}{\sum}} \epsilon_t P_t(\us,G)} = \sum_\us \prod_{a=1}^M w_a(\us_\da) \, e^{\underset{t=1}{\overset{\infty}{\sum}} \epsilon_t \underset{i=1}{\overset{N}{\sum}} \ind(t_i(\us) \ge t+1) } = \sum_\us \prod_{a=1}^M w_a(\us_\da)
\prod_{i=1}^N b(t_i(\us)) \ ,
\eeq
where we recall that $w_a$ is 1 if the $a$-th constraint is satisfied, $0$ otherwise, and we defined:
\beq
b(t) = \exp\left[ \sum_{t'=1}^{t-1} \epsilon_{t'} \right] \ .
\label{eq_epsilon_to_b}
\eeq
We will now introduce a probability measure over the set of solutions $\S(G)$, but at variance with the uniform one defined in (\ref{eq_mu_unbiased}) it will give different weights to the different solutions, this bias being chosen in such a way that $Z(\varepsilon,G)$ becomes the normalizing constant of the measure:
\beq
\mu(\us) = \frac{1}{Z(\varepsilon,G)} \prod_{a=1}^M w_a(\us_\da) \ \prod_{i=1}^N b(t_i(\us)) \ .
\label{eq_mu}
\eeq
The uniform case of (\ref{eq_mu_unbiased}) is recovered by setting $b(t)=1$ for all $t$, i.e. $\varepsilon=0$.

The difficulty in dealing with this measure $\mu$ and in computing $Z(\varepsilon,G)$ comes from the fact that the whitening time $t_i(\us)$ is a complicated and rather implicitly defined function of $\us$, that depends on the variables at an a priori arbitrary distance from $i$. These long-range interactions between the elementary variables $\s_i$ can however be made local in an extended space of variables, with a few steps explained below. This part of our work is very similar to the recent large deviations studies of the bootstrap percolation dynamics, see in particular~\cite{AlBrAsZe13,AlBrAsZe13b,GuSe15}. Let us first emphasize that the whitening times from a given initial configuration $\us$ fulfill the following local equations:
\beq
t_i(\us) = 1 + \underset{a \in \di}{\max}\left[\ind(\s_j = -\s_i \ \forall j \in \dami) \  \underset{j\in\dami}{\min} t_j(\us) \right] \ ,
\label{eq_relti}
\eeq
that follow immediately from the definition of the whitening as a dynamics on $\us^t$ given in (\ref{eq_def_whitening}). Moreover if one considers the set of $N$ equations 
\beq
t_i = 1 + \underset{a \in \di}{\max}\left[\ind(\s_j = -\s_i \ \forall j \in \dami) \  \underset{j\in\dami}{\min} t_j \right] \ ,
\eeq
on the $N$ unknowns $t_i \in \{1,2,\dots,\infty\}$, one can convince oneself that they have for each proper bicoloring $\us$ a single solution, that coincides with $t_i(\us)$. We can thus introduce (redundant) variables $t_i$ on each vertex, their global configuration being denoted $\ut$, and a measure on the new variables $(\us,\ut)$:
\beq
\mu(\us,\ut) = \frac{1}{Z(\varepsilon,G)} \prod_{a=1}^M w_a(\us_\da) \ \prod_{i=1}^N b(t_i) \prod_{i=1}^N \ind\left(t_i = 1 + \underset{a \in \di}{\max}\left[\ind(\s_j = -\s_i \ \forall j \in \dami) \  \underset{j\in\dami}{\min} t_j \right]\right) \ .
\label{eq_mu_st}
\eeq
From the observations made above one concludes that the partition function $Z(\varepsilon,G)$ is the same in (\ref{eq_mu}) and (\ref{eq_mu_st}), and that in the support of $\mu$, i.e. for $(\us,\ut)$ with $\mu(\us,\ut)>0$, the $\ut$ are precisely the whitening times from $\us$. The biased measure on $\us$ originally defined in (\ref{eq_mu}) is thus nothing but the $\us$ marginal of the extended one introduced in (\ref{eq_mu_st}). Note also that the dominant biased path defined in Eq.~(\ref{eq_def_Pt_epsilon}) can be computed with suitable averages under (\ref{eq_mu_st}). The crucial advantage of the formulation in (\ref{eq_mu_st}) with respect to (\ref{eq_mu}) is the locality (in terms of the underlying hypergraph $G$) of the interactions between the variables $\s_i$ and $t_i$. To make this last point more obvious and to reach our final formulation of the problem we shall introduce additional time variables, for each directed edge $a \to i$ from the clause $a$ to the variable $i$ of the hypergraph $G$,
\beq
u_{a \to i}(\us) = \ind(\s_j = -\s_i \ \forall j \in \dami) \  \underset{j\in\dami}{\min} t_j(\us) \ .
\eeq
Their global configuration being denoted $\uu$ we introduce a measure on $(\us,\ut,\uu)$ as
\bea
\mu(\us,\ut,\uu) = \frac{1}{Z(\varepsilon,G)} \prod_{a=1}^M w_a(\us_\da) \ \prod_{i=1}^N b(t_i) 
&& \prod_{i=1}^N \ind(t_i = 1 + \underset{a \in \di}{\max} \, u_{a \to i}) \nonumber \\
&& \prod_{\la a,i \ra} \ind(u_{a \to i} = \ind(\s_j = -\s_i \ \forall j \in \dami) \  \underset{j\in\dami}{\min} t_j ) \ ,
\label{eq_mu_stu}
\eea
where the last product runs over the edges of the bipartite graph representation of $G$. The $t_i$ are in $\{1,2,\dots,\infty\}$ while the $u_{a \to i}$ are in $\{0,1,2,\dots,\infty\}$. Again the $\uu$ are redundant variables, in the sense that the normalization $Z(\varepsilon,G)$ is the same in (\ref{eq_mu}), (\ref{eq_mu_st}) and (\ref{eq_mu_stu}), and the law (\ref{eq_mu}) is the $\us$ marginal of (\ref{eq_mu_stu}).

\begin{figure}
\centerline{\includegraphics[width=13cm]{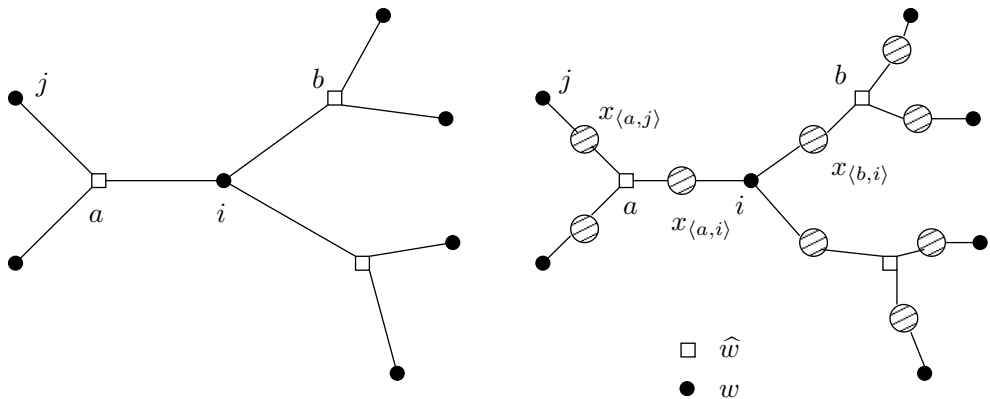}}
\caption{A portion of the bipartite graph representation of a hypergraph $G$ (left panel), and the corresponding factor graph for the measure of Eq.~(\ref{eq_mu_x}) (right panel).}
\label{fig_factor_graph}
\end{figure}

If one puts the variables $\us,\ut,\uu$ at their ``natural'' location, i.e. the $\s_i$ and $t_i$ on node $i$, and $u_{a \to i}$ on the edge $\la a,i \ra$, then the factor graph representing (\ref{eq_mu_stu}) has small loops even if the original hypergraph is a tree. To avoid this problem we shall introduce finally a further redundancy; for each edge $\la a,i \ra$ we define a variable $x_{\la a,i \ra}=(\s_i^a,t_i^a,u_{a\to i})$, the global configuration of all these variables being denoted $\ux$, by copying the original variables $\s_i$ and $t_i$ on each of their neighboring edges $\la a,i \ra$ with $a \in \di$ and setting $\s_i^a=\s_i$, $t_i^a=t_i$. The measure on $\ux$ is defined as
\beq
\mu(\ux) = \frac{1}{Z(\varepsilon,G)} \prod_{a=1}^M \hw(\{x_{\la a,i \ra} \}_{i\in \da})
\prod_{i=1}^N w(\{x_{\la a,i \ra} \}_{a\in \di}) \ ,
\label{eq_mu_x}
\eeq
whose factor graph representation is provided on the right panel of Fig.~\ref{fig_factor_graph}. The variable nodes $x_{\la a,i \ra}$ in this last formulation live on the edges of the original graph $G$, while there are two types of interaction nodes, one on each of the hyperedges $a$ of the original hypergraph, with a weight function  
\beq
\hw(x_1,\dots,x_k) = \ind(\s_1,\dots,\s_k \ \text{n.a.e.}) \prod_{i=1}^k
\ind(u_i = \ind(\s_j = -\s_i \ \forall j \neq i) \, \underset{j \neq i}{\min} \, t_j ) \ ,
\label{eq_def_hw}
\eeq
and one on each of the vertices $i$ of $G$, with the weights
\beq
w(x_1,\dots,x_{l+1}) = \sum_{\s,t} b(t) \left( \prod_{i=1}^{l+1} \ind((\s_i,t_i)=(\s,t)) \right) \, \ind(t=1+\max(u_1,\dots,u_{l+1})) \ .
\label{eq_def_w}
\eeq

To conclude this discussion let us emphasize what we have achieved here: the determination of the potential $\phi(\varepsilon,G)$ of (\ref{eq_phi_pathwise}) has been reduced to the computation of the normalizing constant of the measure $\mu(\ux)$ of Eq.~(\ref{eq_mu_x}), which is a graphical model with more complicated variables than just the Ising spins $\s_i$, but whose interactions have exactly the same topology as the underlying hypergraph $G$. In particular if $G$ is locally tree-like so is the graphical model encoding $\phi(\varepsilon,G)$, hence one can attack this problem with the now standard tools recalled in Sec.~\ref{sec_reminder} (Belief Propagation and the cavity method, possibly including RSB). As these steps are rather technical but conceptually straightforward we explain them in details in Appendix~\ref{app_cavity} and proceed in the main text by presenting the predictions of the cavity method for random regular hypergraphs.

\subsection{The factorized solution on regular hypergraphs}
\label{sec_statmech_regular}

The prediction of the RS cavity method for the free-entropy $\phi(\epsilon,T,k,l)$ defined in Eq.~(\ref{eq_phi_average}) is computed as follows (cf. Appendix \ref{app_cavity} for a complete justification). One has to solve the following set of $4 T$ equations on $4 T$ unknowns denoted $Q_1,\dots,Q_T,R_1,\dots,R_T,\hQ_1,\dots,\hQ_{T-1},\hQ_\infty,\hrho_1,\dots,\hrho_T$, which should be thought of as (cumulative) probability distributions for the whitening times:
\bea
Q_t &=& \sum_{t'=t}^T \left[ (\hrho_{t'} + \hQ_{t'-1})^l - (\hrho_{t'} + \hQ_{t'-2})^l \right] + e^\epsilon  \left[ (\hrho_T + \hQ_\infty)^l - (\hrho_T + \hQ_{T-1})^l \right] \qquad \text{for} \ t \in [1,T] \ ,
\label{eq_reg_Qt}
\\
R_t &=&  (\hrho_{t+1} + \hQ_t)^l + \sum_{t'=t+2}^T \left[ (\hrho_{t'} + \hQ_{t'-1})^l - (\hrho_{t'} + \hQ_{t'-2})^l \right]  + e^\epsilon  \left[ (\hrho_T + \hQ_\infty)^l - (\hrho_T + \hQ_{T-1})^l \right] \ \text{for} \ t \in [1,T-1] \ ,
\nonumber \\
R_T &=& e^\epsilon (\hrho_T + \hQ_\infty)^l \ , \\
\hQ_t &=& \left[Q_1^{k-1} - Q_{t+1}^{k-1} \right] \qquad \text{for} \ t \in [1,T-1] \ , 
\label{eq_reg_hQt} \\
\hQ_\infty &=&  Q_1^{k-1}  \ , \label{eq_reg_hQinfty} \\
\hrho_t &=&  (2^{k-1} - k - 1) Q_1^{k-1} + (k-1)\left[ \sum_{t'=1}^{t-1} R_{t'} (Q_{t'}^{k-2} -Q_{t'+1}^{k-2} ) + R_t Q_t^{k-2}\right]
 \qquad \text{for} \ t \in [1,T] \ . \label{eq_reg_rhot}
\eea
To correctly interpret the terms with $t'=1,2$ in the first line one has to adopt the convention that $\hQ_0=0$ and $\hQ_{-1}=-\hrho_1$. Then the free-entropy is obtained as
\beq
\phi(\epsilon,T,k,l) = \frac{1-l(k-1)}{k} \ln z_{\rm v} \ ,
\label{eq_phi_T_reg}
\eeq
with 
\beq
z_{\rm v} = 2 \sum_{t=1}^T \left( (\hrho_t + \hQ_{t-1} )^{l+1} - (\hrho_t + \hQ_{t-2} )^{l+1} \right) + 2 e^\epsilon \left( (\hrho_T + \hQ_\infty )^{l+1} - (\hrho_T + \hQ_{T-1} )^{l+1} \right) \ . \label{eq_zv_reg}
\eeq
The entropy $s(\theta,T,k,l)$ is obtained by an inverse Legendre transformation on $\phi$, parametrized by $\epsilon$,
\beq
s= \phi - \epsilon \theta \ ,
\label{eq_s_T_reg}
\eeq
where $\theta=P_T$ is the fraction of variables still frozen at time $T$, which is obtained from the solution of the above set of equations as
\beq
\theta = \frac{2 e^\epsilon}{ z_{\rm v}} \left( (\hrho_T + \hQ_\infty )^{l+1} - (\hrho_T + \hQ_{T-1} )^{l+1} \right) \ .
\label{eq_theta_T_reg}
\eeq
Finally the typical time trajectory $P_t$ for the whitening from the configurations selected by this bias at time $T$ (recall the definition given in Eq.~(\ref{eq_def_Pt_epsilon})) reads
\beq
P_t(T,\epsilon,k,l) = \begin{cases}
\frac{2}{z_{\rm v}} \underset{t'=t+1}{\overset{T}{\sum}} \left( (\hrho_{t'} + \hQ_{t'-1} )^{l+1} - (\hrho_{t'} + \hQ_{t'-2} )^{l+1} \right) + \theta & \text{for} \ t \in [0,T-1] \\
\frac{2 e^{\epsilon}}{z_{\rm v}}  \left( (\hrho_T + \hQ_\infty )^{l+1} - (\hrho_T + \hQ_{t-1} )^{l+1} \right) & \text{for} \ t \ge T
\end{cases} \ ,
\label{eq_Pt_regular}
\eeq
where for the second line, i.e. beyond the time horizon $T$, the quantities $Q_t$ and $\hQ_t$ are obtained from the solution of the $4 T$ equations by the additional relations
\beq
\text{for} \ t >T \ , \ \ Q_t = e^\epsilon
\left( (\hrho_T +\hQ_\infty)^l - (\hrho_T +\hQ_{t-2} )^l \right) \ , \qquad
\hQ_{t-2} = \left( Q_1^{k-1} - Q_{t-1}^{k-1} \right) \ .
\eeq

A minimal sanity check is provided by the investigation of the case $\epsilon=0$; indeed this has to correspond, for all $T$, to the unbiased measure over all solutions. One finds that the above set of equations is solved for $\epsilon=0$ by
\beq
Q_1 = \left(2^{k-1} -1 \right)^\frac{l}{1-l(k-1)} \ , \quad
Q_t = Q_1 \tP_{t-1} \ , \quad
R_t = Q_1 \ \ \forall t \ , \quad
\hrho_t = \left(2^{k-1} -2 \right) Q_1^{k-1} \ \ \forall t \ ,
\eeq
where $\tP_t$ is the series defined in (\ref{eq_Ptyp}). Plugging this solution in (\ref{eq_zv_reg},\ref{eq_Pt_regular}) yields
\beq
s=\phi = \ln 2 + \frac{l+1}{k} \ln\left( 1 - \frac{1}{2^{k-1}}\right)
\ , \qquad P_t = \Ptyp_t \ \ \forall t \ ,
\eeq
i.e. the typical results (see Eqs.~(\ref{eq_s_ann},\ref{eq_Ptyp})), as expected.

As already discussed in Sec.~\ref{sec_main_results} the most interesting information to extract from the computation of $P_t$ beyond the time horizon $T$ is whether $P_t \to 0$ as $t\to \infty$ or not. In the first case the whitening ultimately leads to the all-joker states, i.e. the initial solution was unfrozen, in the latter there remains blocked variables forever. One can easily see that $P_t \to 0$ as $t \to \infty$ is equivalent to $Q_t \to 0$ as $t \to \infty$. But for $t>T$, one obtains from the above equations that $Q_t$ evolves according to a simple recursion relation of the type $Q_t = f(Q_{t-1})$, namely
\beq
Q_t = e^\epsilon  \left[ (\hrho_T + Q_1^{k-1})^l - (\hrho_T + Q_1^{k-1} - Q_{t-1}^{k-1})^l \right] \ .
\label{eq_Qt_beyond}
\eeq
This recursion has always a fixed point in $0$; depending on the values of $\theta$ (or equivalently of $\epsilon$) it might be the only one, or they might be a larger one, smaller than $Q_T$. In the latter case the limit of $Q_t$ as $t \to \infty$ is strictly positive, let us call it $Q_*$, and as already said in this case $P_t$ tends to a strictly positive value as $t \to \infty$. At the transition between $Q_*=0$ and $Q_* > 0$, which is discontinuous for $k>2$, one has two additional conditions corresponding to the bifurcation:
\bea
Q_* &=& e^\epsilon  \left[ (\hrho_T + Q_1^{k-1})^l - (\hrho_T + Q_1^{k-1} - Q_*^{k-1})^l \right] \ , \label{eq_Qinfty} \\
1 &=& e^\epsilon l (k-1) Q_*^{k-2} (\hrho_T + Q_1^{k-1} - Q_*^{k-1})^{l-1} \ . \label{eq_Qinfty_deriv}
\eea
These two equations fix the values of $Q_*$ and $\epsilon$ in such a way that $\theta=\ttip(T,k,l)$, the tipping point defined in Sec.~\ref{sec_main_results}.

To conclude this section we note that all that is needed to obtain the results announced in Sec.~\ref{sec_main_results} amounts to solve the $4T$ equations on $4T$ unknowns written in Eqs.~(\ref{eq_reg_Qt}-\ref{eq_reg_rhot}), from which follow all the thermodynamic predictions. For not too large values of $T$ this resolution can easily be performed numerically by iteration (with a little bit of damping to avoid some stability issues), this is for instance how we obtained the curves plotted in Fig.~\ref{fig_softheta_k3l6}. To obtain numerical results for larger values of $T$ we used a Newton-Raphson approach, on an equivalent set of equations with only $2T$ unknowns, see Appendix~\ref{sec_compact_RS} for more details. We shall see in the following subsections some analytical simplifications that arise for $T=1$ on the one hand, and in the limit $T\to\infty$ on the other hand, where a well chosen ansatz allows to close the infinite dimensional set of equations on a small number of unknowns.

\subsection{The $T=1$ case}
\label{sec_res_T1}

\subsubsection{Resolution of the equations}

When the bias between solutions depends on the number of white variables at the first non-trivial time ($T=1$) the RS equations given in (\ref{eq_reg_Qt}-\ref{eq_reg_rhot}) correspond to $4$ equations on the $4$ unknowns $Q_1$, $R_1$, $\hQ_\infty$ and $\hrho_1$. We rewrite them here, renaming these variables $Q$, $R$, $\hQ$, $\hrho$ for simplicity:
\bea
Q &=& \hrho^l + e^\epsilon ((\hrho + \hQ)^l - \hrho^l) \ , \label{eq_T1_Q} \\
R &=& e^\epsilon (\hrho + \hQ)^l \ , \label{eq_T1_R}  \\
\hQ &=& Q^{k-1} \ , \label{eq_T1_hQ}  \\
\hrho &=& (2^{k-1} - k - 1) Q^{k-1} + (k-1) R \, Q^{k-2} \ . 
\label{eq_T1_hrho} 
\eea
The thermodynamic potential $\phi(T=1,\epsilon)$ and its Legendre transform $s(T=1,\theta)$ are then obtained from Eqs.~(\ref{eq_phi_T_reg},\ref{eq_s_T_reg}), with the expressions of $z_{\rm v}$ and $\theta$ obtained by specializing (\ref{eq_zv_reg},\ref{eq_theta_T_reg}) to the case $T=1$:
\beq
z_{\rm v} = 2 \hrho^{l+1} + 2 e^\epsilon \left((\hrho + \hQ)^{l+1} - \hrho^{l+1} \right) \ , \quad 
\theta = \frac{1}{z_{\rm v}} 2 e^\epsilon \left((\hrho + \hQ)^{l+1} - \hrho^{l+1} \right) \ .
\label{eq_T1_thermo}
\eeq

We shall now explain that the set of equations (\ref{eq_T1_Q}-\ref{eq_T1_hrho}) has always a single solution that can be very easily found numerically, a fact which might be not completely obvious at first sight. First we note that it is very easy to eliminate $R$ and $\hQ$ explicitly in terms of the two other unknowns with (\ref{eq_T1_R}) and (\ref{eq_T1_hQ}), which leaves us with two equations on $Q$ and $\hrho$:
\bea
Q &=& \hrho^l + e^\epsilon ((\hrho + Q^{k-1})^l - \hrho^l) \ , \\
\hrho &=& (2^{k-1} - k - 1) Q^{k-1} + (k-1) Q^{k-2} e^\epsilon (\hrho + Q^{k-1})^l \ .
\eea
It turns out to be more convenient to add $\epsilon$ to the set of unknowns and to define a new variable $\psi$ by $\psi=Q^{k-1}/\hrho$. Indeed one can solve explicitly this new set of equations and find:
\bea
e^\epsilon &=& \frac{\psi^{-1} - (2^{k-1} - k - 1)}{(k-1) - ((1+\psi)^l-1)(\psi^{-1} - (2^{k-1} - 2))} \ , \\
Q &=& \psi^\frac{l}{l(k-1)-1} \left[\frac{(k-1)(1+\psi)^l}{(k-1) - ((1+\psi)^l-1)(\psi^{-1} - (2^{k-1} - 2))} \right]^\frac{1}{1-l(k-1)} \ , \\
\hrho &=& \psi^\frac{1}{l(k-1)-1} \left[\frac{(k-1)(1+\psi)^l}{(k-1) - ((1+\psi)^l-1)(\psi^{-1} - (2^{k-1} - 2))} \right]^\frac{k-1}{1-l(k-1)} \ .
\eea
In this way the potential $\phi(\epsilon)$, and also its Legendre transform $s(\theta)$, can be computed explicitly in a parametric way by varying this new variable $\psi$, that evolves monotonously with $\epsilon$ and $\theta$ (one has in particular $\psi^{-1} = 2^{k-1}-2$ for $\epsilon=0$, $\theta=\Ptyp_1$, and $\psi^{-1} = 2^{k-1}-k-1$ for $\epsilon\to-\infty$, $\theta= 0$). We used this very simple numerical procedure in order to obtain the curves $s(\theta)$ of Fig.~\ref{fig_softheta_T1_k6}. The location of the tipping point is determined for generic values of $T$ by the equations (\ref{eq_Qinfty},\ref{eq_Qinfty_deriv}), which become in the case $T=1$, with the simplified notations introduced in (\ref{eq_T1_Q}-\ref{eq_T1_hrho}):
\bea
Q_* &=& e^\epsilon  \left[ (\hrho + \hQ)^l - (\hrho + \hQ - Q_*^{k-1})^l \right] \ , \label{eq_Qinfty_T1} \\
1 &=& e^\epsilon l (k-1) Q_*^{k-2} (\hrho + \hQ - Q_*^{k-1})^{l-1} \label{eq_Qinfty_deriv_T1} \ . 
\eea

Note that the expression of the entropy in $\theta=0$ for $T=1$ is particularly simple; plugging $\psi^{-1} = 2^{k-1}-k-1$ in the above formulas yields indeed:
\beq
s(T=1,\theta=0) = \ln 2 + \frac{l+1}{k} \ln \left(1 - \frac{k+1}{2^{k-1}} \right) \ .
\eeq
As a matter of fact for all the variables to be white in one time-step of the whitening it is necessary that around each hyperedge at least two variables are equal to $+1$ and at least two others equal to $-1$, forbidding $2(k+1)$ configurations out of the $2^k$ possible ones, the expression above is the annealed result for such a constraint (compare with the total entropy of solutions in Eq.~(\ref{eq_s_ann})). This yields an explicit expression for the threshold $l'_1(k)$ where this entropy vanishes:
\beq
l'_1(k) = -1 - \frac{k \ln 2}{\ln \left(1 - \frac{k+1}{2^{k-1}} \right) } \sim \frac{2^{k-1}k \ln 2}{k+1} \qquad \text{as} \ \ k \to \infty \ ,
\eeq
i.e. an asymptotic behavior even smaller than the typical rigidity threshold $\lr(k)$.

\subsubsection{Phase diagram}

\begin{figure}
\includegraphics[width=8cm]{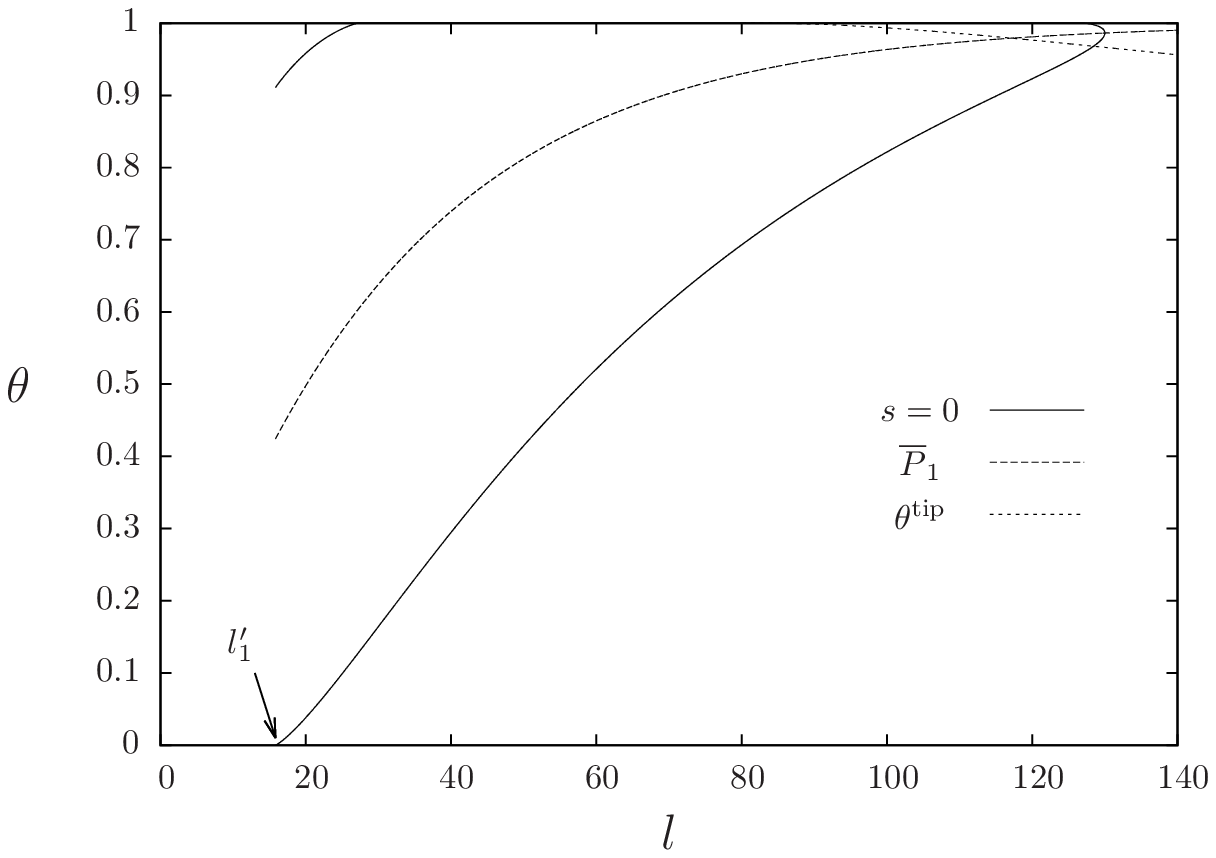}
\hspace{1cm}
\includegraphics[width=8cm]{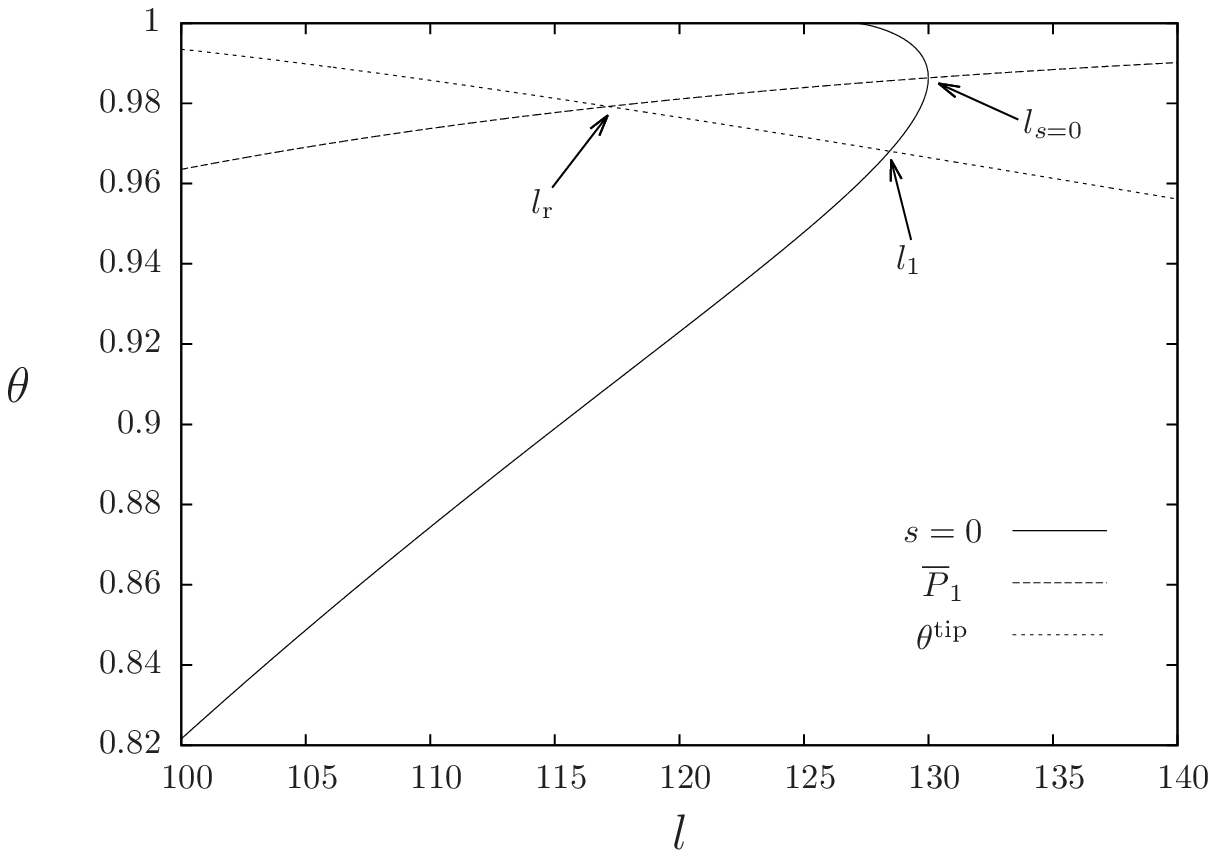}
\caption{Phase diagram for $k=6$ and $T=1$, in the plane $(l,\theta)$, the right panel being an enlargement of the most interesting part for the largest values of $l$. The solid line represents the location of the vanishing of the entropy (recall the shape of $s(\theta)$ plotted in Fig.~\ref{fig_softheta_T1_k6}), the long dashed line the location of its maximum in the typical value $\theta=\Ptyp_1$, while the short dashed line corresponds to the tipping point $\ttip$, configurations being frozen for $\theta > \ttip$, unfrozen for $\theta < \ttip$.}
\label{fig_pdrs_k6T1}
\end{figure}

We present in Figure~\ref{fig_pdrs_k6T1} the phase diagram summarizing our results at $T=1$ for a given value of $k$ (we chose $k=6$ as the smallest generic value, all cases $k\ge 6$ behaving qualitatively in the same way), in the plane $(l,\theta)$ (an equivalent representation can be obtained in the canonical parametrization $(l,\epsilon)$). It has been obtained by extracting from the curves $s(\theta)$ shown in Fig.~\ref{fig_softheta_T1_k6} a few representative points for each $l$, namely the upper and lower limit of the interval on which the entropy is positive (the solid lines denoted $s=0$), the location of the maximum of $s(\theta)$ that corresponds to the typical value $\Ptyp_1$ (long dashed curve), as well as the tipping point $\ttip$ (short dashed curve). Several thresholds values of $l$ can be read on this diagram: the RS prediction for the satisfiability threshold $l_{s=0}$ corresponds to the intersection of the typical curve with the $s=0$ one, the typical rigidity $\lr$ is located at the crossing of the tipping point and the typical curve, while the threshold $l_1(k)$ is the intersection of the tipping point and the entropy vanishing curve. More marginally interesting is the point where the lower limit of the interval of positive entropies reaches $\theta=0$, which gives the threshold $l'_1(k)$.

As already discussed in the introduction to our main results given in Sec.~\ref{sec_main_results}, the most interesting point on this phase diagram is $l_1(k)$, which gives in the RS framework the largest connectivity up to which there exist unfrozen configurations that can be selected by a bias at $T=1$. The numerical values for $l_1(k)$ are given in Table~\ref{table_typical} for small values of $k$. The large $k$ analysis for the threshold $l_1(k)$ that yields (\ref{eq_l1_largek}) being rather technical we defer its presentation to the Appendix~\ref{app_largek_T1}, that contains further results valid at large $k$, in particular a simple explicit formula for $s(T=1,\theta)$.

For what concerns the locked solutions one can visualize on this phase diagram the thresholds $\llm \approx 27.1$ and $\llp \approx 127.1$ as the two points where the curve $s=0$ reaches the axis $\theta=1$, while $\llu \approx 82.8$ corresponds to the point where $\ttip=1$.

\subsection{The fixed points of the whitening process (the limit $T\to \infty$)}
\label{sec_res_Tinfty}

We shall now present our predictions for $s_\infty(\theta)$, the entropy of proper bicolorings having a fraction $\theta$ of frozen variables, that we defined in Eq.~(\ref{eq_s_infty}). We obtained these results by studying the large $T$ limit of $s(T,\theta)$; these two quantities would coincide if the large $N$ and large $T$ limit would commute, which we believe to be almost the case, in a sense that we shall explain shortly afterwards (cf. Sec.~\ref{sec_res_Tfinite}).

As recalled in Sec.~\ref{sec_def_whitening} the fraction of frozen variables in a solution $\us$ of a given (finite size) instance is defined via the fixed point $\us^*$ of the WP/whitening dynamics reached from this solution. A crucial property of these fixed points was proven in~\cite{ManevaMossel05} (where fixed points were called core assignments): on hypergraphs with good expansion properties (typically exhibited by random hypergraphs) $\us^*$ is either the all white configuration, or contain an extensive number of frozen variables (the proof of~\cite{ManevaMossel05} is for $k$-SAT but can be adapted to the bicoloring problem). In terms of the entropy $s_\infty(\theta)$ this dichotomy implies the existence of a constant $\tmin(k,l)>0$ such that $s_\infty$ is well defined only for $\theta=0$ and $\theta \ge \tmin$, or more precisely $s_\infty=-\infty$ in the interval $]0,\tmin[$ between the isolated point $\theta=0$ and the branch $\theta\ge \tmin$. 

\begin{figure}
\includegraphics[width=8cm]{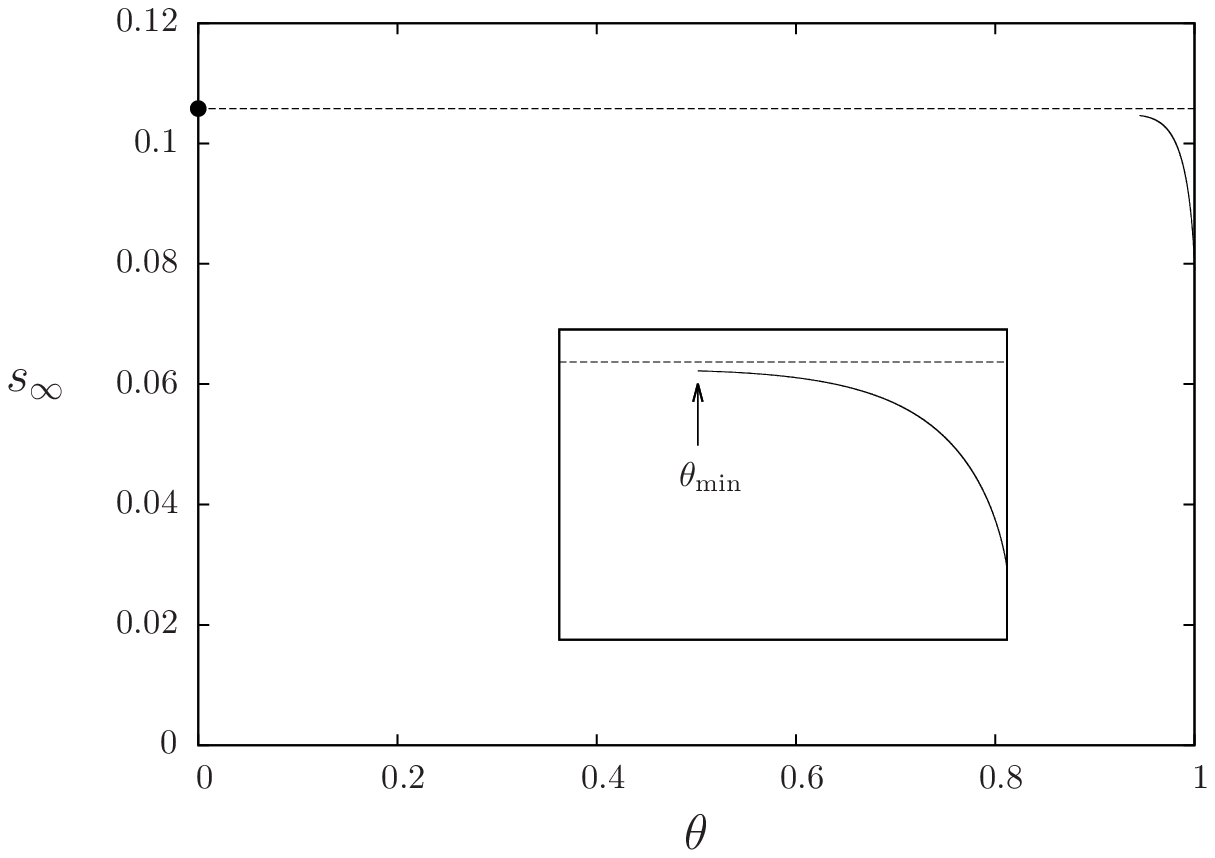}
\hspace{1cm}
\includegraphics[width=8cm]{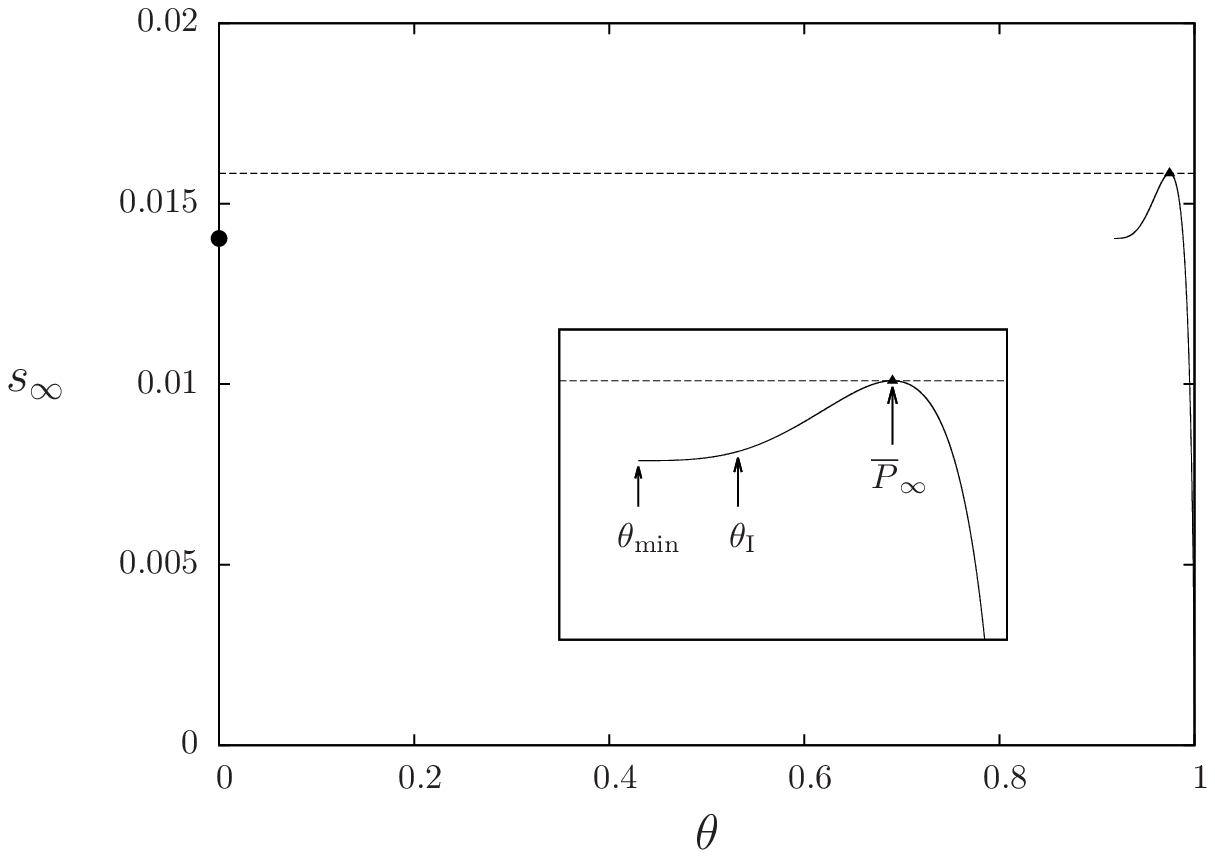}
\caption{The entropy $s_\infty(\theta)$ counting the proper bicolorings with a fraction $\theta$ of frozen variables, for $k=6$, $l=110 < \lr(k)$ (left panel) and $l=127 > \lr(k)$ (right panel). The filled dot represents the entropy of unfrozen solutions, the horizontal dashed line indicates the total entropy of solutions of Eq.~(\ref{eq_s_ann}), the solid line is the branch of the entropy $s_\infty$ for $\theta \ge \tmin$. For $l<\lr$ the typical solutions are unfrozen, the branch with $\theta \ge \tmin$ has an entropy strictly smaller than the total one, while for $l>\lr$ this branch contains the typical solutions, hence the coordinates of its maximum (filled triangle) are the typical values $(\Ptyp_\infty,s)$. In both panels the insets present a magnification of the regime $\theta \ge \tmin$, in the right one an arrow designate $\theta_{\rm I}$, the limit separating frozen solutions of the first and second kind.}
\label{fig_sinftyoftheta}
\end{figure}

\begin{figure}
\includegraphics[width=8cm]{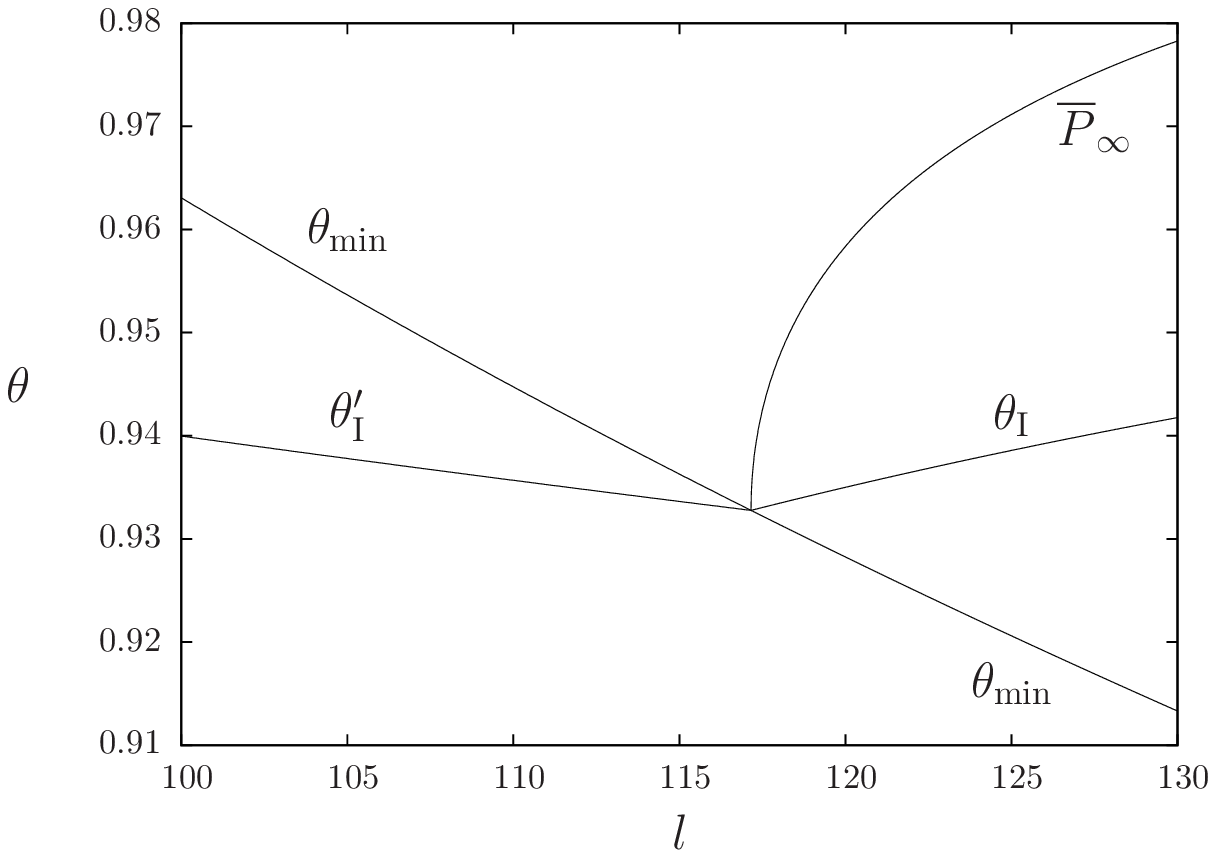}
\hspace{1cm}
\includegraphics[width=8cm]{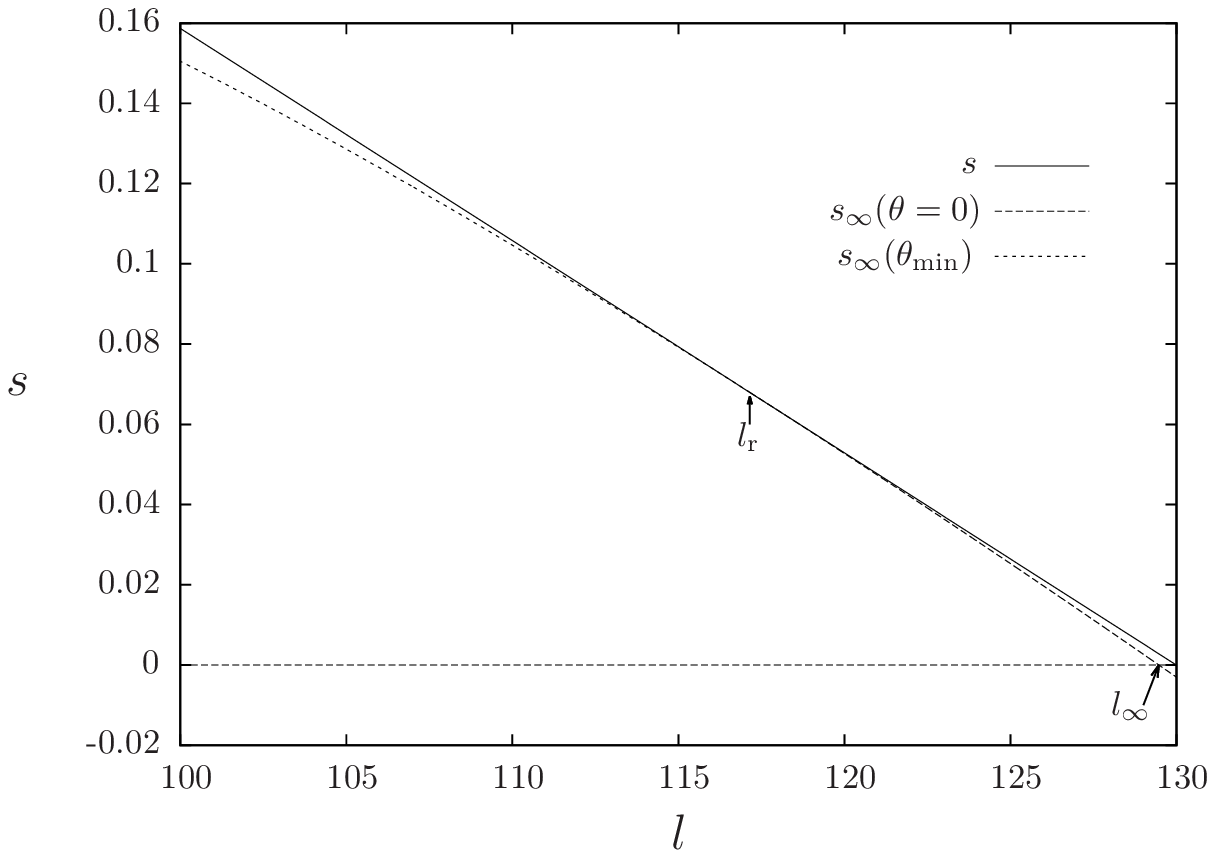}
\caption{Left panel: phase diagram for the fixed points of the whitening in the $(l,\theta)$ plane for $k=6$, displaying the minimal fraction of frozen variables in frozen configurations $\tmin$, the fraction $\Ptyp_\infty$ of frozen variables in typical solutions for $l \ge \lr$, and the threshold $\theta_{\rm I}$ separating frozen solutions of the first and second kind for $l > \lr$ (the meaning of the curve $\theta'_{\rm I}$ for $l<\lr$ will be explained in Sec.~\ref{sec_res_Tfinite}) all these curves meet in $l=\lr$. The curve $\tmin$ reaches 1 at the threshold $\llu$ (not shown on this plot) below which all frozen solutions are locked.
Right panel: the entropies $s_\infty(\theta=0)$ and $s_\infty(\tmin)$, compared to the total entropy of solutions $s$ of Eq.~(\ref{eq_s_ann}), plotted as a function of $l$ for $k=6$. One has $s_\infty(\theta=0)=s$ for $l<\lr$, while $s_\infty(\theta=0)=s_\infty(\tmin)$ for $l>\lr$; the threshold $l_\infty$ is defined by the cancellation of $s_\infty(\theta=0)$.
}
\label{fig_Tinfty_theta_and_s}
\end{figure}

\begin{figure}
\includegraphics[width=8cm]{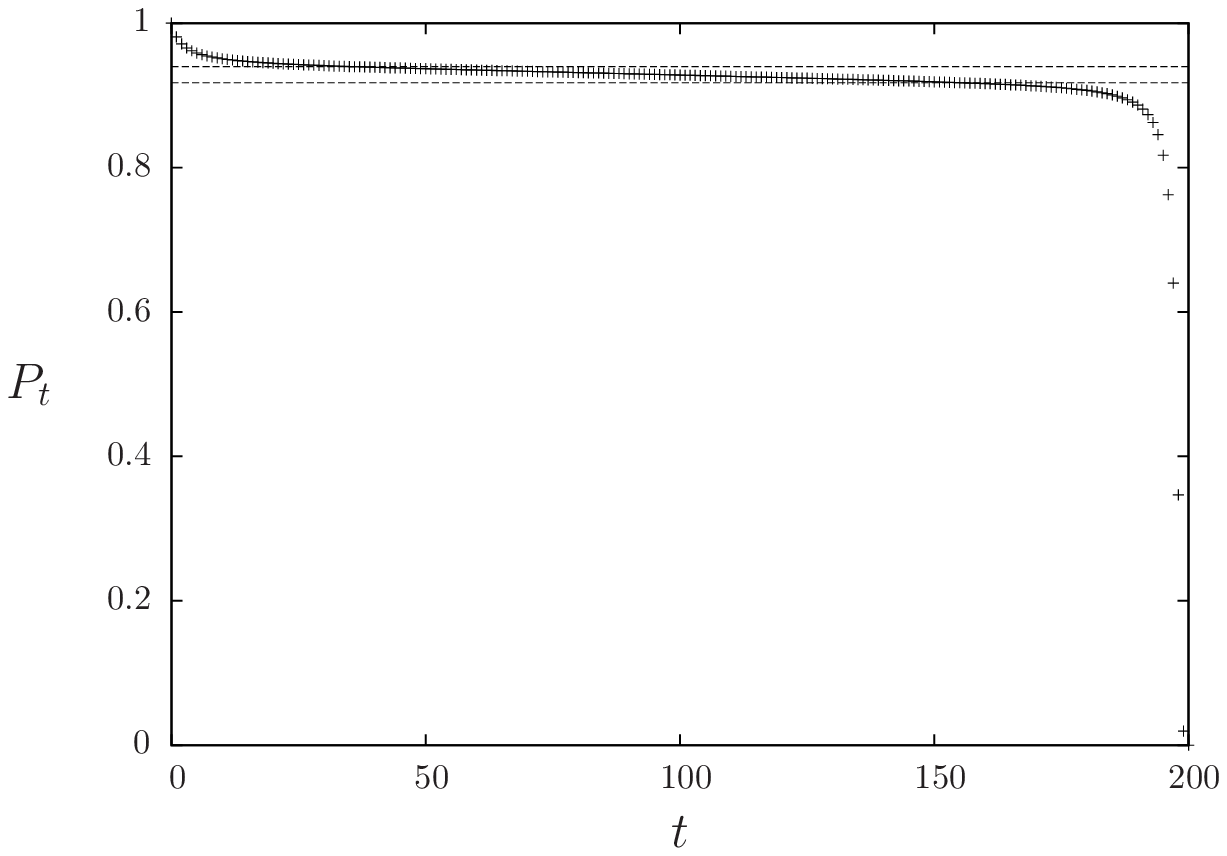}
\hspace{1cm}
\includegraphics[width=8cm]{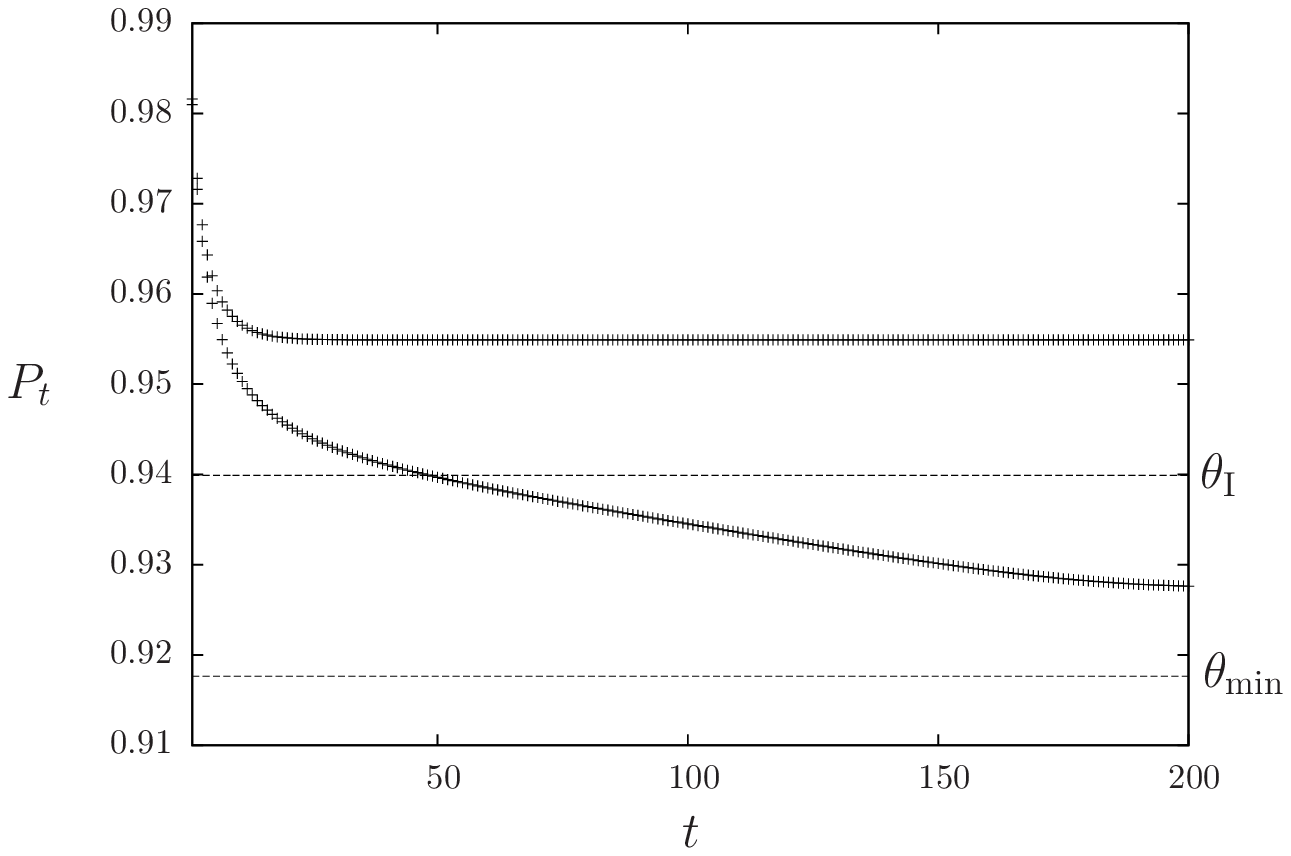}
\caption{The cumulative distributions $P_t$ of the whitening times, for $k=6$ and $l=127>\lr$, with a large time scale fixed to $T=200$. Left panel: unfrozen solution ($\theta=0$), the dashed horizontal lines indicate the fraction of variables that whiten in the three time regimes, according to the analytical predictions of (\ref{eq_zv1}-\ref{eq_zv3}). Right panel : frozen solutions of the first (top) and second kind (bottom).}
\label{fig_Tinfty_Pt}
\end{figure}

It turns out that in the $T\to\infty$ limit the solution of the (infinite dimensional) set of equations (\ref{eq_reg_Qt}-\ref{eq_reg_rhot}) obey some rather simply parametrized ansatz, hence the computation of $s_\infty$ can be reduced to the resolution of finite dimensional sets of equations. The technical details of this reduction, that rely on techniques developed for the bootstrap percolation problem in~\cite{GuSe15}, are presented in Appendix~\ref{app_Tinfty}; in the rest of this subsection we first explain qualitatively our findings (illustrated in Figs.~\ref{fig_sinftyoftheta}, \ref{fig_Tinfty_theta_and_s} and \ref{fig_Tinfty_Pt}), then present the equations that allows to obtain them quantitatively.

Consider first the plots of Fig.~\ref{fig_sinftyoftheta}, which display our predictions of $s_\infty(\theta)$ for two values of $l$, below and above the typical rigidity threshold (left and right panel respectively). For $l<\lr$ the isolated point at $\theta=0$ counts the typical configurations (by definition of $\lr$), hence in this regime $s_\infty(\theta=0,k,l)=s(k,l)$; we found that the non-trivial branch for $\theta \ge \tmin$ has a strictly smaller entropy, starts from $\tmin$ with a strictly negative derivative, and is decreasing and concave. On the other hand for $l > \lr$ the point in $\theta=0$ counts the atypical unfrozen solutions, it has thus an entropy smaller than the total one, $s_\infty(\theta=0,k,l)<s(k,l)$; the non-trivial branch for $\theta \ge \tmin$ has a maximum in the typical point of coordinates $(\Ptyp_\infty,s)$, and is non-concave in the interval $[\tmin,\Ptyp_\infty]$. More precisely, this branch starts at $\tmin$ with a vanishing derivative, and an entropy which coincides with the one of the unfrozen configurations: $s_\infty(\tmin)=s_\infty(\theta=0)$ for $l > \lr$.

In Figure~\ref{fig_Tinfty_theta_and_s} we reported the dependency of some of these quantities with $l$ for a given $k$ (here $k=6$), in particular $\tmin$ is plotted on the left panel and compared to $\Ptyp_\infty$ (which vanishes for $l<\lr$). The right panel summarizes the evolution of the entropies $s_\infty(\tmin)$ and $s_\infty(\theta=0)$ as a function of $l$; in particular the vanishing of the latter yields the threshold $l_\infty(k)$ for the disappearance of all unfrozen solutions (in the RS framework). The numerical values for $l_\infty(k)$ are given in Table~\ref{table_typical} for small values of $k$, and the large $k$ asymptotic expansion of $l_\infty$ was stated in (\ref{eq_linfty_largek}); the derivation of this analytic expansion is explained in Appendix~\ref{sec_app_linfty_largek}.

A more detailed description of the configurations counted by the entropy $s_\infty(\theta)$ is provided by the corresponding distribution of the whitening times $P_t$, which by definition goes to $\theta$ as $t\to\infty$. Let us discuss its behavior in the various cases corresponding to $\theta=0$ (unfrozen solutions) vs $\theta \ge \tmin$ (frozen solutions) on the one hand, $l<\lr$ vs $l>\lr$ on the other hand. 
\begin{itemize}
\item
For $l<\lr$ the unfrozen solutions are typical, and their whitening profile $P_t=\Ptypt$ was displayed in Fig.~\ref{fig_Ptyp}; as long as $l < \lr$ essentially all variables whiten in finite time (with respect to $N$), a divergence occurs when $l \to l_{\rm r}^-$, with a fraction $\Ptyp_\infty(\lr)$ having whitening times growing like $(\lr-l)^{-1/2}$. 
\item
For $l>\lr$ the atypical unfrozen solutions have a peculiar whitening trajectory, presented on the left panel of Fig.~\ref{fig_Tinfty_Pt}: the decay of $P_t$ from 1 to 0 occurs in three steps, in other words the whitening time of the variables fall into three categories each containing a positive fraction of the total number of variables. Calling $T$ the (large) time scale on which the whitening reaches its trivial fixed point, the three whitening processes occur on times $t=O(1)$ for the first, $t/T=O(1)$ for the second, and $t=T-O(1)$ for the third one. Note that in our computations we have to fix this large scale $T$ arbitrarily, because we took the thermodynamic limit $N\to\infty$ in a previous step. In a hypergraph with $N$ large but finite the scale $T(N)$ on which the fixed points are reached will be a growing function of $N$, and we can only conjecture that the whitening trajectories would correspond to the ones we computed, modulo this unknown scale $T(N)$. It turns out that the fraction of variables which whiten in the last regime is nothing but $\tmin$, the minimal fraction of frozen variables in frozen solutions.
\item
By definition the long time limit of $P_t$ is $\theta \ge \tmin >0$ for frozen solutions. We found that the decay of $P_t$ towards $\theta$ occurs in two qualitatively different ways depending on the parameters $(l,\theta)$. For some cases we found this decay to occur on a single regime, with all whitening times $O(1)$ with respect to $T$ (except of course the fraction $\theta$ of infinite ones), let us call the corresponding frozen solutions ``of the first kind''. On the contrary frozen solutions of the second kind have a two time regimes whitening trajectory, with a positive fraction of variables having whitening times $t=O(1)$, and another one $t/T=O(1)$. An illustration of these two kind of whitening trajectories is presented on the right panel of Fig.~\ref{fig_Tinfty_Pt}. We found that for $l<\lr$ all frozen solutions are (typically) of the first kind, while for $l>\lr$ there is a threshold denoted $\theta_{\rm I}$ (also plotted in the phase diagram of Fig.~\ref{fig_Tinfty_theta_and_s}) such that the frozen solutions are of the first kind for $\theta \ge \theta_{\rm I}$, and of the second kind for $\theta \in [\tmin,\theta_{\rm I}]$. When $\theta=\tmin$ the decay of $P_t$ coincides with the first two regimes of the evolution of $P_t$ for unfrozen solutions ($\theta=0$), the difference being the absence of the final rapid decay in the regime $t=T-O(1)$.

\end{itemize}

We present now the formulas describing quantitatively the entropy $s_\infty$ and the associated time evolutions $P_t$, a complete justification being deferred to Appendix~\ref{app_Tinfty}.

\subsubsection{The entropy of unfrozen solutions for $l>\lr$}
\label{sec_Tinfty_whitening}

The entropy of unfrozen solutions in the regime where they are atypical (i.e. $s_\infty(\theta=0)$ for $l>\lr$) is obtained by solving the following set of equations on the (positive) unknowns $Q_1, Q_{\rm i}, Q_{\rm f}, \hrho_{\rm i}, \hrho_{\rm f}$ (see Appendix~\ref{app_Tinfty_unfrozen} for a detailed justification):
\bea
\hrho_{\rm i} &=& (2^{k-1}-2)Q_1^{k-1} \ , \label{eq_Tinfty_mans_1} \\
Q_{\rm i} &=& Q_1 - (\hrho_{\rm i} + Q_1^{k-1} - Q_{\rm i}^{k-1})^l \ , 
\label{eq_Tinfty_mans_2} \\
1 &=& l(k-1)Q_{\rm i}^{k-2} (\hrho_{\rm i} + Q_1^{k-1} - Q_{\rm i}^{k-1})^{l-1}
\ , \label{eq_Tinfty_mans_3} \\
Q_{\rm f} &=& (\hrho_{\rm f} + Q_1^{k-1})^l  - (\hrho_{\rm f} + Q_1^{k-1} - Q_{\rm f}^{k-1})^l \ , \label{eq_Tinfty_mans_4} \\
1 &=& l(k-1)Q_{\rm f}^{k-2} (\hrho_{\rm f} + Q_1^{k-1} - Q_{\rm f}^{k-1})^{l-1} \ , \label{eq_Tinfty_mans_5}
\eea
where in the second line one has to choose $Q_{\rm i}$ as the largest of the solutions of the equation on $[0,Q_1]$. From the solution of this system of equations one obtains the entropy $s_\infty(\theta=0)$ using (\ref{eq_phi_T_reg},\ref{eq_s_T_reg}) with 
\bea
z_{\rm v}&=&z_{{\rm v},1}+z_{{\rm v},2}+z_{{\rm v},3} \ , \qquad \theta=0 \ ,
\label{eq_zv_Tinfty} \\
z_{{\rm v},1} &=& 2 (\hrho_{\rm i} + Q_1^{k-1} - Q_{\rm i}^{k-1})^{l+1}  \ , \label{eq_zv1} \\
z_{{\rm v},2} &=& 2 \frac{(l+1)(l-1)}{l^\frac{l}{l-1} (k-1)^\frac{1}{l-1} (l-k+1)} \left( Q_{\rm i}^\frac{l-k+1}{l-1} - Q_{\rm f}^\frac{l-k+1}{l-1}  \right) 
\ , \label{eq_zv2} \\
z_{{\rm v},3} &=& 2 (\hrho_{\rm f} + Q_1^{k-1})^{l+1}  - 2 (\hrho_{\rm f} + Q_1^{k-1} - Q_{\rm f}^{k-1})^{l+1} \ . \label{eq_zv3}
\eea
As explained above and illustrated in the left panel of Fig.~\ref{fig_Tinfty_Pt}, $P_t$ decays from $P_0=1$ to $P_T=0$ in three regimes, corresponding to $t=O(1)$, $t/T=O(1)$ and $t=T-O(1)$, the fraction of variables whitening during the $i$-th regime being $z_{{\rm v},i}/z_{\rm v}$. 

The system of equations (\ref{eq_Tinfty_mans_1}-\ref{eq_Tinfty_mans_5}) actually admits a single solution which is simpler to find numerically than might seem at first sight. Indeed its resolution amounts to finding the largest solution $u=u(k,l)\in]0,1[$  and the unique solution $v=v(k,l) \in ]0,1[$ of the following equations:
\beq
l = \frac{1}{k-1} \frac{2^{k-1}-1-u^{k-1}}{u^{k-2}(1-u)} \ , \qquad
v =1- \left(1 - \frac{v}{v+l(k-1)(1-v)} \right)^l \ .
\label{eq_uv}
\eeq
Then the five unknowns of (\ref{eq_Tinfty_mans_1}-\ref{eq_Tinfty_mans_5}) are expressed as
\bea
Q_1 &=& (1-u)^\frac{1}{l(k-1)-1} (2^{k-1}-1-u^{k-1})^{-\frac{l}{l(k-1)-1}} \ , \nonumber \\
Q_i &=& u \, Q_1  \ , \nonumber \\
\hrho_{\rm i} &=& (2^{k-1}-2) \, Q_1^{k-1} \ ,  \nonumber \\
\hrho_{\rm f} &=& -Q_1^{k-1} + \beta^\frac{1}{l(k-1)-1} \ , \nonumber \\
Q_{\rm f} &=& v \, \beta^\frac{l}{l(k-1)-1} \ ,
\label{eq_Q1Qf}
\eea
where
\beq
\beta =\frac{1}{v^{k-2}}\frac{1}{v+l(k-1)(1-v)} = \frac{1}{v^{k-1}} \left(1-(1-v)^\frac{1}{l} \right) \ . 
\label{eq_beta}
\eeq
From these expressions the entropy $s_\infty(\theta=0)$ can be deduced unambiguously via (\ref{eq_phi_T_reg},\ref{eq_s_T_reg},\ref{eq_zv_Tinfty}-\ref{eq_zv3}).
One can check that for $l=l_{\rm r}(k)$, i.e. at the rigidity threshold, the entropy thus predicted coincides with the total entropy of solutions of (\ref{eq_s_ann}) (one has indeed $u=v$ and $\beta=(2^{k-1}-1)^{-1}$ when $l=l_{\rm r}(k)$).

\subsubsection{The frozen solutions of the first kind}
\label{sec_Tinfty_nonwhitening}

As explained in Appendix~\ref{app_Tinfty} the computation of the part of $s_\infty(\theta)$ corresponding to frozen solutions of the first kind amounts to determine the four (positive) unknowns $Q_1,Q_{\rm b},\hrho_{\rm b},\hrho_T$, solutions of
\bea
Q_{\rm b} &=& Q_1 - (\hrho_{\rm b} + Q_1^{k-1} - Q_{\rm b}^{k-1})^l \ , 
\label{eq_Qb} \\
\hrho_{\rm b} &=& (2^{k-1}-2) Q_1^{k-1} \ , \\
Q_{\rm b} &=& e^\epsilon \left[ (\hrho_T + Q_1^{k-1})^l- (\hrho_T + Q_1^{k-1} - Q_{\rm b}^{k-1})^l \right] \ , \label{eq_Qb2} \\
\hrho_{\rm b} &=& \hrho_T + (k-1) Q_{\rm b}^{k-2} \left[ (\hrho_{\rm b} + Q_1^{k-1} - Q_{\rm b}^{k-1})^l - e^\epsilon (\hrho_T + Q_1^{k-1} - Q_{\rm b}^{k-1})^l \right] \ ,
\label{eq_hrhob}
\eea
with the additional conditions that $Q_{\rm b}$ is the largest solution on $[0,Q_1]$ of (\ref{eq_Qb}), if $\epsilon < 0$ one has to enforce $\hrho_T < \hrho_{\rm b}$, and moreover one must have
\beq
e^\epsilon l (k-1) Q_{\rm b}^{k-2} (\hrho_T + Q_1^{k-1} - Q_{\rm b}^{k-1})^{l-1} \le 1 \ .
\label{eq_condition_stability_frozen}
\eeq
Once these equations are solved the predictions for the entropy follows from (\ref{eq_phi_T_reg},\ref{eq_s_T_reg}),
with
\bea
z_{\rm v} &=& 2 (\hrho_{\rm b} + Q_1^{k-1} - Q_{\rm b}^{k-1})^{l+1} 
+ 2 e^\epsilon \left[ (\hrho_T + Q_1^{k-1})^{l+1} - (\hrho_T + Q_1^{k-1} - Q_{\rm b}^{k-1})^{l+1} \right] \ , \label{eq_zv_Tinfty_frozen} \\ 
\theta &=& \frac{1}{z_{\rm v}} 2 e^\epsilon \left[ (\hrho_T + Q_1^{k-1})^{l+1} - (\hrho_T + Q_1^{k-1} - Q_{\rm b}^{k-1})^{l+1} \right] \ . \label{eq_theta_Tinfty_frozen}
\eea
Moreover in this regime the cumulative distribution of the whitening times $P_t$ admits a limit when $T\to\infty$ at finite $t$, that reads
\beq
P_t = \theta + \frac{1}{z_{\rm v}} \left[2 (\hrho_{\rm b} + Q_1^{k-1} - Q_{\rm b}^{k-1})^{l+1} - 2 (\hrho_{\rm b} + Q_1^{k-1} - Q_{t-1}^{k-1})^{l+1}  \right] \,
\label{eq_Pt_Tinfty_frozen}
\eeq
where $Q_t$ denotes here the solution of:
\beq
Q_{t+1} = Q_1 - (\hrho_{\rm b} + Q_1^{k-1} - Q_t^{k-1})^l \ .
\eeq

To solve numerically the equations (\ref{eq_Qb}-\ref{eq_hrhob}) we adopted the following procedure. We parametrize $Q_1$ and $Q_b$ with
\beq
Q_1 = (2^{k-1}-1)^{-\frac{l}{l(k-1)-1}} \lambda^\frac{1}{l(k-1)-1} \ , \qquad
Q_{\rm b} = x \, Q_1 \ ,
\eeq
and reduce the first two equations to the determination of the largest solution on $[0,1]$ of
\beq
x = 1 - \lambda \left(1- \frac{x^{k-1}}{2^{k-1}-1} \right)^l \ .
\label{eq_x_lambda}
\eeq
Note the similarity with (\ref{eq_rigidity_typ}). Now $\lambda$ is considered as the parameter to be varied to find all solutions of (\ref{eq_Qb}-\ref{eq_hrhob}); having determined $Q_1$, $Q_{\rm b}$ and $\hrho_{\rm b}$ as a function of $\lambda$ we obtain $\hrho_T$ and $\epsilon$ from the last two equations. This can be done more easily by first eliminating $\epsilon$ with (\ref{eq_Qb2}), hence reducing the problem to a single equation on $\hrho_T$:
\beq
\hrho_T = \hrho_{\rm b} - (k-1) Q_{\rm b}^{k-2} Q_1 + (k-1) Q_{\rm b}^{k-1} \frac{(\hrho_T + Q_1^{k-1})^l}{(\hrho_T + Q_1^{k-1})^l- (\hrho_T + Q_1^{k-1} - Q_{\rm b}^{k-1})^l} \ .
\eeq
For $l>\lr$ the branch of $s_\infty(\theta)$ corresponding to $\theta \ge \theta_{\rm I}$ is described by $\lambda \in [0,\lambda_+]$, where $\lambda_+$ corresponds to a bifurcation in (\ref{eq_x_lambda}) with the solution $x(\lambda)$ disappearing discontinuously; the entropy $s_\infty(\theta)$ thus computed is not concave, hence there exists multiple solutions if viewed as a function of $\epsilon$, but always a single solution when parametrized by $\lambda$ or $\theta$.

For $l<\lr$ the a priori possible range for the parameter $\lambda$ is $[0,1]$; however only a part of this interval yields a solution satisfying all the additional conditions given after (\ref{eq_Qb}-\ref{eq_hrhob}). Increasing $\lambda$ from 0 one first encounters a value $\lambda_-$ above which the condition (\ref{eq_condition_stability_frozen}) is no longer satisfied, this yields the value $\tmin$. If one discards this condition and keeps on increasing $\lambda$ one finds a second threshold $\lambda'_-$ at which $\epsilon$ vanishes, and above which $\epsilon < 0$ with $\rho_{\rm T}>\rho_{\rm b}$, which violates a second requirement. The value of $\theta$ at $\lambda'_-$ is denoted $\theta'_{\rm I}$ and plotted in the left panel of Fig.~\ref{fig_Tinfty_theta_and_s}, its interpretation will be discussed in Sec.~\ref{sec_res_Tfinite}.

\subsubsection{The frozen solutions of the second kind}
\label{sec_Tinfty_nonwhitening_II}

The frozen solutions of the second kind are described by a set of equations which combines some features of the unfrozen solutions, and some of the frozen solutions of the first kind. More explicitly, they are described by the unknowns $Q_1, Q_{\rm i}, Q_{\rm f}, \hrho_{\rm i}, \hrho_{\rm f}$ and $\hrho_T$, which obey the three equations (\ref{eq_Tinfty_mans_1}-\ref{eq_Tinfty_mans_3}), fixing $Q_1,Q_{\rm i}$ and $\hrho_{\rm i}$. They also obey (\ref{eq_Tinfty_mans_5}), which implies a relation between $Q_{\rm f}$ and $\hrho_{\rm f}$, the last two equations determining all the unknown being 
\bea
Q_{\rm f} &=& e^\epsilon \left[ (\hrho_T + Q_1^{k-1})^l- (\hrho_T + Q_1^{k-1} - Q_{\rm f}^{k-1})^l \right] \ ,  \\
\hrho_{\rm f} &=& \hrho_T + (k-1) Q_{\rm f}^{k-2} \left[ (\hrho_{\rm f} + Q_1^{k-1} - Q_{\rm f}^{k-1})^l - e^\epsilon (\hrho_T + Q_1^{k-1} - Q_{\rm f}^{k-1})^l \right] \ ,
\eea
which are the analog of (\ref{eq_Qb2},\ref{eq_hrhob}) for the frozen solutions of the first kind,  with the replacement $(Q_{\rm b},\hrho_{\rm b}) \to (Q_{\rm f},\hrho_{\rm f})$.

The prediction for the entropy then follows from (\ref{eq_phi_T_reg},\ref{eq_s_T_reg}), with
\beq
z_{\rm v}=z_{{\rm v},1}+z_{{\rm v},2}+z_{{\rm v},3} \ , \qquad \theta=\frac{z_{{\rm v},3}}{z_{\rm v}} \ ,
\eeq
where $z_{{\rm v},1}$ and $z_{{\rm v},2}$ are given by (\ref{eq_zv1}) and (\ref{eq_zv2}), and
\beq
z_{{\rm v},3} = 2 e^\epsilon \left[ (\hrho_T + Q_1^{k-1})^{l+1} - (\hrho_T + Q_1^{k-1} - Q_{\rm f}^{k-1})^{l+1} \right] \ .
\eeq
The fraction of variables that whiten on times $t=O(1)$ (resp. $t/T=O(1)$) is given by $z_{{\rm v},1}/z_{\rm v}$ (resp. $z_{{\rm v},2}/z_{\rm v}$).

One can check that the two limits of existence of this type of solution corresponds on the one hand to $\theta=\theta_{\rm I}$, where $(Q_{\rm i},\hrho_{\rm i})=(Q_{\rm f},\hrho_{\rm f})$, which matches the end of the first kind of frozen solutions with the same slope $-\epsilon$, and on the other hand to $\theta=\tmin$, where $\hrho_{\rm f}=\hrho_T$ with $\epsilon=0$, an entropy equal to $s_{\infty}(\theta=0)$ and a value of $\theta$ corresponding to the fraction of variables that whiten in the third regime for the unfrozen solutions.

\subsection{The intermediate case $1<T < \infty$}
\label{sec_res_Tfinite}

\begin{figure}
\includegraphics[width=8cm]{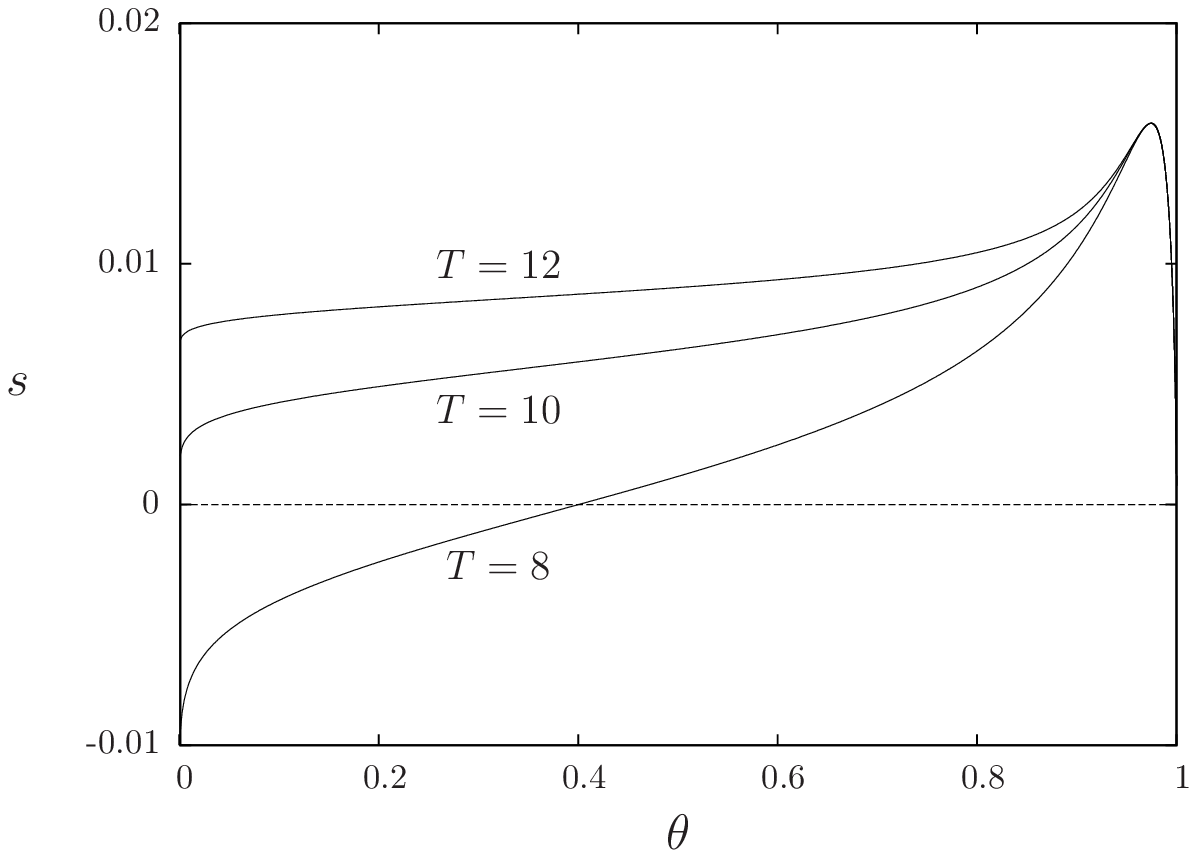}
\hspace{1cm}
\includegraphics[width=8cm]{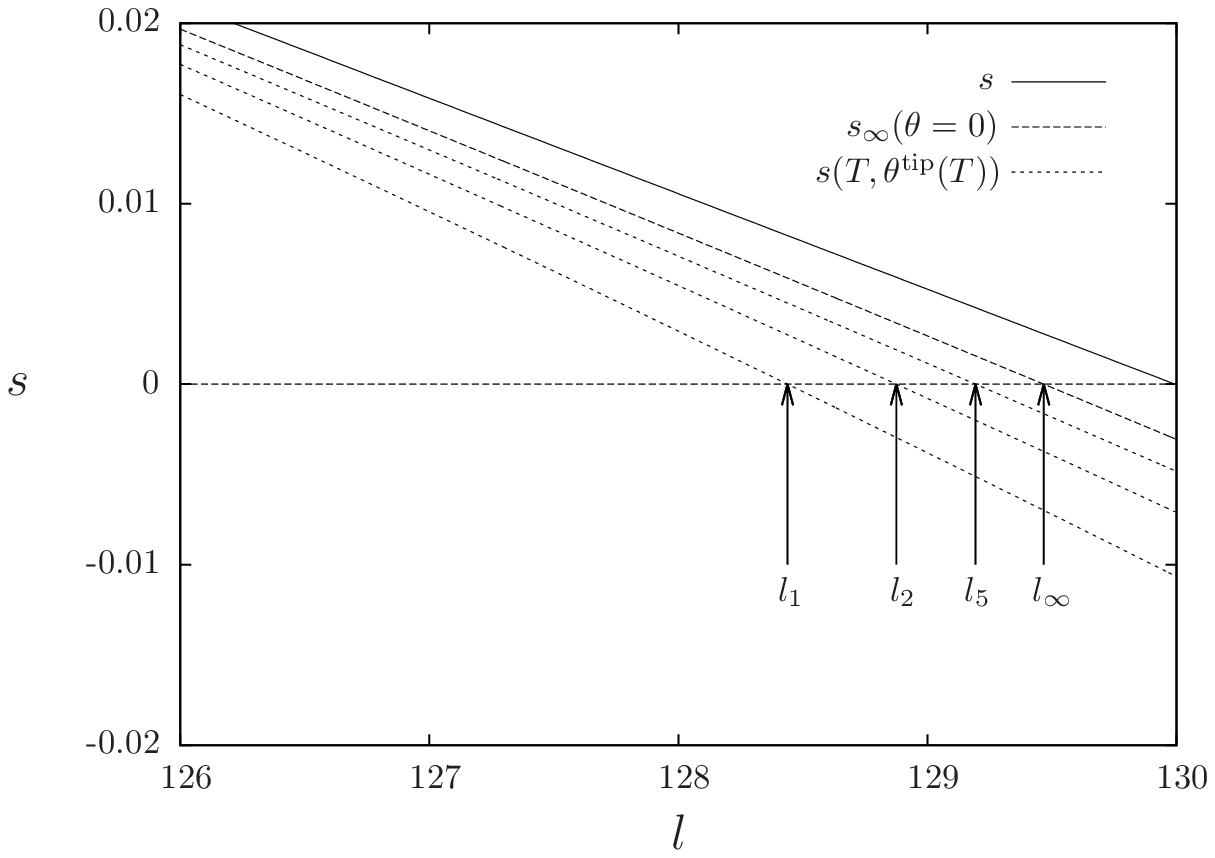}
\caption{Left panel: the entropy $s(T,\theta)$ for $k=6$, $l=127$, and three values of $T$. Right panel: the entropy of the tipping point $s(T,\ttip(T))$ as a function of $l$ for $k=6$, $T=1,2,5$ (from bottom to top); also shown for comparison are the entropy of unfrozen solutions $s_\infty(\theta=0)$ and the entropy of all solutions $s$.}
\label{fig_Tintermediate}
\end{figure}

Let us now discuss the results we obtained for values of $T$ in between the two extreme cases presented above. As the entropy $s_\infty(\theta)$ is rather singular, with a domain $]0,\tmin[$ where it is equal to $-\infty$, and for $l>\lr$ a non-concave branch for $\theta \ge \tmin$, one could expect that for large enough but finite values of $T$ the entropy $s(T,\theta)$ will also have these singular features which would provoke first-order phase transitions in the canonical ensemble parametrized by $\epsilon$. However in such a mean-field model one can explore the metastable and unstable branches of such an entropy by working in the microcanonical ensemble. We have achieved this goal here by following continuously a solution of the RS equations (\ref{eq_reg_Qt}-\ref{eq_reg_rhot}), using the Newton-Raphson iterative method, in which we treated $\epsilon$ as an additional unknown and varying slowly the desired value of $\theta$. In this way we have obtained curves of $s(T,\theta)$ for arbitrary values of $T$, without holes in the range of $\theta$, see the left panel of Fig.~\ref{fig_Tintermediate} for an illustration. In particular we checked that the tipping point $\ttip(T,k,l)$ was well defined (at least for all parameters we investigated), in the sense that whitening trajectories constrained to $P_T=\theta$ ultimately go to their trivial fixed point ($P_t \to 0$ as $t\to\infty$) if and only if $\theta < \ttip(T,k,l)$; the main qualitative features of the phase diagram presented for $T=1$ in Fig.~\ref{fig_pdrs_k6T1} are thus reproduced for larger values of $T$. We computed in this way the thresholds $l_T(k)$, defined through the condition of cancellation of the entropy of the tipping point (see the right panel of Fig.~\ref{fig_Tintermediate} for an example), their values for some $T$ and $k$ being reported in Table~\ref{table_typical}; the large $k$ asymptotics of $l_T(k)$ at fixed $T>1$, which was stated in (\ref{eq_lT_largek}), is explained in some details in Appendix~\ref{app_largek_Tarb}.

\begin{figure}
\includegraphics[width=8cm]{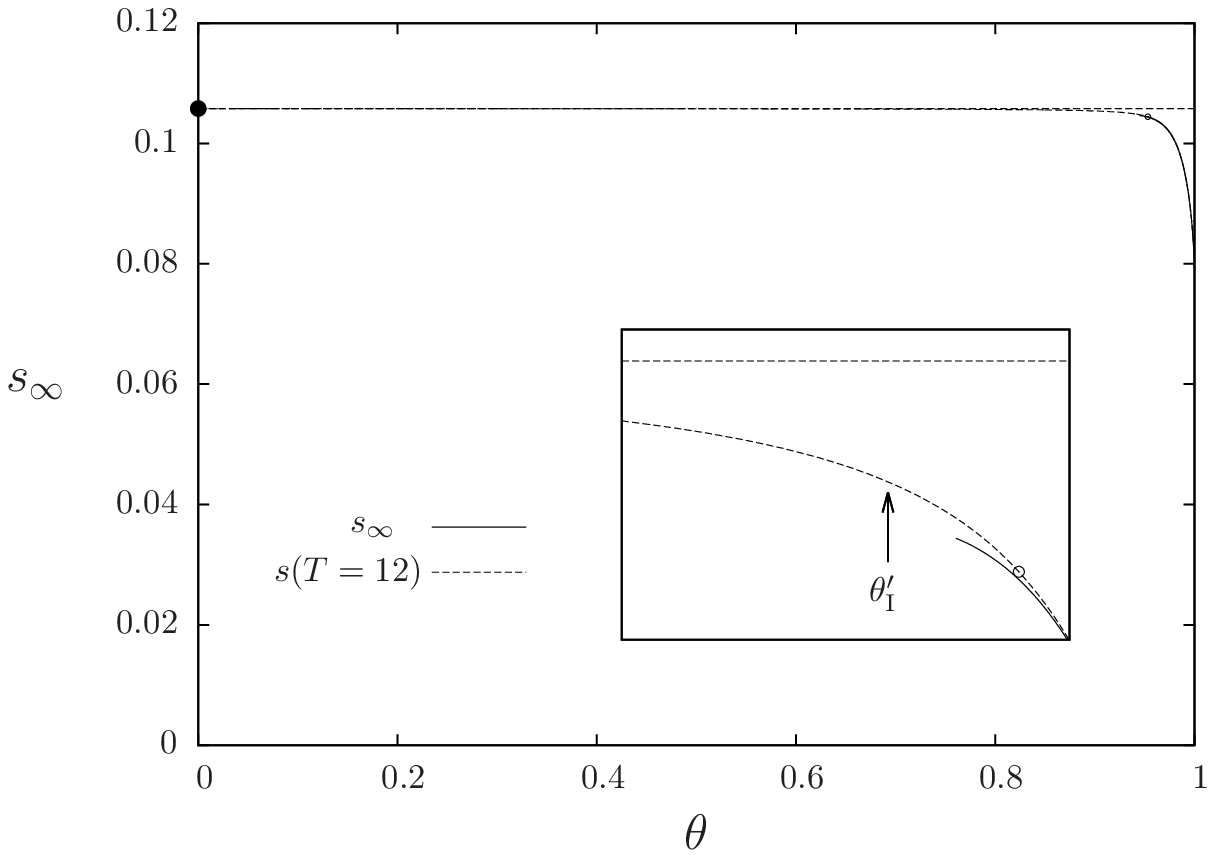}
\hspace{1cm}
\includegraphics[width=8cm]{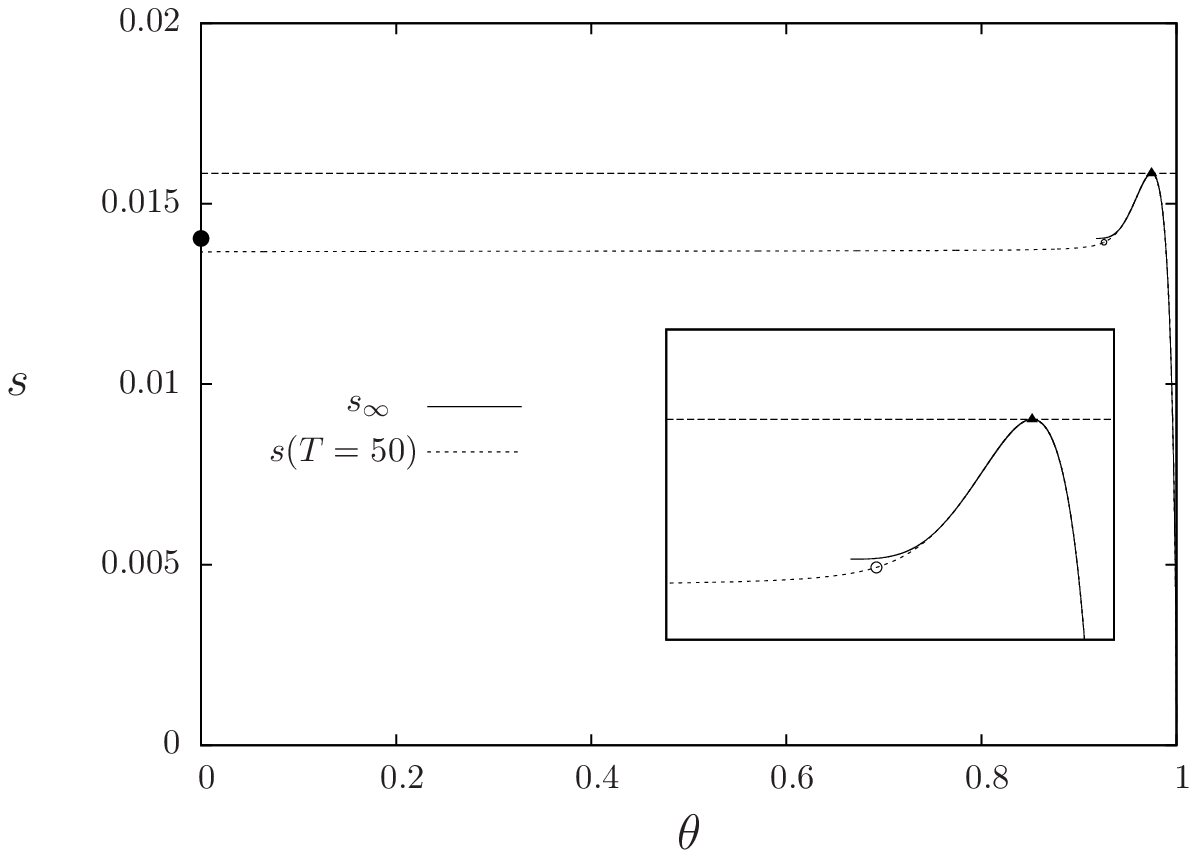}
\caption{Comparison of the entropies $s_\infty(\theta)$ and $s(T,\theta)$ for a large but finite value of $T$. Left panel: $k=6$, $l=110 < \lr$, $T=12$, right panel: $k=6$, $l=127 > \lr$, $T=50$. In both panels the solid line is $s_\infty(\theta)$ (as plotted in Fig.~\ref{fig_sinftyoftheta}), the dashed line $s(T,\theta)$, the two curves are almost superimposed for $\theta \ge \tmin$. The open circle denotes the position of the tipping point on the finite $T$ curve. Insets present magnifications of the curves around $\tmin$.}
\label{fig_compa_s_Tinf_vs_large}
\end{figure}

\begin{figure}
\includegraphics[width=8cm]{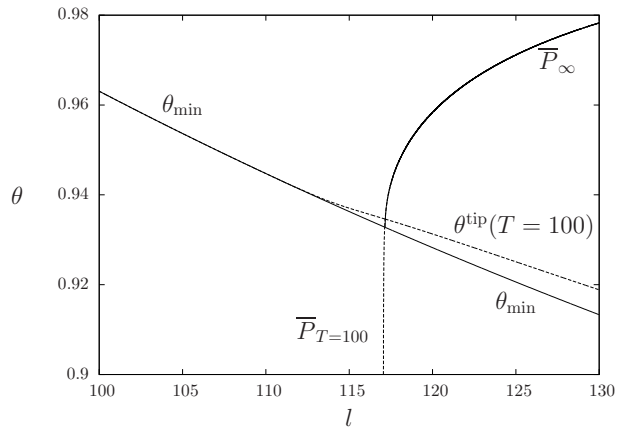}
\caption{The value of $\ttip(T)$ (here $T=100$) as a function of $l$ for $k=6$, compared to its large $T$ limit $\tmin$ (the convergence is much faster for $l<\lr$ than for $l>\lr$). Also shown for comparison are the fraction of frozen variables after $T$ steps of whitening from a typical solution $\overline{P}_T$ and its large $T$ limit $\overline{P}_\infty$.}
\label{fig_pd_largeT}
\end{figure}

Let us now discuss the large but finite $T$ case, and explain some subtleties of the interchange of the limits $N\to\infty$ and $T\to \infty$. In Fig.~\ref{fig_compa_s_Tinf_vs_large} we present the curve of the entropy $s(T,\theta)$ for some large values of $T$, for two values of $l$ above and below the typical rigidity transition, and compare them with the entropy $s_\infty(\theta)$ of fixed points described in the previous section (see in particular Fig.~\ref{fig_sinftyoftheta}). Consider first the case $l > \lr$ (right panel of Fig.~\ref{fig_compa_s_Tinf_vs_large}). One sees that for $\theta \ge \tmin$ these two entropies are almost indistinguishable, however $s(T,\theta)$ has an horizontal branch extending in the interval $[0,\tmin]$ that was forbidden in $s_\infty(\theta)$. A moment of thought reveals that this must indeed be the case: the configurations counted in $s(T,\theta)$ with $\theta \in ]0,\tmin[$ correspond to unfrozen solutions, that would reach the trivial fixed point $\theta=0$ at a slightly later time $T'=T+O(1)$ (recall the distribution of the whitening times for atypical unfrozen solutions plotted on the left panel of Fig.~\ref{fig_Tinfty_Pt}). As these have not reached their fixed point they were discarded in the $s_\infty$ curve. The consistency of this interpretation is ensured by the convergence of $\ttip(T)$ to $\tmin$ as $T$ diverges (see also Fig.~\ref{fig_pd_largeT}), that we have checked from our numerical data at finite $T$, a numerical fit suggests that the limit is reached with corrections of order $1/\sqrt{T}$. We have also seen numerically that $s(T,\ttip(T))$ reaches $s_\infty(\tmin)=s_\infty(\theta=0)$ with corrections of order $1/T$, which should in consequence also be the order of the difference $l_\infty(k)-l_T(k)$ at large $T$ for $k$ fixed.

The case $l<\lr$ is similar (see the left panel of Fig.~\ref{fig_compa_s_Tinf_vs_large}): for $\theta \ge \tmin$ one has $s(T,\theta) \to s_\infty(\theta)$ as $T \to \infty$. There is an interval $\theta \in [\theta'_{\rm I},\tmin]$ in which $\lim_{T\to \infty} s(T,\theta)$ is a non-trivial function, described by the formalism of Sec.~\ref{sec_Tinfty_nonwhitening} if the condition (\ref{eq_condition_stability_frozen}) is relaxed; this branch ends in $\theta'_{\rm I}$ with a vanishing derivative, and is continued for $\theta \le \theta'_{\rm I}$ by an horizontal branch. All the configurations counted in $\lim_{T\to \infty} s(T,\theta)$ for $\theta < \tmin$ are atypical unfrozen solutions that reach their trivial fixed point at a later time $T'>T$, hence do not contribute to $s_\infty(\theta)$; the whitening times in such configurations are either $t=O(1)$ or $t = T \pm O(1)$. As in the case $l > \lr$ this interpretation is confirmed by the convergence of $\ttip(T)$ to $\tmin$ when $T \to \infty$, which is here much faster (see Fig.~\ref{fig_pd_largeT}).

\subsection{The locked solutions ($\theta=1$)}

We give here the formulae to compute the entropy of locked solutions $s_{\rm l}(k,l)$. As already explained this quantity is defined as the limit of $s(T,\theta,k,l)$ when $\theta$ goes to $1$, which should give a result independent of $T$. This is indeed the case, as can be shown by studying the limit $\epsilon \to \infty$ of the solution of the set of equations (\ref{eq_reg_Qt}-\ref{eq_reg_rhot}). A moment of thought reveals that the following ansatz can be used at the leading order in this limit:
\bea
Q_t &=& Q \ (e^{-\epsilon})^\frac{1}{l(k-1)-1} \ , \\
R_t &=& Q \ (e^{-\epsilon})^\frac{1}{l(k-1)-1} \qquad \text{for} \ \ t \in [1,T-1] \ , \\
R_T &=& R \ (e^{-\epsilon})^\frac{1}{l(k-1)-1} \ , \\
\hQ_t &=& 0 \qquad \text{for} \ \ t \in [1,T-1] \ ,\\
\hQ_\infty &=& \hQ \ (e^{-\epsilon})^\frac{k-1}{l(k-1)-1} \ , \\
\hrho_t &=& (2^{k-1}-2) \ Q^{k-1} (e^{-\epsilon})^\frac{k-1}{l(k-1)-1} \qquad \text{for} \ \ t \in [1,T-1] \ , \\
\hrho_T &=& \hrho \ (e^{-\epsilon})^\frac{k-1}{l(k-1)-1} \ ,
\eea
where $Q$, $R$, $\hQ$ and $\hrho$ are solutions of a very simple system of equations:
\bea
Q &=& (\hrho + \hQ)^l - \hrho^l \ , \label{eq_locked_Q} \\
R &=& (\hrho + \hQ)^l \ , \\
\hQ &=& Q^{k-1} \ , \\
\hrho &=& (2^{k-1} - k - 1) \, Q^{k-1} + (k-1) R \, Q^{k-2} \label{eq_locked_rho} \ .
\eea
The latter system has a non-trivial solution if and only if $l>k-1$; this condition can be understood very intuitively, as it translates into $M>N$. Indeed each bicoloring constraint can block at most one of its variables, hence the number of hyperedges must be larger than the number of vertices for a locked solution in which all variables are frozen to potentially exist. From the solution of (\ref{eq_locked_Q}-\ref{eq_locked_rho}) one obtains the entropy of the locked solutions as
\beq
s_{\rm l}(k,l) = \frac{1-l(k-1)}{k} \ln \left(2 ((\hrho + \hQ)^{l+1} - \hrho^{l+1}) \right) \ .
\eeq
Computing numerically this quantity one finds it to be positive on the interval $l \in [\llm(k),\llp(k)]$, the numerical values of these two thresholds are presented in Table~\ref{table_typical}, and their large $k$ asymptotic expansions (obtained similarly as the one developed in the Appendices) in Eqs.~(\ref{eq_llm_largek},\ref{eq_llp_largek}). The threshold $\llu(k)$ where $\ttip(T)=1$ for all values of $T$ (or $\tmin=1$ in the $T\to\infty$ limit) is obtained by complementing the system (\ref{eq_locked_Q}-\ref{eq_locked_rho}) with the following condition (derived from (\ref{eq_Qinfty_deriv}) in the $\epsilon \to \infty$ limit):
\beq
l(k-1) Q^{k-2} \hrho^{l-1} = 1 \ ,
\eeq
see again Table~\ref{table_typical} and Eq.~(\ref{eq_llu_largek}) for the numerical values and asymptotic expansion.

\section{The effects of RSB}
\label{sec_rsb}

Let us summarize the main lines of the computations presented up to now. We have defined in Sec.~\ref{sec_ld} an entropy $s(T,\theta)$ counting the number of solutions of the hypergraph bicoloring problem that have a fraction $\theta$ of still frozen variables after $T$ steps of the whitening dynamics, and its Legendre transform $\phi(T,\epsilon)$. We then showed in Sec.~\ref{sec_statmech_fg} how the latter could be expressed in terms of a graphical model with the same local structure as the original hypergraph, the price to be paid for maintaining this locality structure being the introduction of more complicated variables. At this point we applied the machinery of the cavity method to this extended graphical model, using the replica symmetric (RS) ansatz, i.e. assuming that the biased measure over solutions obeyed the long-range decorrelation properties recalled in the unbiased case in Sec.~\ref{sec_reminder}.

It is clear that the RS ansatz cannot be correct for all values of the parameters $(k,l,T,\epsilon)$: as a matter of fact the prediction of negative entropies is self-contradictory, and we know that for $\epsilon=0$ the computation reduces to the unbiased case which undergoes dynamic and condensation transitions at $l_{\rm d}$ and $l_{\rm c}$. We thus also applied the 1RSB version of the cavity method to the extended graphical model parametrized by $(T,\epsilon)$ to assess the domain of validity of our RS results; there is no major conceptual difficulty in doing this, the recipes of the 1RSB cavity method have been recalled in Sec.~\ref{sec_reminder}, we will thus state the results of these computations without writing explicitly the corresponding equations. 

\begin{figure}
\includegraphics[width=9cm]{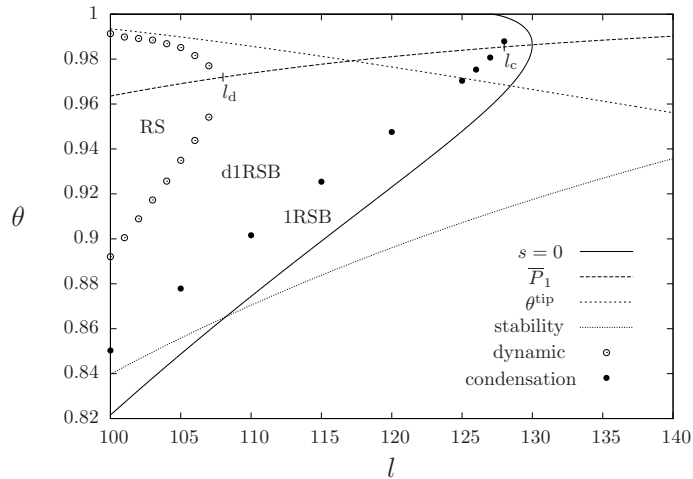}
\caption{The phase diagram of the biased measure for $k=6$, $T=1$, completing the one of Fig.~\ref{fig_pdrs_k6T1} with the effects of replica symmetry breaking, computed with 1RSB populations of $10^6$ cavity fields.}
\label{fig_pdwithrsb_k6T1}
\end{figure}

We have summarized our findings in the phase diagram of Fig.~\ref{fig_pdwithrsb_k6T1}, which repeats the RS phase diagram of Fig.~\ref{fig_pdrs_k6T1} for $k=6$, $T=1$, and adds the additional results of the 1RSB investigations. The phase diagram is plotted in the $(l,\theta)$ plane for clarity, however it should be emphasized that we tested the properties of the graphical model parametrized by $\epsilon$, the translation in terms of $\theta$ is made according to the RS relationship between these two conjugated variables. In the part of the phase diagram labeled RS, and delimited by the ``dynamic'' transition points, the only solution of the 1RSB equations with Parisi parameter $m=1$ is the RS one. In this regime, that intersects the typical line $\Ptyp_T$ at $l_{\rm d}$, the point-to-set correlation length of the biased measure is finite, a single pure state describes most of its relevant solutions. In the regime labeled d1RSB (for ``dynamic 1RSB'') there is a non-trivial solution to the 1RSB equations with $m=1$, that has a strictly positive complexity; in this case the RS prediction for the thermodynamic quantities $\phi(\epsilon)$ and $s(\theta)$ is expected to be still the correct one. This part of the phase diagram ends at the condensation transition points (that intersect the typical $\theta$ at $l_{\rm c}$), the crossing of which leads to a true 1RSB regime in which $\Sigma(m=1)<0$. In the latter the RS thermodynamic predictions are no longer exact, a better estimate would require the determination of the parameter $m$ such that $\Sigma(m)=0$. As it should the line of vanishing of the RS entropy is always in the 1RSB condensed part of the phase diagram. For completeness we have also drawn the line of local stability of the RS solution inside the 1RSB space of solutions, below which the RS solution cannot be correct; for the parameters displayed in Fig.~\ref{fig_pdwithrsb_k6T1} this line is always irrelevant, this continuous transition being always preceded by the discontinuous one at the border between the RS and d1RSB phases. 

As could be anticipated the RS prediction for the threshold $l_1(k)$ up to which one can bias the measure over solutions according to the results of a single step of the whitening in such a way to make typical solutions unfrozen is not exact; it was indeed defined as the intersection of the tipping point line with the vanishing of the RS entropy, which is always in the condensed part of the phase diagram where the RS predictions are wrong. We can however conclude here that $l_1(k=6) \ge 125$, the last value of $l$ for which the tipping point is in the uncondensed phase, which is certainly greater than $\lfloor \lr(k=6) \rfloor =117$, the largest value of $l$ for which typical solutions are unfrozen. It should be in principle possible to set up a large $k$ expansion of the 1RSB equations of the biased measure over solutions; our hope, based on the known results for the unbiased measure, is that the condensed phase becomes negligible in this limit, and hence that the large $k$ expansion of $l_1(k)$ given in (\ref{eq_l1_largek}) is indeed the correct one. A similar reasoning justifies our expectation that $l_\infty(k)$ coincides with the freezing threshold $l_{\rm f}(k)$ in the large $k$ limit, with the asymptotic expansion presented in (\ref{eq_linfty_largek}).

As a final remark on the effects of replica symmetry breaking we draw the attention of the reader to the fact that $\ttip$ always lies in the d1RSB or 1RSB phase. A moment of thought reveals that this must be so: by definition for $\theta \ge \ttip$ the typical solutions of the biased measure are frozen, hence the solution of the $m=1$ 1RSB equations of the extended graphical model have to contain hard fields, and hence cannot be equal to the RS solution. This intuition can actually be confirmed explicitly by a (rather technical) computation starting from the extended 1RSB equations.

\section{Numerical experiments}
\label{sec_numerics}

\begin{figure}
\begin{center}
\includegraphics[width=8cm]{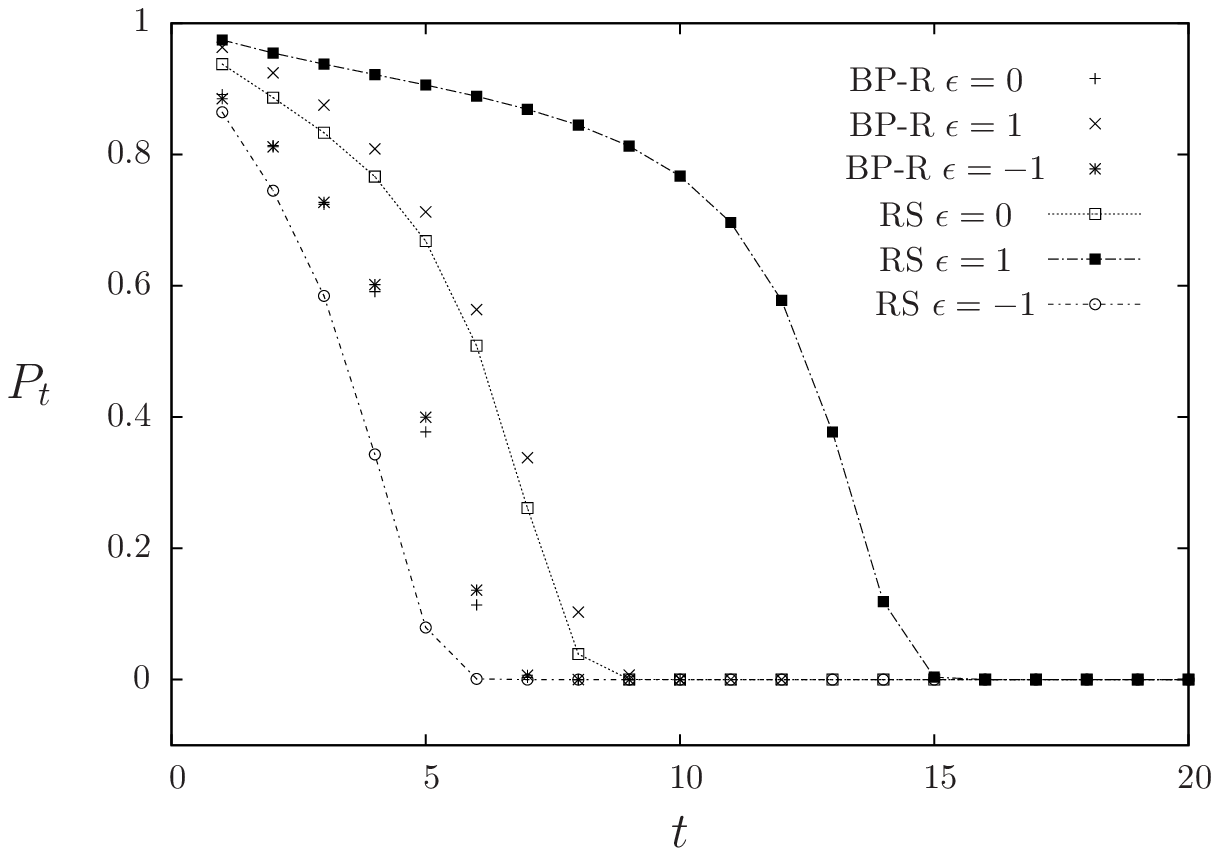}
\hspace{1cm}
\includegraphics[width=8cm]{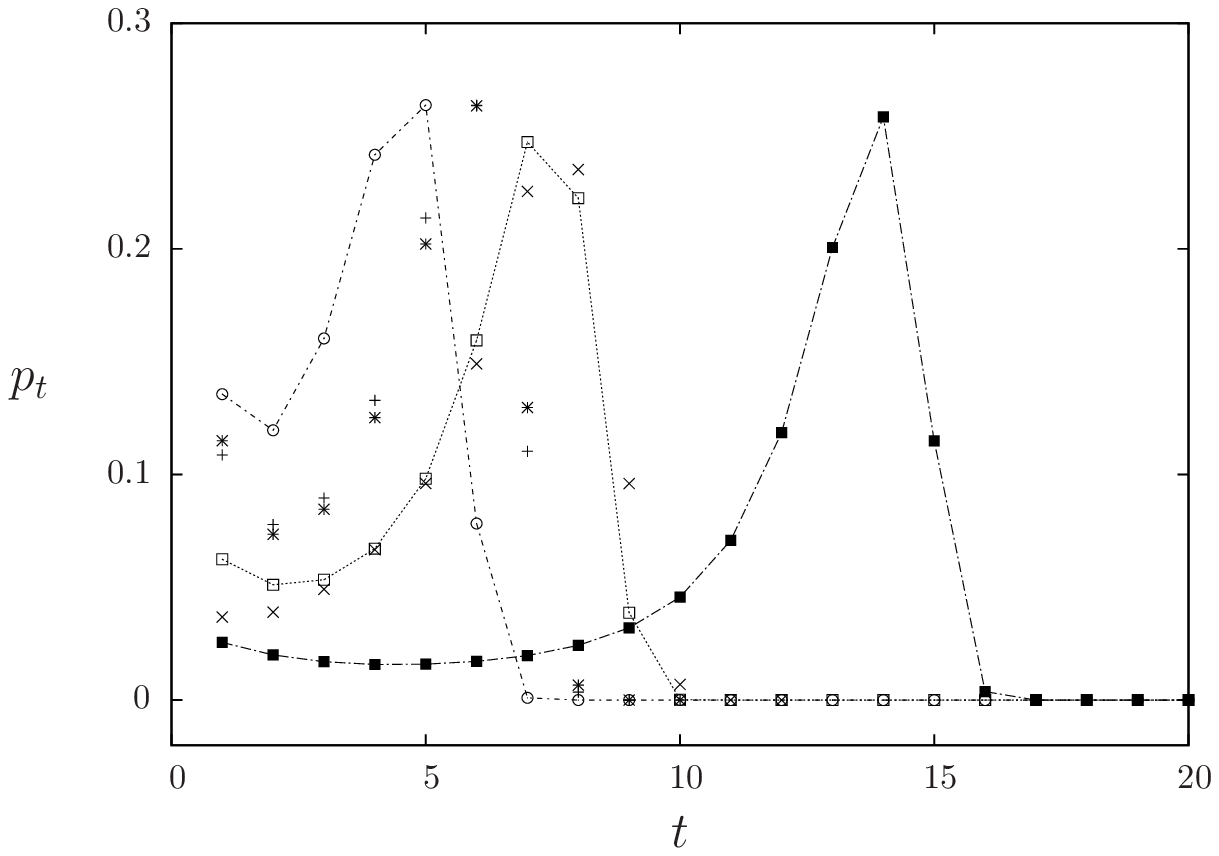} 
\caption{Cumulative distribution $P_t$ (left) and distribution $p_t=P_{t-1}-P_t$ (right) of whitening times in solutions obtained by BP+reinforcement on a hypergraph of size $N=10^4$, $k=4$ and $l=17$ with $\epsilon=-1,0,1$ and $T=1$ (points), compared to the theoretical RS predictions of Sec.~\ref{sec_res_T1} (points joined by lines); the same symbols are used in both panels.}
\label{fig_numerics_reinforcement}
\end{center}
\end{figure}

We have performed some numerical experiments inspired by the analysis of the large deviations properties of the whitening process. The first one allowed us to demonstrate the possibility of tuning the velocity of the whitening, in other words to adapt some algorithms in order to construct solutions with various whitening time distributions (within a certain range). An ideal way of doing this would be to sample configurations according to the biased measure defined in Eq.~(\ref{eq_mu}); it is of course impossible in general to perform such a sampling in polynomial time, and one has to resort to indirect approximate methods, like simulated annealing~\cite{KirkpatrickGelatt83} or some decimation process, for instance using BP computed marginals~\cite{Allerton,RiSe09}. We used a variant of this last idea called soft-decimation or reinforcement~\cite{BraunsteinZecchina06}. More precisely, we solved the single instance BP equations of Eq.~(\ref{eq_BP_singlesample}) iteratively, according to
\begin{equation}
\eta_{i \to a}^{n+1} = f(\{\heta_{b \to i}^n\}_{b\in \dima}) \ , \qquad
\heta_{a \to i}^{n+1} = \hf(\{\eta_{j \to a}^n\}_{j\in \dami}) \ ,
\end{equation}
where $n$ is a discrete time index, with a reinforcement procedure in which an external prior that depends on the corresponding marginal calculated at the previous iteration is dynamically applied. Specifically, Eq.~(\ref{eq_f}) at iteration $n$ is multiplied by a site-dependent prior $p^n(\sigma)$ that is dynamically updated as follows:
\begin{equation}
p^n(\sigma) = \frac{1}{z_{\rm v}} \left[p^{n-1}(\sigma)\right]^{\Gamma_n}
\sum_{t' \geq 1} b(t')\left(
\prod_{i=1}^{l+1}\left[\hrho_i(\sigma,t')+
\hQ_i(\sigma,t'-1)\right] -
\prod_{i=1}^{l+1}\left[\hrho_i(\sigma,t')+
\hQ_i(\sigma,t'-2)\right]\right) \ ,
\end{equation}
where for simplicity we kept implicit the $n$-dependence of the messages in the right hand side. We take initially $p^{0}(\sigma) = \frac12+\sigma z$ with $z$ a site-dependent small noise term to break the up-down symmetry, and let the intensity of this reinforcement term grow linearly during the iterations, according to $\Gamma_n = n \Gamma_{1}$, progressively increasing the external prior and thus selecting a single solution $\us$ of the original problem on which the measure becomes concentrated. It should be noted that the equations with the site-dependent noise term but without the reinforcement procedure, starting with random initial messages, did not converge on the regimes we tested. However, the reinforcement procedure did converge nevertheless on the regimes we report, producing solutions to the corresponding bicoloring problem. The whitening profile $P_t(\us,G)$ for the solution obtained can then be easily computed, some results are presented in Fig.~\ref{fig_numerics_reinforcement}, where the reinforced BP has been applied to the measure biased at $T=1$. As this figure demonstrates we managed to find solutions that are more frozen (i.e. that have larger whitening times) than the ones obtained by the usual BP+reinforcement algorithm by using a biasing parameter $\epsilon > 0$. Using a parameter $\epsilon <0$ allowed us to find solutions which whiten more at the first time step (in agreement with the choice $T=1$), yet the effect on $P_t$ at larger times is much less effective than in the case $\epsilon >0$. As could be expected neither the usual algorithm nor its biased version samples typical solutions of the uniform or biased measure (\ref{eq_mu}) with the same value of $\epsilon$; as explained above such an uniform sampling is an extremely difficult task. Surprisingly enough it turns out that the reinforced BP with a bias towards more frozen solutions discovers configurations whose distribution of whitening times is rather close to the theoretical prediction for uniformly drawn solutions.

\begin{figure}
\begin{center}
\includegraphics[width=8cm]{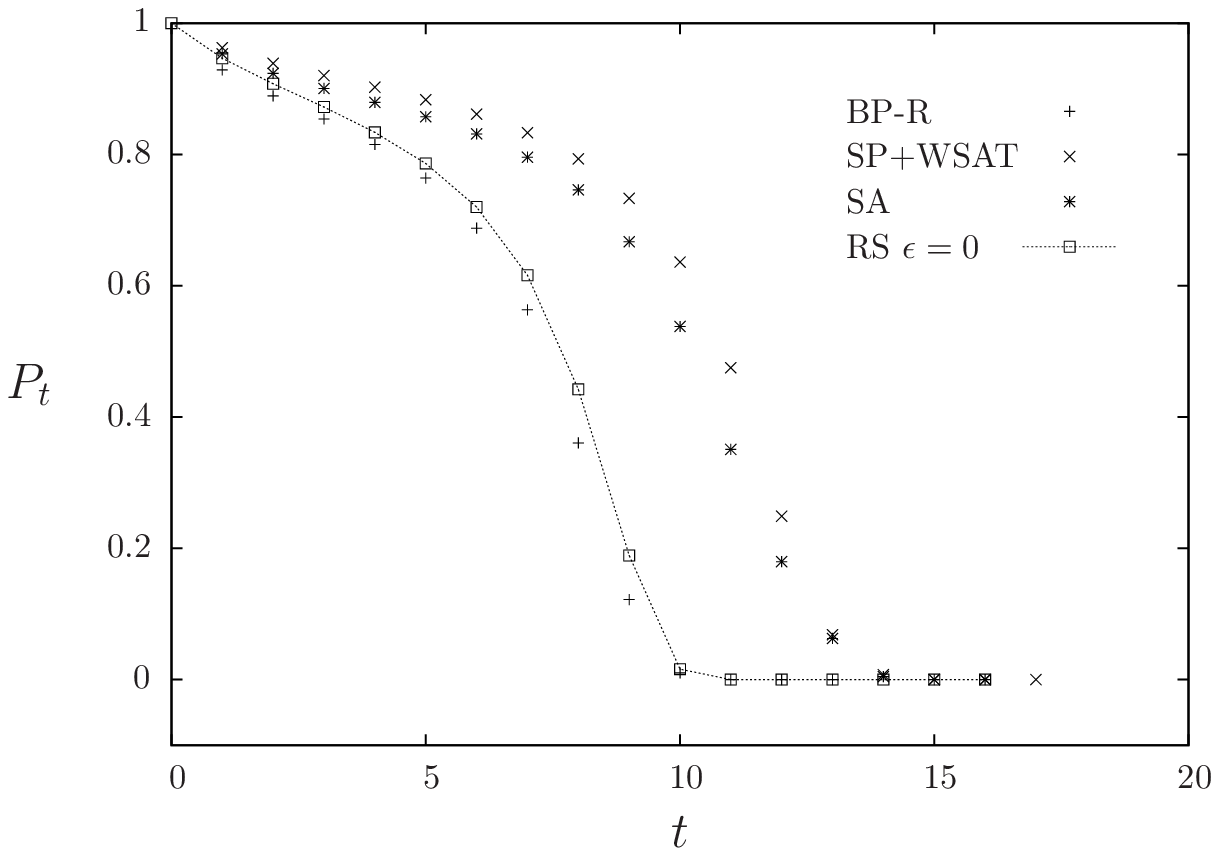}
\hspace{1cm}
\includegraphics[width=8cm]{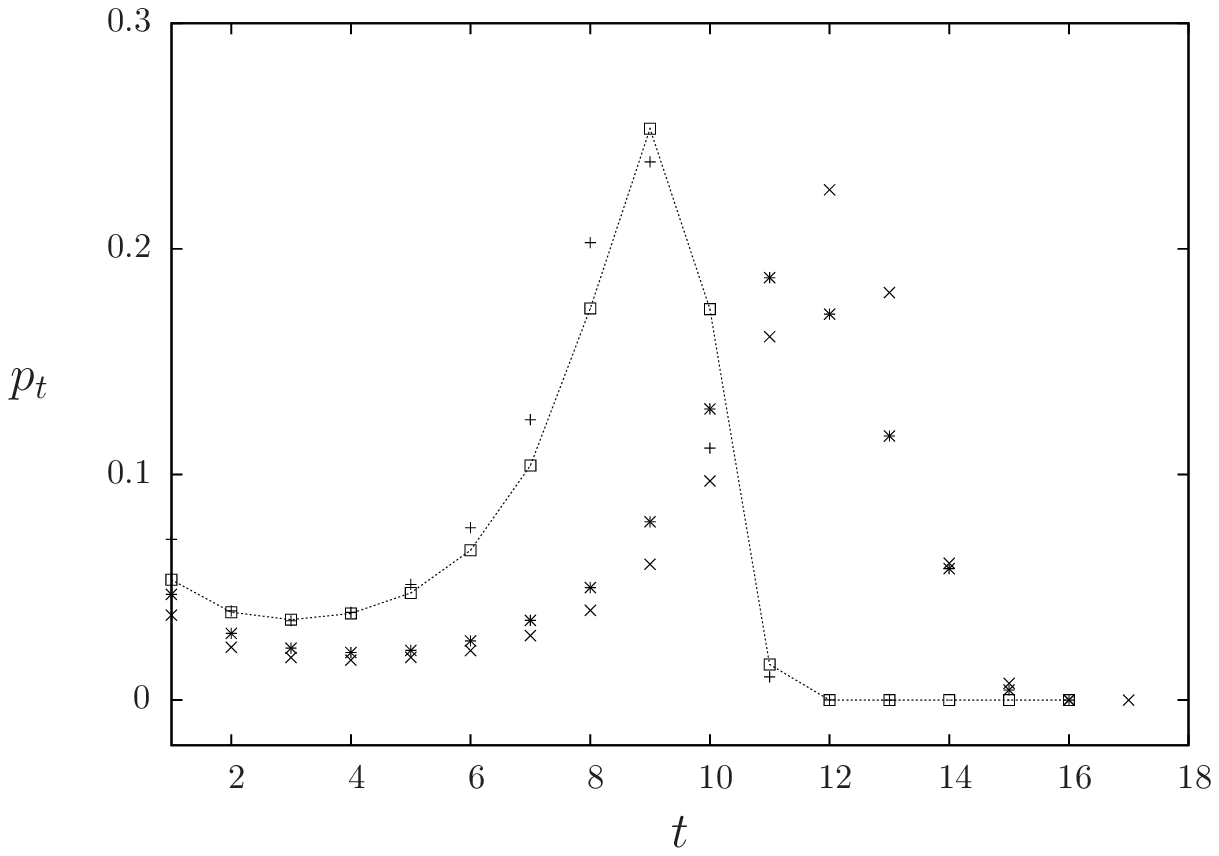} 
\caption{Cumulative distribution $P_t$ (left) and distribution $p_t=P_{t-1}-P_t$ (right) of whitening times in a hypergraph of size $N=10^4$, $k=4$ and $l=18$, for solutions obtained by three algorithms: BP+reinforcement~\cite{BraunsteinZecchina06} (averaged over 50 samples), SP+WalkSAT algorithm~\cite{CastellaniNapolano03} (20 samples), and Simulated Annealing (SA)~\cite{KirkpatrickGelatt83} (5 samples). For comparison we plotted the theoretical prediction for typical solutions (points joined by lines); the same symbols are used in both panels.}
\label{fig_numerics_algos}
\end{center}
\end{figure}

In a second set of experiments, whose results are displayed in Fig.~\ref{fig_numerics_algos}, we used three algorithms of various types (reinforced BP~\cite{BraunsteinZecchina06}, Survey Propagation+Walksat~\cite{CastellaniNapolano03}, and Simulated Annealing~\cite{KirkpatrickGelatt83}), and measured the distribution of whitening times for the solutions discovered by these algorithms. In this case we observed that the solutions found by reinforced BP have whitening profiles rather close to the theoretical one for uniformly chosen solutions, while the two other algorithms discover substantially more frozen solutions. 

\section{Conclusions and perspectives}
\label{sec_conclu}

In this paper we have presented a detailed characterization of the solutions of the bicoloring (or NAESAT) problem on regular random hypergraphs in terms of their fraction of frozen variables, both after a finite number of steps $T$ of the whitening algorithm and in their fixed point, which allowed us in particular to determine the location of the freezing transition above which all solutions are frozen. The asymptotic (at large $k$) scale of this transition, as well as its finite $T$ variants, is larger than the typical rigidity one, and rather close to the satisfiability transition. Let us sketch some possible directions for future research this work suggests.

A natural extension of this paper would be to generalize it to other constraint satisfaction problems, satisfiability and coloring in particular; we expect that the results obtained here for the bicoloring are generically valid for a large class of random CSPs, and that the asymptotic scalings at large $k$ will be the same for the satisfiability problem, as already explained in the introduction. The choice of the bicoloring problem we made was motivated by the technical simplifications it presents, on three aspects: the dynamics of its whitening process is equivalent to the warning propagation (or directional whitening) one, the RS solution for typical configurations is trivial (symmetric under the reversal of spins), and it can be studied on regular hypergraphs (with a factorized ansatz). Studying other CSPs will require to face new technical difficulties in absence of these three properties.

Most of the computations presented in this paper were performed using the cavity method in its simplest version, namely the replica symmetric one, even if on a biased measure over solutions which capture indirectly one aspect of replica symmetry breaking (the existence of hard fields). We only briefly discussed in Sec.~\ref{sec_rsb} the effects of replica symmetry breaking on the biased measure itself, one could pursue further in this direction and perform a 1RSB estimation of the thresholds $l_T(k)$ that we computed at the RS level. We expect however that the large $k$ asymptotics we derived would not be modified at their leading order.

We studied the dynamics of the whitening process, in the space of extended configurations $\{-1,1,0\}^N$, always initialized in solutions of the bicoloring problem, thus exploring a restricted part of this space. However the whitening can be defined in a larger subspace of $\{-1,1,0\}^N$, and a connection between its fixed points (not necessarily reached from a solution of the original problem) and the survey propagation algorithm~\cite{MezardParisi02,BraunsteinMezard05} was unveiled in~\cite{ManevaMossel05,BraunsteinZecchina04}: the complexity computed by SP (i.e. the 1RSB cavity method with $m=0$) counts the number of such fixed points. Our computation is thus distinct, as it concentrates on the fixed points reached from valid initial configurations, but the two perspectives might be combined, for instance to estimate how much weight SP gives to the ``spurious'' fixed points not reachable from solutions of the original CSP (the proof of the satisfiability threshold for NAESAT in~\cite{DiSlSu13_naeksat} is actually based on such an analysis for densities of constraints around the condensation regime).

In our finite $T$ computations we exploited the possibility to constrain a dynamical process in its $T$ first time steps, and yet to compute its trajectory at all later times, which led us to the definition of the tipping point according to the final state reached by the trajectory. We believe this idea could be applied to other similar dynamical processes, in particular the bootstrap percolation problem on random graphs~\cite{AlBrAsZe13,AlBrAsZe13b,GuSe15}.

Let us finally mention two major open challenges of a more algorithmic flavor. The solutions of CSPs found by heuristic algorithms seem to always be unfrozen; one can admit that frozen solutions are very hard to find because they contain an extensive number of frozen variables~\cite{ManevaMossel05} that should be set consistently in a collective way, inducing very strong correlations that seem hard to handle efficiently by polynomial time algorithms. This leaves open the questions of why and which unfrozen solutions are easier to find: to enlighten these points, in particular for stochastic local search algorithms, it would be necessary to study the energy landscape ``seen from an unfrozen solution'', and understand if it can possess a larger ``basin of attraction'' that would drive the local search towards it. 

The existence of unfrozen solutions at much larger densities than the rigidity threshold could in principle motivate some hope to break the algorithmic barrier which at present lies at the clustering/rigidity threshold (for large $k$). A promising perspective on this challenge, largely unexplored at the moment except for~\cite{BaInLuSaZe15,BaInLuSaZe15_long}, lies in the study of biased measures over solutions. As we have shown a properly chosen bias can turn an atypical property of the solutions sampled in the uniform measure (being unfrozen) into a typical property of the biased one, for rather large scale of densities even when the biasing function induces interactions of fixed range $T$. To put it as a joke it ``only'' remains to find the bias which makes the property of the solutions ``easy to find by some algorithm'' typical (or to prove that no such bias can exist).

\acknowledgments
We warmly thank
Victor Bapst,
Amin Coja-Oghlan,
Allan Sly,
Riccardo Zecchina and
an anonymous referee
for useful discussions and suggestions.
AB and LD acknowledge support by Fondazione CRT under the initiative “La Ricerca dei Talenti”. LD acknowledges the European Research Council for grant n. 267915.
\appendix

\section{Technical details on the cavity treatment}
\label{app_cavity}

\begin{figure}
\centerline{\includegraphics[width=16cm]{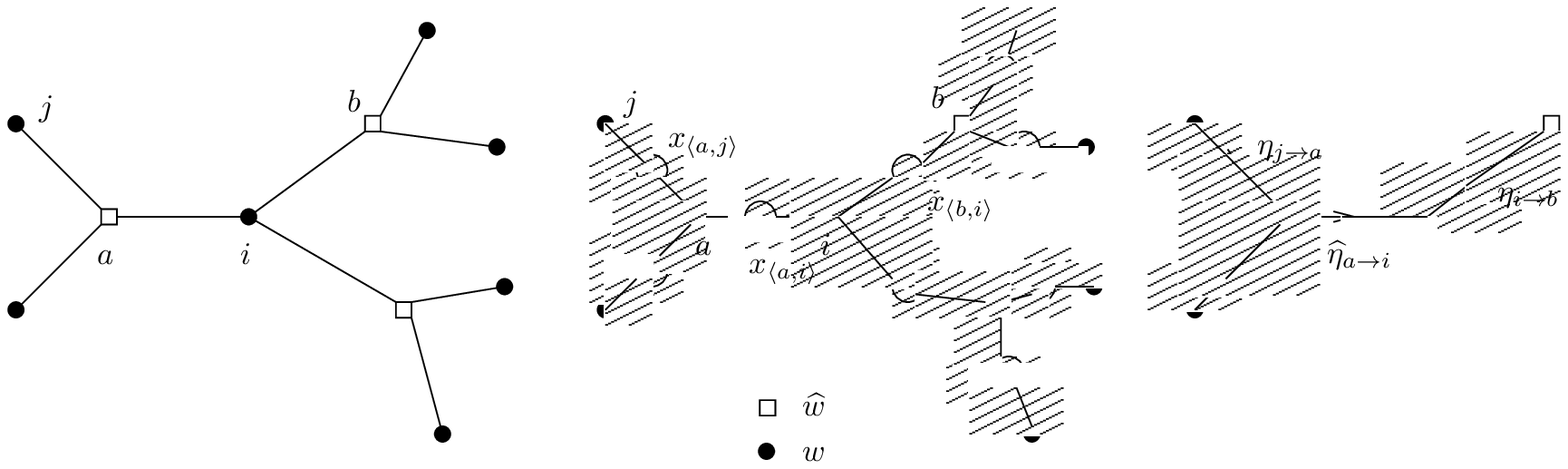}}
\caption{A portion of the bipartite graph representation of a hypergraph $G$ (left panel), the corresponding factor graph for the measure of Eq.~(\ref{eq_mu_x}) (center panel), and the location of the Belief Propagation messages (right panel).}
\label{fig_factor_graph_app}
\end{figure}

In this Appendix we give the details of the treatment of the graphical model introduced in Sec.~\ref{sec_statmech_fg} (see in particular Eq.~(\ref{eq_mu_x}) and Fig.~\ref{fig_factor_graph}), filling the gaps between these definitions and the results stated in Sec.~\ref{sec_statmech_regular} for the replica symmetric version of the computation on the regular hypergraphs.

\subsection{Simplifying the Belief Propagation equations}

Consider the factor graph depicted on the center panel of Fig.~\ref{fig_factor_graph_app}, with variables $x_{\la a,i \ra}$ on the edges of $G$ and interactions on both types of vertices of the bipartite graph representation of $G$. We shall now write the Belief Propagation equations for the messages exchanged between variable and interaction nodes, that would be the exact marginals (for amputated graphs) of (\ref{eq_mu_x}) if $G$ were a tree, as briefly explained in Sec.~\ref{sec_reminder} on the simpler case of the unbiased measure over proper bicolorings of $G$. Note first that all the variables nodes of the factor graph have degree two, hence the message from interaction $i$ to variable $\la a,i \ra$ is the same as the message from $\la a,i \ra$ to interaction $a$, we denote $\eta_{i \to a}(x_{\la a,i \ra})$ their common value. Similarly we denote $\heta_{a \to i}(x_{\la a,i \ra})$ the two messages flowing in the opposite direction, see the right panel of Fig.~\ref{fig_factor_graph_app} for an illustration. These messages obey the Belief Propagation equations
\beq
\eta_{i \to a} = f(\{\heta_{b \to i}\}_{b\in \dima}) \ , \qquad
\heta_{a \to i} = \hf(\{\eta_{j \to a}\}_{j\in \dami}) \ ,
\label{eq_BP_singlesample}
\eeq
where the functions $f$ and $\hf$ are easily found from the definition of the interaction nodes in (\ref{eq_def_hw},\ref{eq_def_w}) to be
\beq
\eta(\s,t,u) = \frac{1}{z} b(t) \sum_{u_1,\dots,u_l} \heta_1(\s,t,u_1) \dots 
\heta_l(\s,t,u_l) \, \ind(t=1+\max(u,u_1,\dots,u_l))
\label{eq_f}
\eeq
and
\bea
\heta(\s,t,u) = &&\frac{1}{\hz} \sum_{x_1,\dots,x_{k-1}} \eta_1(\s_1,t_1,u_1) \dots
\eta_{k-1}(\s_{k-1},t_{k-1},u_{k-1}) \, \ind(\s,\s_1,\dots,\s_{k-1} \ \text{n.a.e.})
\nonumber \\ &&
\ind(u=\ind(\s_1=\dots=\s_{k-1}=-\s)\min(t_1,\dots,t_{k-1})) \nonumber
\\ &&
\prod_{i=1}^{k-1} \ind(u_i=\ind(\s=-\s_i , \s_j =-\s_i \ \forall j\neq i)\min(t, \{t_j\}_{j\neq i})) \ ,
\label{eq_fhat}
\eea
the constants $z$ and $\hz$ ensuring the normalization of the messages. It is understood here that for a generic hypergraph the number of incoming messages $k-1$, $l$ can change from node to node, and we relabeled the messages in a more compact way. We adopt the convention that a summation on $x$ means a summation on $(\s,t,u)$, and when the domain of temporal summation is not precised it is understood that $t$ runs over $\{1,2,\dots,\infty\}$ (with an explicit infinity) and $u$ over $\{0,1,2,\dots,\infty\}$. 

The Bethe free-entropy prediction for $\phi(\varepsilon,G) = \frac{1}{N} \ln Z(\varepsilon,G)$ is obtained from the solution of the BP equations as
\beq
\phi(\varepsilon,G) = 
-\frac{1}{N} \sum_{\la a,i \ra} \ln z_{\rm e}(\eta_{i \to a},\heta_{a \to i}) +
\frac{1}{N} \sum_{a=1}^M \ln z_{\rm c}(\{\eta_{i \to a} \}_{i \in \da}) +
\frac{1}{N} \sum_{i=1}^N \ln z_{\rm v}(\{\heta_{a \to i} \}_{a \in \di}) \ ,
\label{eq_phi_Bethe}
\eeq
where the partition function for the three contributions are defined as
\bea
z_{\rm e}(\eta,\heta) &=& \sum_{x} \eta(x) \, \heta(x) \ ,
\label{eq_ze} \\
z_{\rm c}(\eta_1,\dots,\eta_k)&=&\sum_{x_1,\dots,x_k} \eta_1(x_1) \dots
\eta_k(x_k) \, \hw(x_1,\dots,x_k) \ ,
\label{eq_zc} \\
z_{\rm v}(\heta_1,\dots,\heta_{l+1}) &=&
\sum_{x_1,\dots,x_{l+1}} \heta_1(x_1) \dots \heta_{l+1}(x_{l+1}) \,
w(x_1,\dots,x_{l+1})
\label{eq_zv}
\eea
the last two lines being written for an hyperedge of degree $k$ and a variable of degree $l+1$ respectively.

The $(t,u)$ dependency of the messages $\eta$ and $\heta$ is not arbitrary and allows for some simplifications. Consider first Eq.~(\ref{eq_f}); it is clear that if $u>t-1$ then $\eta(\s,t,u)$ vanishes, while as long as $u<t-1$ the value of $\eta(\s,t,u)$ is actually independent of $u$. One has thus to consider only two cases for each value of $t$,
\bea
\eta(\s,t,t-1) &=& \frac{1}{z} b(t) \sum_{u_1,\dots,u_l} \heta_1(\s,t,u_1) \dots 
\heta_l(\s,t,u_l) \, \ind(\max(u_1,\dots,u_l)\le t-1) \ , \\
\eta(\s,t,u<t-1) &=& \frac{1}{z} b(t) \sum_{u_1,\dots,u_l} \heta_1(\s,t,u_1) \dots 
\heta_l(\s,t,u_l) \, \ind(\max(u_1,\dots,u_l) = t-1) \ .
\eea
The equation (\ref{eq_fhat}) can also be simplified. The important point to notice here is that among the $k$ variables $(\s,\s_1,\dots,\s_{k-1})$ at most one can be in the value opposite to all others, hence at most one of the times $(u,u_1,\dots,u_{k-1})$ can be different from $0$. If $u \ge 1$ this imposes the value of $(\s_1,\dots,\s_{k-1})$ and as a consequence $\heta(\s,t,u)$ is independent of $t$; if $u=0$ one has to distinguish two cases, either at least two of the spins in $(\s_1,\dots,\s_{k-1})$ are equal to $-\s$, or exactly one spin has this property. Thus the non-trivial values of $\heta$ are given by:
\bea
\heta(\s,\bullet,u\ge 1) &=& \frac{1}{\hz} \sum_{t_1,\dots,t_{k-1}} \eta_1(-\s,t_1,0) \dots
\eta_{k-1}(-\s,t_{k-1},0) \, \ind(u=\min(t_1,\dots,t_{k-1})) \ , \\
\heta(\s,t,0) &=& \frac{1}{\hz} \sum_{\substack{I\subset [1,k-1] \\ 2 \le |I| \le k-2 }} \prod_{i \in I} \left(\sum_{t'} \eta_i(-\s,t',0) \right) \prod_{i \notin I} \left( \sum_{t'} \eta_i(\s,t',0) \right) \\ && +  \frac{1}{\hz}
\sum_{i=1}^{k-1} \sum_{t_i} \sum_{\{t_j\}_{j \neq i}} \eta_i(-\s,t_i,\min(t,\{t_j\}_{j\neq i})) \prod_{j\neq i} \eta_j(\s,t_j,0) \ .
\eea

To deal with the minimization and maximization operations in these equations it is easier to introduce cumulative distributions. We shall represent $\eta(\s,t,u)$ by two quantities $Q(\s,t)$ and $R(\s,t)$, defined by
\beq
Q(\s,t) = \sum_{t' \ge t} \eta(\s,t',0) \ , \qquad 
R(\s,t) = \sum_{t'} \eta(\s,t',t) \ .
\label{eq_def_QR}
\eeq
Similarly $\heta(\s,t,u)$ is encoded by $\hQ(\s,t)$ and $\hrho(\s,t)$, with
\beq
\hQ(\s,t) = \sum_{u=1}^t \heta(\s,\bullet,u) \ , \qquad
\hrho(\s,t) = \heta(\s,t,0) \ .
\eeq

These quantities can be interpreted as follows; the $Q(\s,t)$ and $R(\s,t)$ summarizing the message $\eta_{i \to a}$ are the probabilities, in the amputated factor graph where $\dami$ has been removed, that $\s_i=\s$, $t_i \ge t$ and $\s_i$ is not forced by the $k-1$ other spins of clause $a$ (for $Q$), and that $\s_i=\s$ is forced to its original value by the other spins of $a$ until time $t$ (for $R$). Similarly $\hQ(\s,t)$ and $\hrho(\s,t)$ represent, for a message $\heta_{a \to i}$, the cavity probabilities that $\s_i=\s$ and that the other spins of $a$ forces initially the value of $\s_i$ but frees if before time $t$ (for $\hQ$), and that $\s_i=\s$ and its whitening time $t_i=t$ is not imposed by clause $a$ (for $\hrho$).

The BP equations (\ref{eq_f},\ref{eq_fhat}) are then found to be equivalently expressed in terms of these quantities as:
\bea
Q(\s,t) &=& \frac{1}{z} \sum_{t'\ge t} b(t') \left[ \prod_{i=1}^l (\hrho_i(\s,t') +\hQ_i(\s,t'-1) ) - \prod_{i=1}^l (\hrho_i(\s,t') +\hQ_i(\s,t'-2) )
\right] \ , \label{eq_Q_first} \\
R(\s,t) &=& \frac{1}{z} b(t+1) \prod_{i=1}^l (\hrho_i(\s,t+1) +\hQ_i(\s,t) ) 
\label{eq_R}
\\ &+& \frac{1}{z} \sum_{t'\ge t+2} b(t') \left[ \prod_{i=1}^l (\hrho_i(\s,t') +\hQ_i(\s,t'-1) ) - \prod_{i=1}^l (\hrho_i(\s,t') +\hQ_i(\s,t'-2) )
\right] \ , \nonumber \\
\hQ(\s,t) &=&\frac{1}{\hz} \left[ 
\prod_{i=1}^{k-1} Q_i(-\s,1) - \prod_{i=1}^{k-1} Q_i(-\s,t+1)
\right] \ , \label{eq_hQ} \\
\hrho(\s,t) &=& \frac{1}{\hz} 
\sum_{i=1}^{k-1} \left[  
R_i(-\s,t) \prod_{\substack{j=1 \\ j \neq i}}^{k-1} Q_j(\s,t) 
+ \sum_{t'=1}^{t-1} R_i(- \s,t') 
\left(  \prod_{\substack{j=1 \\ j \neq i}}^{k-1} Q_j( \s,t') 
- \prod_{\substack{j=1 \\ j \neq i}}^{k-1} Q_j( \s,t'+1)
\right) \right] 
 \label{eq_hrho} \\ &+& \frac{1}{\hz} 
\sum_{\substack{I \subset [1,k-1] \\ 2 \le |I| \le k-2}} 
\prod_{i \in I} Q_i(-\s,1) \prod_{i \notin I} Q_i(\s,1) \ .
 \nonumber
\eea
By convention one has to interpret $\hQ(\s,0)=0$ and $\hQ(\s,-1)=-\hrho(\s,1)$ to make sense of the terms with $t'=1,2$ in the first line. This form of the BP equations is definitely simpler than the original ones as all messages now depend on a single time index instead of two.

The Bethe free-entropy of (\ref{eq_phi_Bethe}) can be expressed with this parametrization of the $\eta$'s in terms of $Q,R$ and of the $\heta$'s in terms of $\hQ,\hrho$. Indeed the three contributions given in (\ref{eq_ze},\ref{eq_zc},\ref{eq_zv}) read respectively:

\bea
z_{\rm e} &=& \sum_{\s,t} \left[  
(Q(\s,t)-Q(\s,t+1)) \hrho(\s,t) 
+ R(\s,t) (\hQ(\s,t)-\hQ(\s,t-1)) \right] \ ,
\label{eq_ze_cumul} \\
z_{\rm c} &=& \sum_{\s,t}
\sum_{i=1}^k \left[  R_i(\s,t) \left(
\prod_{\substack{j=1 \\ j\neq i}}^k Q_j(-\s,t) - 
\prod_{\substack{j=1 \\ j\neq i}}^k Q_j(-\s,t+1) \right)
\right] 
+\sum_{\substack{I \subset [1,k] \\ 2 \le |I| \le k-2}} \prod_{i \in
  I}Q_i(+,1) \prod_{i \notin I} Q_i(-,1) \ ,
\label{eq_zc_cumul} \\
z_{\rm v} &=& \sum_{\s,t} b(t) \left( \prod_{i=1}^{l+1} (\hrho_i(\s,t)
  + \hQ_i(\s,t-1) ) -  \prod_{i=1}^{l+1} (\hrho_i(\s,t) +
  \hQ_i(\s,t-2) ) \right)  \ .
\label{eq_zv_cumul}  
\eea
Note that the normalizations of the messages can be chosen arbitrarily, they indeed compensate in the final result for $\phi$ given in (\ref{eq_phi_Bethe}): each message $\eta_{i \to a}$ (resp. $\heta_{a \to i}$) appears once in $z_{\rm e}$ and once with the opposite sign in $z_{\rm c}$ (resp. in $z_{\rm v}$).

Let us finally explain how to compute the average whitening path in the biased measure over solutions, first defined in (\ref{eq_def_Pt_epsilon}). It corresponds to an average over the different vertices $i$ of the cumulative distribution $P_t^{(i)}$ for the whitening time $t_i(\us)$ when the initial configuration $\us$ is drawn from the biased law (\ref{eq_mu}). The marginal distribution of $t_i$ under this law is easily expressed in terms of the Belief Propagation messages. For a vertex $i$ receiving the messages $\heta_1,\dots,\heta_{l+1}$ the probability that its whitening time is $\ge t+1$ reads
\beq
P_t = \frac{1}{z_{\rm v}} \sum_{\s,t'\ge t+1} b(t') \left( \prod_{i=1}^{l+1} (\hrho_i(\s,t') + \hQ_i(\s,t'-1) ) -  \prod_{i=1}^{l+1} (\hrho_i(\s,t') + \hQ_i(\s,t'-2) ) \right) \ ,
\label{eq_Pt_cumul}
\eeq
which satisfies the normalization condition $P_0=1$.

\subsection{Finite time horizon}
\label{app_horizon}

The Belief Propagation messages we have treated above are infinite dimensional, as the possible values of the whitening times are unbounded (in the thermodynamic limit); fortunately one can close the BP equations on a finite dimensional projection of the messages provided the biasing function $b(t)$ becomes constant beyond some time horizon $T$. For simplicity we shall only consider explicitly the case
\beq
b(1)=b(2)=\dots=b(T)=1 \ , \qquad b(t)=e^\epsilon \ \ \forall \ t \ge T+1 \ ,
\label{eq_standard}
\eeq
corresponding to the choice of $\varepsilon$ defined in (\ref{eq_phi_T}) that is required for the computation of the entropy $s(T,\theta)$, the translation in terms of the biasing function $b(t)$ being given in (\ref{eq_epsilon_to_b}). 

As $b(t)$ is constant for $t \ge T+1$, one finds from (\ref{eq_R}) and (\ref{eq_hrho}) that the functions $R$ and $\rho$ become independent of time beyond the time horizon $T$, namely $R(\s,t)=R(\s,T)$ and $\hrho(\s,t)=\hrho(\s,T)$ for all $t \ge T$. One can then simplify the remaining equations to obtain:
\bea
Q(\s,t) &=& \frac{1}{z} \sum_{t'=t}^T \left[ \prod_{i=1}^l (\hrho_i(\s,t') +\hQ_i(\s,t'-1) ) - \prod_{i=1}^l (\hrho_i(\s,t') +\hQ_i(\s,t'-2) )
\right] \label{eq_Q_horizon} \\
&+&  \frac{e^\epsilon}{z} \left[ \prod_{i=1}^l (\hrho_i(\s,T) +\hQ_i(\s,\infty) ) - \prod_{i=1}^l (\hrho_i(\s,T) +\hQ_i(\s,T-1) )
\right] \ \ \text{for} \ \ t \in [1,T] \ ,
\nonumber \\
Q(\s,t) &=& \frac{e^\epsilon}{z} \left[ \prod_{i=1}^l (\hrho_i(\s,T) +\hQ_i(\s,\infty) ) - \prod_{i=1}^l (\hrho_i(\s,T) +\hQ_i(\s,t-2) )
\right] \ \ \text{for} \ \ t \ge T+1 \ , \label{eq_Q_horizon_beyond} \\
R(\s,t) &=& \frac{1}{z} \prod_{i=1}^l (\hrho_i(\s,t+1) +\hQ_i(\s,t) )
+ \frac{1}{z} \sum_{t'=t+2}^T \left[ 
\prod_{i=1}^l (\hrho_i(\s,t') +\hQ_i(\s,t'-1) ) 
- \prod_{i=1}^l (\hrho_i(\s,t') +\hQ_i(\s,t'-2) )
\right]  \nonumber \\
&+&  \frac{e^\epsilon}{z} \left[ 
\prod_{i=1}^l (\hrho_i(\s,T) +\hQ_i(\s,\infty) ) 
- \prod_{i=1}^l (\hrho_i(\s,T) +\hQ_i(\s,T-1) )
\right] \ \ \text{for} \ \ t \in [1,T-1] \ , \label{eq_R_horizon} \\ 
R(\s,t) &=& \frac{e^\epsilon}{z} \prod_{i=1}^l (\hrho_i(\s,T) +\hQ_i(\s,\infty) ) \ \ \text{for} \ \ t \ge T \ .
\label{eq_RT_horizon}
\eea
The equations on $\hQ$ and $\hrho$ given in (\ref{eq_hQ},\ref{eq_hrho}) remain the same, except that one has to interpret $\hQ(\s,\infty)$ above as
\beq
\hQ(\s,\infty) =\frac{1}{\hz}
\prod_{i=1}^{k-1} Q_i(-\s,1) \ .
\label{eq_hQ_infty}
\eeq

To summarize these finite time horizon simplifications, the messages $\eta(\s,t,u)$ can now be encoded with the $4 T$ reals $\{Q(\s,1),\dots ,Q(\s,T), R(\s,1),\dots , R(\s,T) \}$ with $\s=\pm 1$, and the messages $\heta(\s,t,u)$ are represented similarly by $\{\hQ(\s,1),\dots ,\hQ(\s,T-1), \hQ(\s,\infty), \hrho(\s,1),\dots , \hrho(\s,T) \}$. Indeed the BP recursions between the two type of messages are closed on these finite-dimensional projections, that are enough to compute the Bethe prediction for the thermodynamic potential $\phi(T,\epsilon,G)$. One obtains indeed the following simplified forms of (\ref{eq_ze_cumul},\ref{eq_zc_cumul},\ref{eq_zv_cumul}):
\bea
z_{\rm e} &=& \sum_\s \left[  
\sum_{t=1}^{T-1} (Q(\s,t)-Q(\s,t+1)) \hrho(\s,t) 
+ Q(\s,T) \hrho(\s,T) \right.  \label{eq_ze_horizon}
\\  && \left. 
+ \sum_{t=1}^{T-1} R(\s,t) (\hQ(\s,t)-\hQ(\s,t-1)) + R(\s,T) (\hQ(\s,\infty)-\hQ(\s,T-1)) \right] \ , \nonumber \\
z_{\rm c} &=& \sum_\s
\sum_{i=1}^k \left[ \sum_{t=1}^{T-1} R_i(\s,t) \left(
\prod_{\substack{j=1 \\ j\neq i}}^k Q_j(-\s,t) - 
\prod_{\substack{j=1 \\ j\neq i}}^k Q_j(-\s,t+1) \right)
+ R_i(\s,T) \prod_{\substack{j=1 \\ j\neq i}}^k Q_j(-\s,T)
\right] \label{eq_zc_horizon} \\
&& +\sum_{\substack{I \subset [1,k] \\ 2 \le |I| \le k-2}} \prod_{i \in I} Q_i(+,1) \prod_{i \notin I} Q_i(-,1) \ ,
\nonumber \\
z_{\rm v} &=& \sum_\s \sum_{t=1}^T \left( \prod_{i=1}^{l+1} (\hrho_i(\s,t) + \hQ_i(\s,t-1) ) -  \prod_{i=1}^{l+1} (\hrho_i(\s,t) + \hQ_i(\s,t-2) ) \right) 
\label{eq_zv_horizon} \\
&& + e^\epsilon \sum_\s
\left( \prod_{i=1}^{l+1} (\hrho_i(\s,T) + \hQ_i(\s,\infty) ) -  \prod_{i=1}^{l+1} (\hrho_i(\s,T) + \hQ_i(\s,T-1) ) \right) \ .
\nonumber
\eea
Let us finally simplify the expression (\ref{eq_Pt_cumul}) for the cumulative distribution of the whitening times. For $t\in[1,T-1]$, one has
\beq
P_t = \frac{1}{z_{\rm v}} \sum_\s \sum_{t'= t+1}^T \left( \prod_{i=1}^{l+1} (\hrho_i(\s,t') + \hQ_i(\s,t'-1) ) -  \prod_{i=1}^{l+1} (\hrho_i(\s,t') + \hQ_i(\s,t'-2) ) \right) + \theta \ ,
\label{eq_Pt_small_horizon}
\eeq
where $\theta=P_T$, the marginal probability that a whitening time is $\ge T+1$, reads
\beq
\theta=\frac{e^\epsilon}{z_{\rm v}} \sum_\s
\left( \prod_{i=1}^{l+1} (\hrho_i(\s,T) + \hQ_i(\s,\infty) ) -  \prod_{i=1}^{l+1} (\hrho_i(\s,T) + \hQ_i(\s,T-1) ) \right) \ .
\label{eq_theta_horizon}
\eeq
It is also possible to express the values of $P_t$ beyond the time horizon $T$; this is particularly interesting to determine the probability of a variable remaining frozen forever (as the limit $t\to\infty$ of $P_t$). From (\ref{eq_Pt_cumul}) one finds for $t \ge T$: 
\beq
P_t = \frac{e^{\epsilon}}{z_{\rm v}} \sum_\s \left( \prod_{i=1}^{l+1} (\hrho_i(\s,T) + \hQ_i(\s,\infty) ) -  \prod_{i=1}^{l+1} (\hrho_i(\s,T) + \hQ_i(\s,t-1) ) \right) \ .
\eeq
The value of the messages beyond the time horizon $T$ is determined by (\ref{eq_Q_horizon_beyond}) and (\ref{eq_hQ}).

\subsection{Explicit time clipping}

The reader might be worried by the approach followed above, namely working with unbounded time variables, obtaining BP equations on infinite dimensional messages, and finally, in the case of a finite time horizon (i.e. a biasing function $b(t)$ becoming independent of $t$ for $t \ge T+1$), closing this infinite hierarchy of equations on a finite number of parameters for each message. Actually, a seemingly more direct and safer road can be taken to compute thermodynamic quantities in this case. Indeed one can clip from the beginning of the computation the domain of times involved, replacing Eq.~(\ref{eq_mu_stu}) by
\bea
\mu(\us,\ut,\uu) &=& \frac{1}{Z(T,\epsilon,G)} \prod_{a=1}^M w_a(\us_\da) \ \prod_{i=1}^N e^{\epsilon \delta_{t_i,\infty}} 
\prod_{i=1}^N \ind(t_i = \Theta_T(1 + \underset{a \in \di}{\max} \, u_{a \to i}))
\\ &&
\prod_{\la a,i \ra} \ind(u_{a \to i} = \ind(\s_j = -\s_i \ \forall j \in \dami) \  \min(T,\underset{j\in\dami}{\min} t_j) )
 \ ,
\eea
with 
\beq
\Theta_T(t)= \begin{cases} t & \text{if} \ t \in [1,T] \\ \infty & \text{otherwise}
\end{cases} \ .
\eeq
In other words one has projected the whitening time variables onto $t_i \in \{1,\dots,T,\infty\}$, the infinity value meaning any time $\ge T+1$ in the original computation, and the variables $u_{a\to i}$ are now in $\{0,1,\dots,T\}$. For any finite $T$ the variables of the factor graph are in a finite domain, the same steps as done above then leads directly to the set of equations obtained in App.~\ref{app_horizon}. We followed this more indirect road because it has an important advantage: it allows the computation of the distribution of the whitening times beyond the time horizon $T$, which is not possible if one clips the time variables from the outset.

\subsection{The regular graph case}

Assuming all hyperedges have degree $k$, all vertices have degree $l+1$, one can look for a factorized solution of the BP equations with all messages equal, and that furthermore respect the $\us \leftrightarrow -\us$ symmetry. This gives a simplified set of equations, where for compactness we denote $Q_t$ the common value of $Q_{i\to a}(\s,t)$ (and similarly for other quantities). From Eqs.~(\ref{eq_hQ},\ref{eq_hrho},\ref{eq_Q_horizon},\ref{eq_R_horizon},\ref{eq_RT_horizon},\ref{eq_hQ_infty}), one obtains $4T$ equations for the $4T$ unknowns $Q_1,\dots,Q_T,R_1,\dots,R_T,\hQ_1,\dots,\hQ_{T-1},\hQ_\infty,\hrho_1,\dots,\hrho_T$:

\bea
Q_t &=& \frac{1}{z} \sum_{t'=t}^T \left[ (\hrho_{t'} + \hQ_{t'-1})^l - (\hrho_{t'} + \hQ_{t'-2})^l \right] + \frac{e^\epsilon}{z}  \left[ (\hrho_T + \hQ_\infty)^l - (\hrho_T + \hQ_{T-1})^l \right] \qquad \text{for} \ t \in [1,T] 
\label{eq_reg_Qt_app}
\\
R_t &=& \frac{1}{z} (\hrho_{t+1} + \hQ_t)^l + \frac{1}{z} \sum_{t'=t+2}^T \left[ (\hrho_{t'} + \hQ_{t'-1})^l - (\hrho_{t'} + \hQ_{t'-2})^l \right]  + \frac{e^\epsilon}{z}  \left[ (\hrho_T + \hQ_\infty)^l - (\hrho_T + \hQ_{T-1})^l \right] \ \text{for} \ t \in [1,T-1]
\nonumber \\
R_T &=& \frac{e^\epsilon}{z} (\hrho_T + \hQ_\infty)^l \\
\hQ_t &=& \frac{1}{\hz} \left[Q_1^{k-1} - Q_{t+1}^{k-1} \right] \qquad \text{for} \ t \in [1,T-1] \\
\hQ_\infty &=& \frac{1}{\hz} Q_1^{k-1}  \\
\hrho_t &=& \frac{1}{\hz} \left\{ (2^{k-1} - k - 1) Q_1^{k-1} + (k-1)\left[ \sum_{t'=1}^{t-1} R_{t'} (Q_{t'}^{k-2} -Q_{t'+1}^{k-2} ) + R_t Q_t^{k-2}\right]
\right\} \qquad \text{for} \ t \in [1,T] \label{eq_reg_rhot_app}
\eea

The free-entropy averaged over the $l+1$-regular $k$-uniform hypergraph reads thus in the RS ansatz,
\beq
\phi(T,\epsilon,k,l) = - (l+1) \ln z_{\rm e} + \frac{l+1}{k} \ln z_{\rm c} + \ln z_{\rm v} \ ,
\eeq
with the expression of the partial partition functions obtained by simplifying (\ref{eq_ze_horizon},\ref{eq_zc_horizon},\ref{eq_zv_horizon}) with all messages equal:
\bea
z_{\rm e} &=&2 \sum_{t=1}^{T-1} (Q_t -Q_{t+1}) \hrho_t + 2 Q_T \hrho_T
+ 2 \sum_{t=1}^{T-1} R_t (\hQ_t - \hQ_{t-1}) + 2 R_T (\hQ_\infty - \hQ_{T-1}) \ ,
\\
z_{\rm c} &=& (2^k - 2 - 2k) Q_1^k + 2k \sum_{t=1}^{T-1} R_t (Q_t^{k-1} - Q_{t+1}^{k-1} )+ 2k R_T Q_T^{k-1} \ ,
\\
z_{\rm v} &=& 2 \sum_{t=1}^T \left( (\hrho_t + \hQ_{t-1} )^{l+1} - (\hrho_t + \hQ_{t-2} )^{l+1} \right) + 2 e^\epsilon \left( (\hrho_T + \hQ_\infty )^{l+1} - (\hrho_T + \hQ_{T-1} )^{l+1} \right)  \ .
\eea
We have the parametric expression of the entropy as
\beq
s(T,\theta,k,l)= \phi(T,\epsilon,k,l) - \epsilon \, \theta \ ,
\eeq
where the fraction $\theta=P_T$ of variables which are still frozen at time $T$ is (cf. (\ref{eq_theta_horizon}))
\beq
\theta = \frac{2 e^\epsilon}{ z_{\rm v}} \left( (\hrho_T + \hQ_\infty )^{l+1} - (\hrho_T + \hQ_{T-1} )^{l+1} \right) \ .
\label{eq_theta_reg_app}
\eeq

In the set of equations (\ref{eq_reg_Qt_app}-\ref{eq_reg_rhot_app}) the normalization constants $z$ and $\hz$ can be chosen arbitrarily, because as already said above the final expressions of $\phi$ and $\theta$ turn out to be independent of this choice. In the main text we used the convention $z=\hz=1$; if a solution of these equations is to be sought for by iteration it is on the contrary advisable for stability reasons to fix them by imposing a given value to one of the messages, for instance $Q_1=1$. Let us make a few other comments on these formula. The expression of $\phi$ given above is variational, in the sense that its derivatives with respect to the $Q_t$'s, $R_t$'s, $\hQ_t$'s and $\hrho_t$ vanish when the equations (\ref{eq_reg_Qt_app}-\ref{eq_reg_rhot_app}) are satisfied. This allows to check that the expression of $\theta$ given in (\ref{eq_theta_reg_app}) coincides with the derivative of $\phi$ with respects to $\epsilon$, as it should. Moreover one can show that if (\ref{eq_reg_Qt_app}-\ref{eq_reg_rhot_app}) are satisfied, then $z_{\rm v} = z \, z_{\rm e}$ and $z_{\rm c} = \hz \, z_{\rm e}$. In particular if one solves the RS equations with the convention $z=\hz=1$ on has $z_{\rm e}=z_{\rm c}=z_{\rm v}$ and the free-entropy simplifies into
\beq
\phi = \frac{1-l(k-1)}{k} \ln z_{\rm v} \ ,
\eeq
which is the expression we used in the main text.

Finally in the regular case the fraction of variables that have a whitening time $\ge t+1$, with $t \in [0,T-1]$, reads
\beq
P_t = \frac{2}{z_{\rm v}} \sum_{t'=t+1}^T \left( (\hrho_{t'} + \hQ_{t'-1} )^{l+1} - (\hrho_{t'} + \hQ_{t'-2} )^{l+1} \right) + \theta \ ,
\eeq
as obtained easily from (\ref{eq_Pt_small_horizon}). For times beyond the horizon $T$,
\beq
P_t = \frac{2 e^{\epsilon}}{z_{\rm v}}  \left( (\hrho_T + \hQ_\infty )^{l+1} - (\hrho_T + \hQ_{t-1} )^{l+1} \right) \ ,
\eeq
which can be computed for $t \ge T+1$ with in this case
\beq
Q_t = \frac{e^\epsilon}{z} 
\left( (\hrho_T +\hQ_\infty)^l - (\hrho_T +\hQ_{t-2} )^l \right) \ , \qquad
\hQ_{t-2} = \frac{1}{\hz} \left( Q_1^{k-1} - Q_{t-1}^{k-1} \right) \ .
\eeq
This concludes the justification of the expressions given in Sec.~\ref{sec_statmech_regular}.

\section{The large $k$ limit of the $T=1$ results}
\label{app_largek_T1}

In this appendix we shall give some technical details on the fate of the results presented in Sec.~\ref{sec_res_T1} in the limit $k\to\infty$, our main goal being the justification of Eq.~(\ref{eq_l1_largek}) that gives the asymptotic expansion of the threshold $l_1(k)$ for the existence of unfrozen solutions that can be made thermodynamically relevant by a bias on the whitening state after a single time step.

Before entering in the core of the computations let us make some remarks and define some notations that will be useful for all the large $k$ expansions performed below. We have already recalled in Sec.~\ref{sec_reminder} the known results about the various phase transitions undergone by the uniform measure over solutions (i.e. the typical ones), and shown that all their thresholds share a leading asymptotic behavior in the large $k$ limit, proportional to $2^k$. Hence in the following we shall denote $l(k)=2^{k-1} \, \hl(k)$ for a sequence of degrees depending on $k$ (we will often keep this dependency understood when there is no risk of confusion) and work with reduced degrees $\hl$. We assume here that $\hl(k)$ does not vary exponentially with $k$, otherwise this would change the leading $2^k$ behavior of $l(k)$; for the ease of the discussion we will say that a series in $k$ is ``slow'' if it has no exponential behavior at large $k$, $\hl(k)$ being an example of a slow series. More generically we shall have to deal with asymptotic expansions of $k$ dependent quantities, hierarchically organized in three levels as expansions in exponentials in $k$, powers of $k$, and (iterated) logarithms of $k$. To simplify notations we will denote $\hO(\bullet)$ a series in $k$ whose asymptotic expansion is equal to the one of $\bullet$ at the exponential level, discarding their power-law behavior, and $\tO(\bullet)$ will have the same meaning at the level of powers of $k$, discarding logarithmic contributions. The notations $\ho$ will mean instead ``negligible at the exponential level in the expansion at large $k$''. 

\subsection{The large $k$ limit of the entropy function}

We have to solve the equations (\ref{eq_T1_Q}-\ref{eq_T1_hrho}) in the limit $k\to\infty$, with $l =2^{k-1}\, \hl$ under the assumption that $\hl$ and $\epsilon$ are slow functions of $k$. A moment of thought reveals that the solution of these equations have to scale as
\bea
Q &=& \frac{1}{2} \left(1 + \frac{1}{l} Q' + \ho \left(\frac{1}{l}\right) \right) \ , \\
R &=& R' + \hO \left(\frac{1}{l}\right) \ , \\
\hQ &=& \frac{1}{2^{k-1}} \left(1 + \frac{1}{l} \hQ' + \ho \left(\frac{1}{l}\right) \right) \ , \\
\hrho &=& 1 + \frac{1}{l} \hrho' + \ho \left(\frac{1}{l}\right) \ ,
\eea
where the primed quantities are slowly depending on $k$. Inserting this ansatz in the equations (\ref{eq_T1_Q}-\ref{eq_T1_hrho}) gives, neglecting exponentially small corrections,
\bea
\frac{1}{2} &=& e^{\hrho'} \left(1+e^\epsilon \left(e^\hl - 1 \right) \right) \ , \\
R' &=& e^{\epsilon+\hl+\hrho '} \ , \\
\hQ' &=& (k-1) Q' \ , \\
\hrho' &=& (k-1) Q' - \hl (k+1) + 2 \hl (k-1) R' \ ,
\eea
which can be solved to give explicitly $Q'$, $R'$, $\hQ'$ and $\hrho'$ as a function of $\epsilon$ and $\hl$.

Consider now the expression of $z_{\rm v}$ given in (\ref{eq_T1_thermo}). We shall prove that $z_{\rm v} = 1 + \frac{1}{l} z'_{\rm v} + \ho \left(\frac{1}{l}\right)$, with $z'_{\rm v} = \hO(1)$. Indeed
\bea
z_{\rm v} &=& 2 \hrho^{l+1} + 2 e^\epsilon \left((\hrho + \hQ)^{l+1} - \hrho^{l+1} \right) \\
&=& 2 \hrho^l \left(1 + \frac{1}{l} \ln (\hrho^l ) + \ho \left(\frac{1}{l}\right)\right) 
+ 2 e^\epsilon \left(
(\hrho + \hQ)^l \left(1 + \frac{1}{l} \ln ((\hrho+\hQ)^l ) + \ho \left(\frac{1}{l}\right)\right)
- \hrho^l \left(1 + \frac{1}{l} \ln (\hrho^l) + \ho \left(\frac{1}{l}\right)\right)
\right) \nonumber \\
&=& 2 \hrho^l + 2 e^\epsilon \left((\hrho + \hQ)^l - \hrho^l \right) + \frac{1}{l}
\left[ 2 \hrho^l \ln (\hrho^l) +  2 e^\epsilon (\hrho + \hQ)^l \ln ((\hrho + \hQ)^l)
- 2 e^\epsilon \hrho^l \ln (\hrho^l)
\right] + \ho\left(\frac{1}{l} \right) \ . \nonumber
\eea
One recognizes in the first term the (exact) equation on $Q$, namely (\ref{eq_T1_Q}), while the terms which are explicitly of order $1/l$ can be computed at the lowest order. This yields the announced form for $z_{\rm v}$, with
\beq
z'_{\rm v} = Q' + 2 \hrho' e^{\hrho'} + 2 e^\epsilon (\hrho' + \hl) e^{\hrho' + \hl} - 2 e^\epsilon \hrho' e^{\hrho'} \ .
\eeq
With the expression of $\phi$ in terms of $z_{\rm v}$ given in (\ref{eq_phi_T_reg}), this scaling of $z_{\rm v}$ yields $\phi = - \frac{k-1}{k} z'_{\rm v} + \ho\left(\frac{1}{l} \right) $. Inserting the explicit solution of $Q'$ and $\hrho'$ as a function of $\epsilon$ and $\hl$, and treating similarly the expression of $\theta$ gives, within corrections exponentially small in $k$:
\beq
\phi = \ln 2 - \hl \ \frac{k+1}{k}  + \ln \left(1 + e^\epsilon (e^\hl -1) \right) \ , 
\qquad \theta = \frac{e^\epsilon (e^\hl -1)}{1 + e^\epsilon (e^\hl -1)} \ .
\label{eq_thermo_T1_largek}
\eeq
The parametric relation between $\epsilon$ and $\theta$ can here be inverted, and yields finally the asymptotic expression for $s(\theta)$:
\beq
s(T=1,\theta,k,l=2^{k-1} \, \hl) = \ln 2 - \hl \ \frac{k+1}{k}  - \theta \ln \theta - (1-\theta) \ln(1-\theta) + \theta \ln \left( e^\hl -1 \right) + \hO \left(\frac{1}{l}\right) \ ,
\label{eq_softheta_T1_largek}
\eeq
which is exact in the large $k$ limit up to exponentially small corrections. Note that it has the form of the binary entropy tilted by an affine function of $\theta$. It is easy to check that the maximum of $s$ is reached in the typical value $\theta = 1 - e^{-\hl}$, where the entropy is equal to $\ln 2 - \frac{\hl}{k}$. 

\subsection{The tipping point}

As discussed in the main text the configurations counted in the entropy $s(T,\theta,k,l)$ ultimately whiten if and only if $\theta<\ttip(T,k,l)$, which defines this tipping point threshold. In the canonical computation this translates into a critical value for the conjugated parameter $\epsilon$, which is determined by the two equations (\ref{eq_Qinfty},\ref{eq_Qinfty_deriv}) for generic values of $T$, or (\ref{eq_Qinfty_T1},\ref{eq_Qinfty_deriv_T1}) in the case $T=1$, in which $Q_*$ and $\epsilon$ are to be treated as unknowns. In the large $k$ limit we shall make the ansatz $Q_*=\frac{1}{2}(1-Q'_*)+ \hO \left(\frac{1}{l}\right)$, where $Q'_*$ is assumed to vanish slowly. The equations (\ref{eq_Qinfty_T1},\ref{eq_Qinfty_deriv_T1}) thus become, neglecting exponentially small corrections,
\bea
\frac{1}{2}(1-Q'_*) &=& e^{\epsilon+\hl+\hrho'} \left(1-e^{-\hl (1-Q'_*)^{k-1}} \right) \ , \\
\frac{1}{2} &=& e^\epsilon \hl (k-1) (1-Q'_*)^{k-2} e^{\hrho' + \hl - \hl (1-Q'_*)^{k-1}} \ .
\eea
Using the expression of $\hrho'$ obtained previously one can solve these two equations and obtain $Q'_*$ and $\epsilon$ as a function of $k$ and $\hl$, in particular
\beq
1-Q'_* = \left(\frac{w_k}{\hl} \right)^\frac{1}{k-1} \ , 
\eeq
where $w_k$ is a series depending only on $k$, given implicitly as the solution of $e^{w_k} = 1+ (k-1) w_k$. One can obtain similarly the value of $\epsilon$ at the tipping point, and inserting the latter in the formula (\ref{eq_thermo_T1_largek}) for $\theta$ yields
\beq
\ttip(T=1,k,l=2^{k-1}\hl) = (1-e^{-\hl}) \left(\frac{1}{\hl}\right)^{\frac{1}{k-1}} w'_k \ , \qquad \text{with} \ \ \
w'_k = \left(1 + \frac{1}{(k-1) w_k} \right) w_k^{\frac{1}{k-1}} \ .
\label{eq_ttip_T1_largek}
\eeq
Again this expression is exact within exponentially small corrections. It is not too difficult to check that $\ttip$ crosses its typical value $(1-e^{-\hl})$ when $l$ reaches the typical rigidity threshold $\lr$ (expanding Eq.~(\ref{eq_rigidity_typ}) at the exponentially dominant order).

\subsection{The asymptotic behavior of the threshold $l_1(k)$}

We recall that the threshold $l_1(k)$ is defined by the cancellation of the entropy of the tipping point, and is interpreted in the RS framework as the limit of existence of unfrozen solutions in the typical configurations sampled by the measure biased according to the state of the whitening process after $T=1$ time steps. We want to determine here its asymptotic scaling at large $k$, and justify the statement made in Eq.~(\ref{eq_l1_largek}). Combining the expressions (\ref{eq_softheta_T1_largek}) and (\ref{eq_ttip_T1_largek}) of the entropy and the tipping point we have an (implicit) determination of $\hl_1(k)=l_1(k)/2^{k-1}$ as the solution of
\beq
0 = \ln 2 - \hl \left(\frac{1}{k} + 1-\theta \right) - \theta \ln \theta - (1-\theta) \ln(1-\theta) + \theta \ln (1-e^{-\hl}) \ ,
\label{eq_l1_exponential}
\eeq
where for clarity we abbreviated in $\theta$ the expression $\ttip(k,\hl)$ given in (\ref{eq_ttip_T1_largek}). This determination of $\hl_1(k)$ is exact within exponentially small corrections, but still rather implicit. To obtain the leading terms in the asymptotic expansion one can further simplify the equation. A moment of thought reveals that the transition occurs on an intermediate scale between the typical rigidity and the satisfiability, namely $\ln k \ll \hl_1(k) \ll k$, and that on this scale $\ttip(k,\hl) \to 1$; moreover it is easy to see that $w_k$ behaves logarithmically at large $k$. Expanding (\ref{eq_ttip_T1_largek}) we thus obtain
\beq
\ttip(T=1,k,l=2^{k-1}\hl) = 1 - \frac{1}{k} \left( \ln \hl - \ln w_k - \frac{1}{w_k} \right) + \tO\left(\frac{1}{k^2}\right) \ ,
\eeq
where we recall that $\tO$ hides (at most) logarithmic contributions in $k$. Replacing the latter expansion in (\ref{eq_l1_exponential}) gives
\beq
0 = \ln 2 - \frac{\hl}{k} \left(1 + \ln \hl - \ln w_k - \frac{1}{w_k} \right) + \tO\left(\frac{1}{k}\right) \ .
\label{eq_l1_polynomial}
\eeq
Solving this (still implicit) equation neglecting the correction term gives an expression for $\hl_1(k)$ which is exact up to relative corrections of dominant order $1/k$. Noting that $w_k = \ln k + O(\ln \ln k)$,
one can finally gives explicitly the asymptotic expansion of the solution of (\ref{eq_l1_polynomial}):
\beq
\hl_1(k) = \frac{k \ln 2}{\ln k - 2 \ln\ln k + 1 - \ln 2} \left(1+O\left( \frac{\ln \ln k}{(\ln k)^2}  \right)\right) \ ,
\eeq
from which follows the truncated expansion of Eq.~(\ref{eq_l1_largek}).

\section{The fixed points of the whitening (large $T$ limit)}
\label{app_Tinfty}

In this appendix we shall explain with more details the computations that led us to the predictions for the entropy $s_\infty(\theta)$ presented in Sec.~\ref{sec_res_Tinfty}, via a study of the limit $T\to\infty$ of the RS entropy $s(T,\theta)$.

\subsection{A more compact equivalent form of the RS equations}
\label{sec_compact_RS}

As a first step we shall rewrite the equations (\ref{eq_reg_Qt}-\ref{eq_reg_rhot}), that have $4T$ unknowns, into an equivalent form with only half of the unknowns. Obviously one can eliminate very easily the variables $\hQ_t$ using (\ref{eq_reg_hQt},\ref{eq_reg_hQinfty}); with a little bit more of work one can also eliminate the $R_t$'s, considering the differences $R_t-R_{t+1}$ for two consecutive times. This yields an equivalent set of $2T$ equations on the $2T$ unknowns $Q_1,\dots,Q_T,\hrho_1,\dots,\hrho_T$ :
\bea
Q_t - Q_{t+1} &=& (\hrho_t + Q_1^{k-1} - Q_t^{k-1})^l - (\hrho_t + Q_1^{k-1} - Q_{t-1}^{k-1})^l \qquad \text{for} \ t \in [1,T-1] \label{eq_Et} \\
Q_T &=& (\hrho_T + Q_1^{k-1} - Q_T^{k-1})^l - (\hrho_T + Q_1^{k-1} - Q_{T-1}^{k-1})^l + e^\epsilon \left[ (\hrho_T + Q_1^{k-1})^l - (\hrho_T + Q_1^{k-1} - Q_T^{k-1})^l\right] \label{eq_ET} \\
\hrho_1 &=& (2^{k-1}-2) Q_1^{k-1} + (k-1)Q_1^{k-2} (\hrho_2^l-\hrho_1^l) \label{eq_F1} \\
\hrho_t - \hrho_{t+1} &=& (k-1)Q_{t+1}^{k-2} \left[(\hrho_{t+1} + Q_1^{k-1} - Q_{t+1}^{k-1})^l - (\hrho_{t+2} + Q_1^{k-1} - Q_{t+1}^{k-1})^l \right]
\qquad \text{for} \ t \in [1,T-2] \label{eq_Ft} \\
\hrho_{T-1} - \hrho_T &=& (k-1)Q_T^{k-2} (1-e^\epsilon) (\hrho_T + Q_1^{k-1} - Q_T^{k-1})^l 
\ , \label{eq_FT}
\eea
where in the first line for $t=1$ one has to interpret by convention $Q_0$ in such a way that 
\beq
\hrho_1 + Q_1^{k-1} - Q_0^{k-1} = 0 \ .
\label{eq_E0}
\eeq
From these equations one can deduce that the $Q_t$'s are decreasing with $t$ (which is obvious from their definition given in Eq.~(\ref{eq_def_QR})), and that the $\hrho_t$'s are increasing (resp. decreasing) with $t$ if $\epsilon > 0$ (resp. if $\epsilon <0$).

This equivalent set of equations will be used in the following for the analytical computations; note that it is also more practical for a numerical resolution by the Newton-Raphson method. Indeed the latter involves the inversion of a matrix of order the number of equations and unknowns, hence reducing this from $4T$ to $2T$ allows to explore numerically larger values of this parameter.

\subsection{A dynamical system point of view}
\label{sec_dynamical}

In the $T\to\infty$ limit the set of equations written above can be solved self-consistently with different ansatz, one corresponding to the unfrozen solutions ($\theta=0$), two other ones corresponding to the frozen solutions of the two kinds. Before stating these ansatz we shall reinterpret the equations in terms of a discrete time dynamical system, that will give a more intuitive view on the form of the solutions; similar considerations were developed in~\cite{GuSe15} to study the large deviation properties of the bootstrap percolation dynamics.

Let us consider $Q_1$ as a fixed parameter in the equations (\ref{eq_Et}-\ref{eq_FT}), and concentrate on the structure of the dependencies between the time-dependent unknowns $Q_t,\hrho_t$. The equation (\ref{eq_Et}) can be rewritten as $Q_{t+1} = f_Q(Q_t,Q_{t-1},\hrho_t)$, while (\ref{eq_Ft}) amounts to $\hrho_{t+2}=f_\hrho(\hrho_{t+1},\hrho_t,Q_{t+1})$, where $f_Q$ and $f_\hrho$ are two functions. Introducing a four-dimensional vector $w_t$ defined by
\beq
w_t=\begin{pmatrix} Q_{t\phantom{+1}} \\ Q_{t+1} \\ \hrho_{t+1} \\ \hrho_{t\phantom{+1}} \end{pmatrix} \ ,
\eeq
the equations above are seen to be equivalent to a simple recursion on $w_t$, namely $w_{t+1} = R(w_t)$ where the map $R$ acts on four-dimensional vectors according to:
\beq
R \begin{pmatrix} Q_{\phantom +} \\ Q_+ \\ \hrho_+ \\ \hrho_{\phantom +} \end{pmatrix} =
\begin{pmatrix} Q_+ \\ f_Q(Q_+,Q,\hrho_+)  \\ f_\hrho(\hrho_+,\hrho,Q_+) \\ \hrho_+ \end{pmatrix}
\ .
\eeq
Finally the equations (\ref{eq_ET},\ref{eq_F1},\ref{eq_FT},\ref{eq_E0}) that were not yet used in this rewriting induce boundary conditions on this recursion, which can be expressed as two conditions on the vectors $w_0$ and two other conditions on $w_T$. The solution of the RS equations is thus such that it goes from the two-dimensional manifold in which $w_0$ is constrained to the two-dimensional manifold of $w_T$ by $T$ iterations of the map $R$. When $T\to\infty$ it is natural to expect that these iterations will spend most of the time close to fixed-points of $R$, otherwise repeated iterations of $R$ would make the dynamical system flow very quickly far away from the authorized region in its four-dimensional space.

The fixed points of $R$ have a very simple structure: they correspond to $w$ in the two-dimensional manifold with $Q=Q_+$, $\hrho=\hrho_+$. Studying the Jacobian of $R$ at such a fixed point one discovers (after a short computation that we do not detail further) that there are two trivial marginal directions (eigenvectors of the Jacobian with eigenvalue 1), corresponding to the invariances $Q \to Q + \delta Q$, $\hrho \to \hrho + \delta \hrho$ of the manifold of fixed points. Besides there is an eigenvector in the $(Q,Q_+)$ direction with eigenvalue $\mu$, and another eigenvector in the $(\hrho,\hrho_+)$ direction with eigenvalue $1/\mu$, with
\beq
\mu = l(k-1)Q^{k-2} (\hrho + Q_1^{k-1} - Q^{k-1})^{l-1} \ .
\label{eq_def_mu}
\eeq
There is thus a line of marginal fixed points corresponding to $\mu=1$, fixed-points which are stable in the $Q$ direction and unstable in the $\rho$ direction ($\mu<1$), and fixed-points which are unstable in the $Q$ direction and stable in the $\rho$ direction ($\mu>1$).

We shall now consider different ansatz for the solution of the RS equations in the large $T$ limit and see how they nicely fit in this dynamical system perspective.

\subsection{The unfrozen solutions for $l>\lr$}
\label{app_Tinfty_unfrozen}

\begin{figure}
\includegraphics[width=8cm]{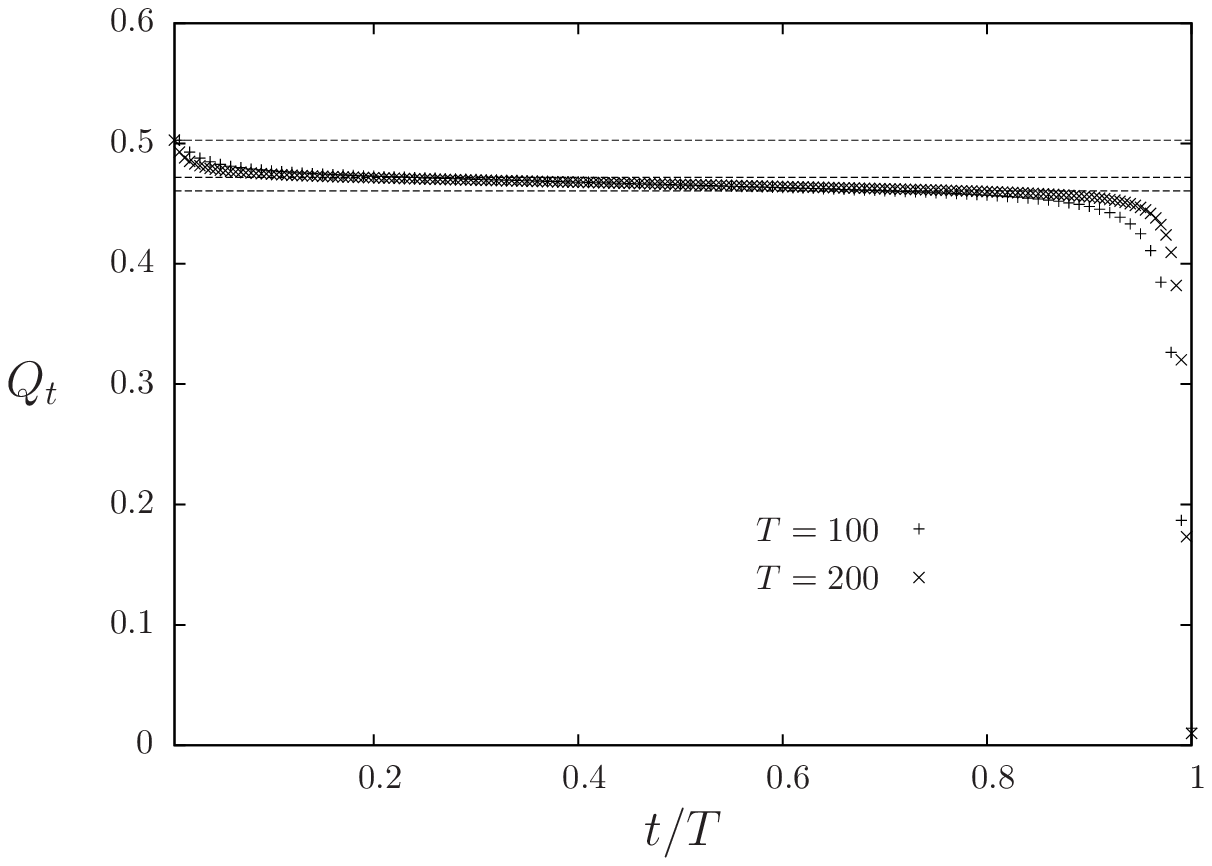}
\hspace{1cm}
\includegraphics[width=8cm]{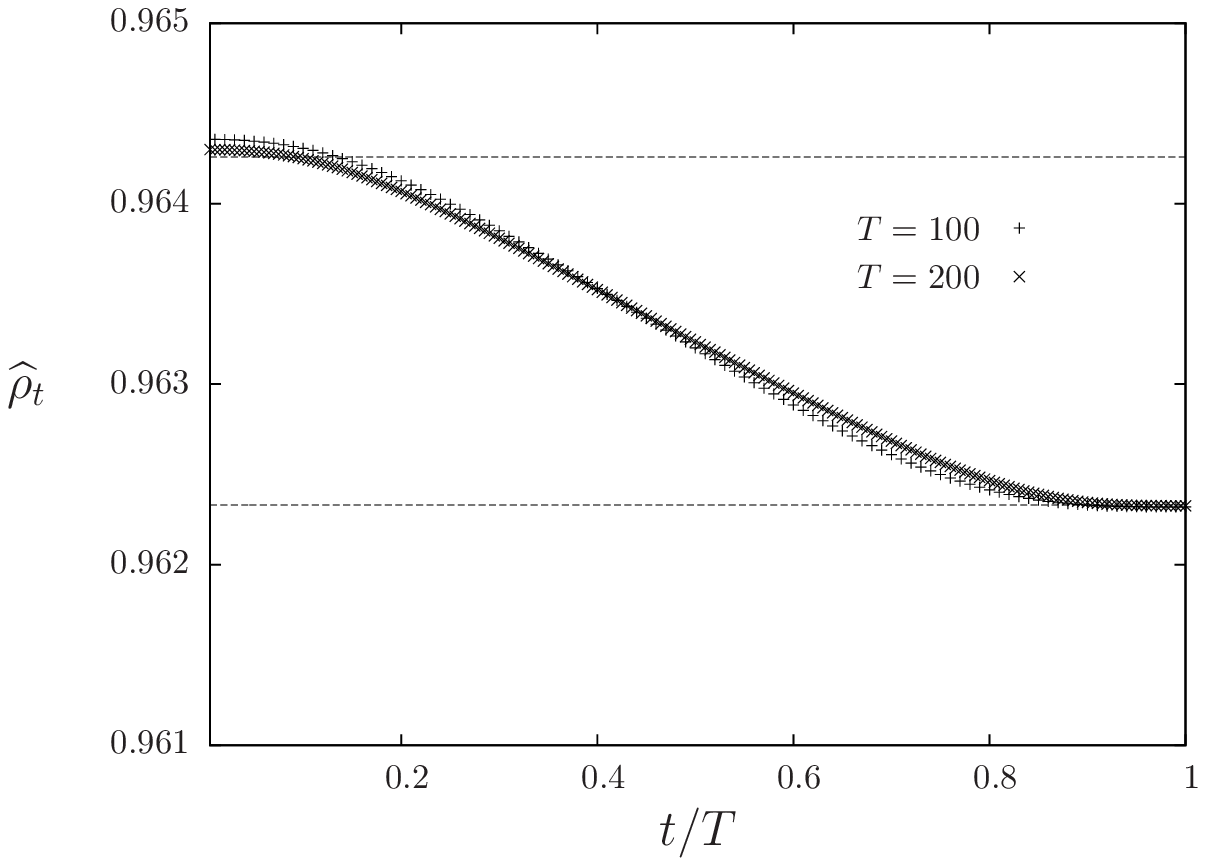}
\caption{The solution $(Q_t,\hrho_t)$ for $k=6$, $l=127 > \lr$, $e^\epsilon=0$, $T=100$ and $T=200$. Left panel : $Q_t$, the dashed lines from top to bottom are $Q_1$, $Q_{\rm i}$ and $Q_{\rm f}$ obtained from the analytical results at $T\to\infty$, cf. Eqs.~(\ref{eq_Tinfty_mans_1}-\ref{eq_Tinfty_mans_5}). Right panel : $\hrho_t$, the dashed lines representing $\hrho_{\rm i}$ (top) and $\hrho_{\rm f}$ (bottom).}
\label{fig_Tinfty_mans}
\end{figure}

\begin{figure}
\includegraphics[width=8cm]{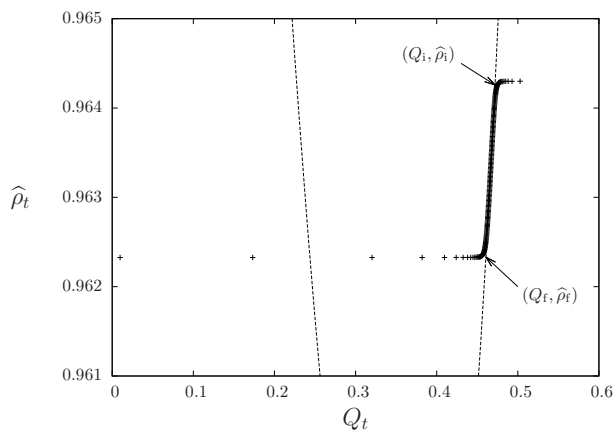}
\caption{The solution $(Q_t,\hrho_t)$ for $k=6$, $l=127$, $e^\epsilon=0$, $T=200$, plotted parametrically as $\hrho_t$ as a function of $Q_t$. The dashed line indicating the marginal condition $\mu=1$, the arrows point to the beginning and end of the scaling regime, $(Q_{\rm i},\hrho_{\rm i})$ and $(Q_{\rm f},\hrho_{\rm f})$ respectively, computed analytically from Eqs.~(\ref{eq_Tinfty_mans_1}-\ref{eq_Tinfty_mans_5}).}
\label{fig_Tinfty_mans_parametric}
\end{figure}

We consider first the solution of the RS equations that describe the isolated point at $\theta=0$ in the $s_\infty(\theta)$ curve for $l>\lr$. We present in Fig.~\ref{fig_Tinfty_mans} the results of a numerical resolution of the RS equations for two large values of $T$, with $e^\epsilon=0$ in order to impose the constraint $\theta=0$. It is seen on these curves that, apart from values of $t$ close to $1$ and close to $T$, the evolution of $Q_t$ and $\hrho_t$ seems to be governed by smooth scaling functions of a reduced time $t/T$; moreover the deviations from this scaling regime only affects the behavior of $Q_t$, not of $\hrho_t$.

More quantitatively one can assume the existence of two scaling functions $Q(s)$ and $\hrho(s)$, for $s \in ]0,1[$, such that $Q(s) = \lim Q_{t=s T}$, $\hrho(s) = \lim \hrho_{t=s T}$. Plugging this ansatz in the equations (\ref{eq_Et},\ref{eq_Ft}) and expanding in powers of $1/T$ one realizes that this ansatz is consistent only if the two scaling functions satisfy the following condition for all $s \in ]0,1[$:
\beq
1 = l(k-1)Q(s)^{k-2} (\hrho(s) + Q_1^{k-1} - Q(s)^{k-1})^{l-1} \ .
\eeq
Comparing with (\ref{eq_def_mu}) one realizes that this condition means that the flow of the dynamical system follows the line of marginal fixed points ($\mu=1$), in agreement with the heuristic discussion of Sec.~\ref{sec_dynamical}. A numerical confirmation of this fact is provided in Figure~\ref{fig_Tinfty_mans_parametric}, where the same data as in Fig.~\ref{fig_Tinfty_mans} is plotted in the plane $(Q,\hrho)$: most of the points fall indeed on the line of marginal fixed points, apart from the finite $t$ and finite $T-t$ regime.

To further characterize the solution of the RS equations in the large $T$ regime we need to study more precisely these two regimes that match the boundary conditions with the scaling regime. Let us denote $Q_{\rm i,f}$ and $\hrho_{\rm i,f}$ the initial and final values of the scaling functions, i.e.
\beq
Q_{\rm i} = \lim_{s \to 0} Q(s) \ , \qquad 
Q_{\rm f} = \lim_{s \to 1} Q(s) \ , \qquad
\hrho_{\rm i} = \lim_{s \to 0} \hrho(s) \ , \qquad 
\hrho_{\rm f} = \lim_{s \to 1} \hrho(s) \ .
\label{eq_Qif}
\eeq
By inspection of the numerical results of Fig.~\ref{fig_Tinfty_mans} and \ref{fig_Tinfty_mans_parametric}, one is led to assume that in these matching regimes only $Q_t$ is rapidly varying, we shall thus take $\hrho_t \to \hrho_{\rm i}$ for $t=O(1)$ and $\hrho_t \to \hrho_{\rm f}$ for $t=T-O(1)$. 
These assumptions allow to simplify the equation (\ref{eq_Et}) that becomes a simple recursion for a time dependent series. More precisely, considering first the ansatz on the initial time regime $t=O(1)$, one deduces from (\ref{eq_F1}) that $\hrho_{\rm i} = (2^{k-1}-2)Q_1^{k-1}$ (which justifies the condition stated in (\ref{eq_Tinfty_mans_1})), while a telescopic summation of (\ref{eq_Et}) yields
\beq
Q_{t+1} = Q_1 - (\hrho_{\rm i} + Q_1^{k-1} - Q_t^{k-1})^l \ .
\eeq
A matching of the $t\to\infty$ limit of this regime with the behavior of the scaling function $Q(s)$ as $s\to 0$ gives the condition of Eq.~(\ref{eq_Tinfty_mans_2}). Let us turn now to the consequences of the ansatz in the final regime $t=T-O(1)$. The equation (\ref{eq_FT}) shows that, if $\epsilon \neq 0$, one has to have $Q_T \to 0$, which corresponds indeed to the unfrozen solutions with $\theta=0$ we are describing here. A telescopic summation of (\ref{eq_Et}) from $T$ to $t$ then yields in this regime:
\beq
Q_{t+1} =(\hrho_{\rm f} + Q_1^{k-1})^l  - (\hrho_{\rm f} + Q_1^{k-1} - Q_t^{k-1})^l \ .
\eeq
This has to be iterated backwards from a neighborhood of $Q_T=0$, and match the end of the scaling regime (i.e. $Q(s)$ with $s\to 1$); this consistency condition yields (\ref{eq_Tinfty_mans_4}). Finally the two remaining equations (\ref{eq_Tinfty_mans_3},\ref{eq_Tinfty_mans_5}) traduces the marginality of the stability of the fixed points $(Q_{\rm i},\hrho_{\rm i})$ and $(Q_{\rm f},\hrho_{\rm f})$ of the dynamical system, which is necessary for the consistency of the scaling assumption in the regime $s\in]0,1[$. This concludes the justification of the system of equations (\ref{eq_Tinfty_mans_1}-\ref{eq_Tinfty_mans_5}).

What remains to be explained is the expression given in (\ref{eq_zv_Tinfty}-\ref{eq_zv3}) for the thermodynamic quantity $z_{\rm v}$. The three terms distinguished in (\ref{eq_zv1},\ref{eq_zv2},\ref{eq_zv3}) corresponds to the contributions of the three time regimes ($t=O(1)$, $t/T=O(1)$ and $t=T-O(1)$ respectively) to the summations in (\ref{eq_zv_reg}). The first (resp. last) is easily obtained by a telescopic summation, with $\hrho_t$ taken equal to $\hrho_{\rm i}$ (resp. $\hrho_{\rm f}$) independently of $t$. The computation of the contribution $z_{{\rm v},2}$ of the regime $t/T=O(1)$ requires a little bit of additional work. The reader might think that one should first determine explicitly the reduced time dependence of the scaling functions $Q(s)$ and $\hrho(s)$ before attempting this computation; fortunately this step can be avoided by a judicious use of the marginality condition, as revealed by the following lines:
\bea
z_{{\rm v},2} &=& 2 \sum_t \left( (\hrho(t/T) + Q_1^{k-1} - Q(t/T)^{k-1} )^{l+1} - (\hrho(t/T) + Q_1^{k-1} - Q((t-1)/T)^{k-1} )^{l+1} \right) \\
&=& 2 \frac{1}{T}\sum_t (l+1)(k-1)(-Q'(t/T)) Q(t/T)^{k-2} (\hrho(t/T) + Q_1^{k-1} - Q(t/T)^{k-1} )^l
\\
&=& 2 (l+1)(k-1) \int_0^1 \dd s \ \frac{- \dd Q}{\dd s} Q(s)^{k-2} (\hrho(s) + Q_1^{k-1} - Q(s)^{k-1})^l \\
&=& 2 (l+1)(k-1) \int_{Q_{\rm f}}^{Q_{\rm i}} \dd Q \ Q^{k-2} (\hrho(Q) + Q_1^{k-1} - Q^{k-1})^l \ .
\eea
In the last equation $\hrho(Q)$ is defined implicitly by the marginality condition $1 = l(k-1)Q^{k-2} (\hrho(Q) + Q_1^{k-1} - Q^{k-1})^{l-1}$, eliminating it yields a simple integral on $Q$ that can be computed (safely assuming $l>k-1$) to obtain the expression given in Eq.~(\ref{eq_zv2}). The three contributions to $z_{\rm v}$ yields the relative fraction of variables that whiten in the three time regimes, as illustrated on the left panel of Fig.~\ref{fig_Tinfty_Pt}.

\subsection{The frozen solutions of the first kind}
\begin{figure}
\includegraphics[width=8cm]{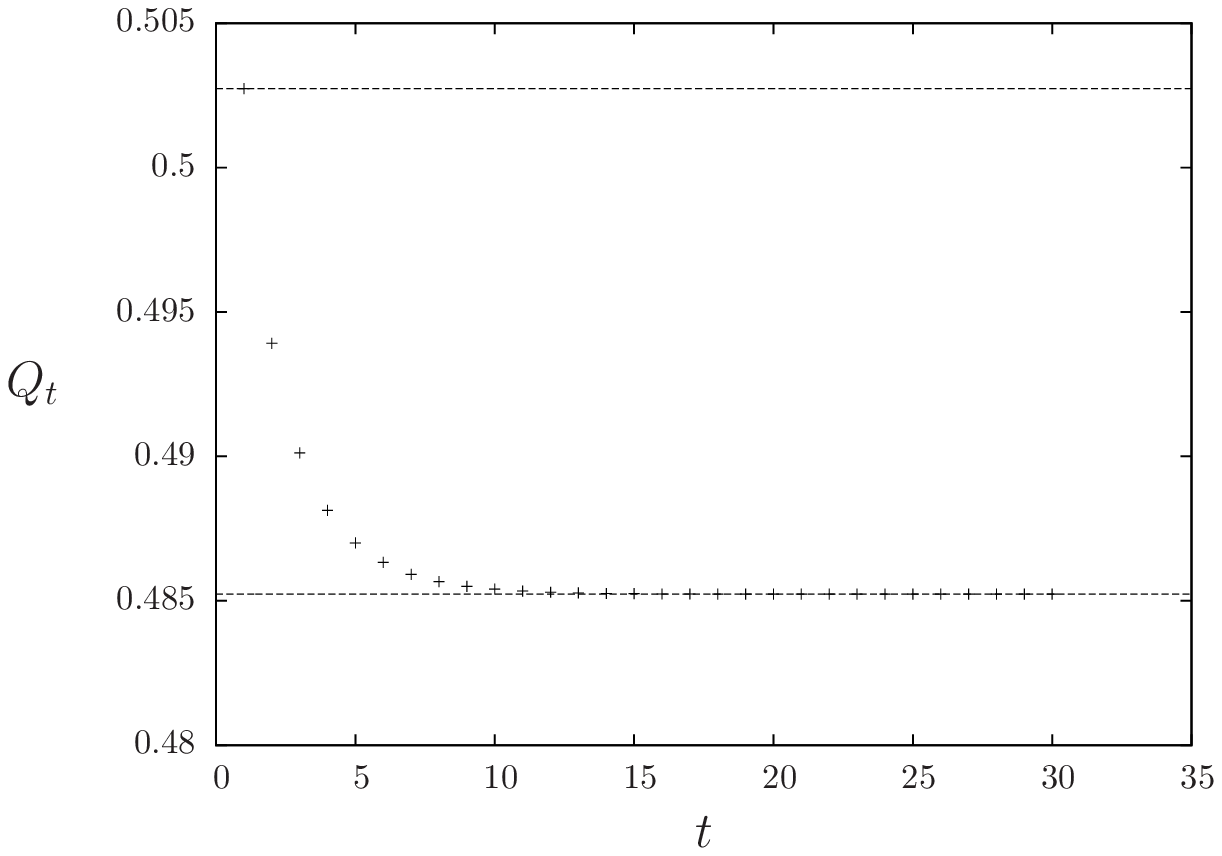}
\hspace{1cm}
\includegraphics[width=8cm]{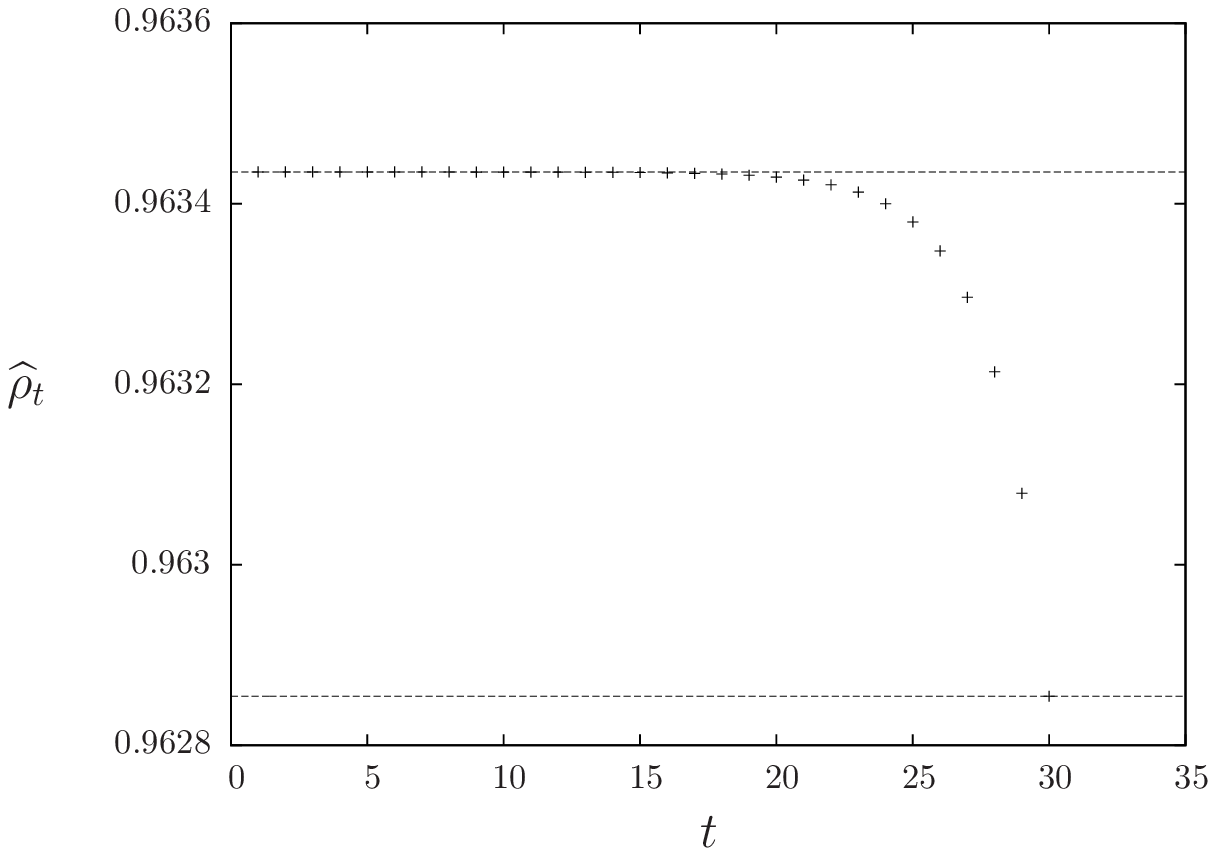}
\caption{The solution of the RS equations for $k=6$, $l=127$, $e^\epsilon=0.95$, $T=30$. Left panel : $Q_t$, the dashed horizontal lines indicate the values of $Q_1$ (top) and $Q_{\rm b}$ (bottom) obtained from the analytical predictions in the $T\to\infty$ limit (cf. Eqs.~(\ref{eq_Qb}-\ref{eq_hrhob})). Right panel : $\hrho_t$, the dashed lines representing $\hrho_{\rm b}$ (top) and $\hrho_T$ (bottom).}
\label{fig_Tinfty_nmans}
\end{figure}

We describe now the ansatz describing the branch of $s_\infty(\theta)$ corresponding to frozen solutions of the first kind. We present in Fig.~\ref{fig_Tinfty_nmans} the results of a numerical resolution of the RS equations for a large but finite value of $T$, for a choice of $\epsilon$ that would lead to this branch in the $T\to\infty$ limit. It is quite apparent on these curves that $(Q_t,\hrho_t) \approx (Q_{\rm b},\hrho_{\rm b})$, some ``bulk'' values, for most of the times $t$: in terms of the dynamical system the vector $w$ remains indeed close to a fixed point of the map $R$. The deviations from these values, that are necessary for the boundary conditions to be satisfied, occur in an asymmetric way: $Q_t$ deviates from $Q_{\rm b}$ when $t$ is of order 1 (with respect to $T$), while the deviations of $\hrho_t$ from its bulk value happen when $t=T-O(1)$. Let us show now quantitatively that this ansatz does indeed close the equations in the large $T$ limit, and how the values of $(Q_{\rm b},\hrho_{\rm b})$ are fixed.

Consider first the regime where $t$ is finite with respects to $T$, and assume that in the large $T$ limit $\hrho_t \to \hrho_{\rm b}$ independently of $t$. Then the equations (\ref{eq_Et}) can be summed telescopically, to give in this finite $t$ regime a simple recursion relation for the series $Q_t$, namely
\beq
Q_{t+1} = Q_1 - (\hrho_{\rm b} + Q_1^{k-1} - Q_t^{k-1})^l \ .
\label{eq_iterates_Q}
\eeq
Moreover (\ref{eq_F1}) implies in this ansatz that $\hrho_{\rm b} = (2^{k-1}-2) Q_1^{k-1} $. Taking the limit $t\to\infty$ of this recursion (i.e. after the $T\to\infty$ limit) one finds that $Q_{\rm b}$ is the fixed point of the recursion (\ref{eq_iterates_Q}) started from the initial value $Q_1$. A short study of this equation reveals that this fixed point is the largest one on the interval $[0,Q_1]$, that it is stable under iterations, and that this condition of stability is equivalent to $\mu<1$, where $\mu$ was defined from the study of the stability of the fixed-points of $R$ in Eq.~(\ref{eq_def_mu}). As explained there the stability in the $Q$ direction implies an instability in the $\hrho$ direction, which is apparent on the curves of Fig.~\ref{fig_Tinfty_nmans} and allows the matching with the boundary conditions at $t=T$ by a deviation of $\hrho_t$ from its bulk value for $t=T-O(1)$.

Indeed in this final regime $t=T-O(1)$, assuming that $Q_t \to Q_{\rm b}$, a telescopic summation of (\ref{eq_Ft}) yields a (backwards) recursion relation for the $\rho$, namely
\beq
\hrho_t = \hrho_T + (k-1) Q_{\rm b}^{k-2} \left[ (\hrho_{t+1} + Q_1^{k-1} - Q_{\rm b}^{k-1})^l - e^\epsilon (\hrho_T + Q_1^{k-1} - Q_{\rm b}^{k-1})^l \right] \ .
\eeq
For consistency $\hrho_{\rm b}$ must be the (stable) fixed-point of these backward iterates started from a neighborhood of $\hrho_T$. In addition (\ref{eq_ET}) gives
\beq
Q_{\rm b} = e^\epsilon \left[ (\hrho_T + Q_1^{k-1})^l- (\hrho_T + Q_1^{k-1} - Q_{\rm b}^{k-1})^l \right] \ .
\eeq
This concludes the justification of the 4 equations (\ref{eq_Qb}-\ref{eq_hrhob}) on $Q_1$, $Q_{\rm b}$, $\rho_{\rm b}$ and $\hrho_T$ given in Sec.~\ref{sec_Tinfty_nonwhitening}. As we saw above $\hrho_t$ is decreasing (resp. increasing) with $t$ if $\epsilon<0$ (resp. $\epsilon>0$), hence the additional condition $\hrho_T < \hrho_{\rm b}$ if $\epsilon <0$ (if $\epsilon >0$ one has necessarily $\hrho_T > \hrho_{\rm b}$ ). The condition (\ref{eq_condition_stability_frozen}) ensures that we are describing a fixed point of the whitening process; indeed it corresponds to the stability of the fixed point $Q_{\rm b}$ under the iterations beyond the time horizon $T$ according to (\ref{eq_Qt_beyond}). Finally the thermodynamic predictions of Eqs.~(\ref{eq_zv_Tinfty_frozen},\ref{eq_theta_Tinfty_frozen},\ref{eq_Pt_Tinfty_frozen}) are easily obtained from (\ref{eq_zv_reg},\ref{eq_theta_T_reg},\ref{eq_Pt_regular}) by plugging this ansatz and performing the sums telescopically.

Note finally that one can check that the boundary conditions do not allow for the reverse situation of a fixed point $(Q_{\rm b},\hrho_{\rm b})$ which is stable in the $\rho$ direction, unstable in the $Q$ one. 

\subsection{The frozen solutions of the second kind}

\begin{figure}
\includegraphics[width=8cm]{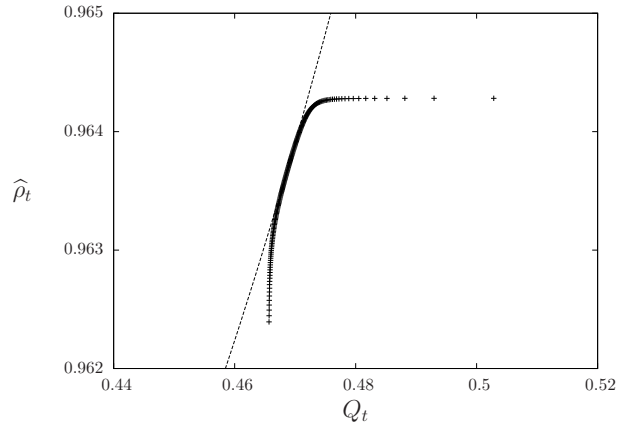}
\caption{The solution $(Q_t,\hrho_t)$ for $k=6$, $l=127$, $T=200$, plotted parametrically as $\hrho_t$ as a function of $Q_t$, for a frozen solution of the second kind. The dashed line indicating the marginal condition $\mu=1$.}
\label{fig_Tinfty_Qrho_II}
\end{figure}

We finally explain the structure of the solution of the RS equations that describe the frozen solutions of the second kind, which is a mixture of the two ansatz developed above (for these reasons our explanations will be quicker here). We show in Fig.~\ref{fig_Tinfty_Qrho_II} the flow in the $(Q,\rho)$ plane of the solution corresponding to the whitening times distribution plotted as the lower curve in the right panel of Fig.~\ref{fig_Tinfty_Pt}. One can see on Fig.~\ref{fig_Tinfty_Qrho_II} the existence of a scaling regime described by smooth functions $Q(s),\hrho(s)$ for $s=t/T \in ]0,1[$, that follows the line $\mu=1$ of marginal fixed points. We define the initial and final values of these regimes, $Q_{\rm i,f}$ and $\hrho_{\rm i,f}$ as in Eq.~(\ref{eq_Qif}). It is preceded for $t=O(1)$ by a regime where $Q_t$ varies with $\hrho_t$ essentially constant and equal to $\hrho_i$; these two first regimes are the same as the ones describing the unfrozen solutions. The last regime for $t=T-O(1)$ has on the contrary the same behavior as the one of the frozen regime of the first kind: $Q_t$ remains constant and equal to $Q_{\rm f}$, while $\hrho_t$ varies between $\hrho_{\rm f}$ and $\hrho_T$. Reproducing the reasonings explained above in the three time-regimes leads to the set of equations presented in Sec.~\ref{sec_Tinfty_nonwhitening_II}.

\subsection{The large $k$ expansion of the entropy $s_\infty(\theta=0)$ and of the threshold $l_\infty(k)$}
\label{sec_app_linfty_largek}

We shall present here the asymptotic expansion at large $k$ for the entropy of unfrozen solutions $s_\infty(\theta=0)$, that led us to the prediction (\ref{eq_linfty_largek}) for the behavior of the threshold $l_\infty(k)$. The notations and parts of the strategy will be the same as in the asymptotic expansion of the $T=1$ results described in Appendix~\ref{app_largek_T1}, in particular we consider the degree $l$ of the hypergraphs to scale with $k$ as $l=2^{k-1} \hl$, with $\hl$ varying slowly with $k$, in the regime $\ln k < \hl < k \ln 2$ between the typical rigidity and the satisfiability thresholds.

A short study of the equations (\ref{eq_uv}) on $u,v$ and of the expressions of the solutions of (\ref{eq_Tinfty_mans_1}-\ref{eq_Tinfty_mans_5}) given in (\ref{eq_Q1Qf}) reveals that $Q_1,Q_{\rm i}$ and $Q_{\rm f}$ all tend to $1/2$ in this limit, with corrections which are exponentially small for $Q_1$:
\beq
2 Q_1 = 1 + \frac{1}{l}\left[ 
\frac{\hl}{k-1} (1+u^{k-1}) + \frac{1}{k-1} \ln\left(\frac{1-u}{2} \right)
\right] + \hO\left(\frac{1}{l^2} \right) \ ,
\label{eq_Q1_expansion}
\eeq
while the approach to $1/2$ is slow for $Q_{\rm i}$ and $Q_{\rm f}$. These considerations allows to expand the contributions (\ref{eq_zv1}-\ref{eq_zv3}) to $z_{\rm v}$ from the three time regimes as follows:
\bea
z_{{\rm v},1} &=& 2 (Q_1 - Q_{\rm i})^\frac{l+1}{l} = 2 (Q_1 - Q_{\rm i}) +
\frac{1}{l} 2 (Q_1 - Q_{\rm i}) \ln (Q_1 - Q_{\rm i}) + \hO\left(\frac{1}{l^2} \right) \ ,
\\
z_{{\rm v},2} &=& 2 (Q_{\rm i} - Q_{\rm f}) + \frac{1}{l} \left[ 
2 (k-2) (Q_{\rm f} \ln Q_{\rm f} - Q_{\rm i} \ln Q_{\rm i} )
+ 2 (Q_{\rm i} - Q_{\rm f}) (k-1-\ln(l(k-1)))
\right]
+ \hO\left(\frac{1}{l^2} \right) \ ,
\\
z_{{\rm v},3} &=& 2 Q_{\rm f} + \frac{1}{l} \left[ 
2 (\hrho_{\rm f} + Q_1^{k-1})^l \ln \left((\hrho_{\rm f} + Q_1^{k-1})^l  \right)
+ 2 (\hrho_{\rm f} + Q_1^{k-1} - Q_{\rm f}^{k-1})^l \ln \left(l(k-1) Q_{\rm f}^{k-2} \right)
\right]
+ \hO\left(\frac{1}{l^2} \right) \ .
\eea
In the terms which are explicitly of order $1/l$ we can insert the expressions of $Q_1$ and the other quantities alike at their lowest order, neglecting exponentially small corrections; summing these three contributions to $z_{\rm v}$, using the expansion (\ref{eq_Q1_expansion}) of $Q_1$ and recalling that $s_\infty = - l \frac{k-1}{k} \ln z_{\rm v}$ at the leading order gives us, after some simplifications, the following expression of the entropy of unfrozen solutions, exact up to exponentially small corrections:
\bea
s_\infty(\theta=0,k,l=2^{k-1}\hl) &=& \ln 2 - \frac{\hl}{k} (1+u^{k-1}) - \ln(1-u) 
- \frac{(k-1)^2}{k} u \nonumber \\ &+& \left(\frac{v}{\hl (k-1) (1-v)} \right)^\frac{1}{k-1} \left(  \frac{(k-1)^2}{k} +  \frac{k-1}{k v} \ln(1-v) \right) + \hO \left(\frac{1}{l}\right) \ ,
\eea
where here $u=u(k,\hl)$ and $v=v(k,\hl)$ can be taken as solutions of simplified equations obtained from (\ref{eq_uv}) where we get rid of their exponentially small corrections, namely
\beq
\hl (k-1) u^{k-2} (1-u) = 1 \ , \qquad 1-v = \exp\left(- \frac{v}{(k-1)(1-v)} \right) \ .
\label{eq_uv_simplified}
\eeq
We note then that this leading order of $v=v(k)$ does not depend on $\hl$, and it is actually related to the series $w_k$ used in Sec.~\ref{app_largek_T1} above, according to $w_k = \frac{v}{(k-1)(1-v)}$. We can thus rewrite a slightly simpler expression for $s_\infty(\theta=0)$:
\bea
s_\infty(\theta=0,k,l=2^{k-1}\hl) &=& \ln 2 - \frac{\hl}{k} (1+u^{k-1}) - \ln(1-u) 
- \frac{(k-1)^2}{k} u  \nonumber \\ &+& \left(\frac{w_k}{\hl} \right)^\frac{1}{k-1} \frac{k-1}{k} \left( k-1 - \frac{1}{k-1} - w_k \right) + \hO \left(\frac{1}{l}\right) \ .
\eea

An asymptotic expansion of $l_\infty(k)$, exact within exponentially small corrections, can be obtained by imposing $s_\infty=0$ in the above expression. This is still a rather implicit determination, we shall thus simplify further this expression dropping the corrections of order smaller than $1/k$:
\bea
s_\infty(\theta=0,k,l=2^{k-1}\hl) &=& \ln 2 - \frac{\hl}{k} (1+u^{k-1}) - \ln(1-u) + \frac{(k-1)^2}{k} (1-u) - w_k + \ln\left(\frac{w_k}{\hl} \right) 
\nonumber \\ 
&+& \frac{1}{k} \left[\frac{1}{2} \left(\ln\left(\frac{w_k}{\hl} \right)\right)^2 - w_k \ln\left(\frac{w_k}{\hl} \right) + w_k - \ln\left(\frac{w_k}{\hl} \right) -1  
\right] + \tO \left(\frac{1}{k^2}\right) \ .
\label{eq_sinfty_simplified}
\eea
To make the dominant terms more apparent we remark now that the first line in (\ref{eq_sinfty_simplified}) can be rewritten as
\beq
\ln 2 - \frac{\hl}{k} (1+u^{k-1}) + (k-2) \ln(u) + \frac{k-1}{k \, \hl \, u^{k-2}} - \ln \left(1+ \frac{1}{(k-1) w_k}\right) \ ,
\eeq
using the equations obeyed by $u$ and $w_k$. Furthermore, the study of the equation (\ref{eq_uv_simplified}) on $u(k,\hl)$ reveals that when $k$ and $\hl$ diverges this quantity behaves as $u = 1 - \frac{1}{k \, \hl}$, and hence that $u^{k-1}$ tends to 1 in this limit. This shows that the equivalent of the entropy at the lowest order is
\beq
s_\infty(\theta=0,k,l=2^{k-1}\hl) \sim \ln 2 - 2 \frac{\hl}{k} \ ,
\eeq
thus the vanishing of the entropy occurs when $\hl_\infty(k) = \frac{k \ln 2}{2}$, which justifies our claim of Eq.~(\ref{eq_linfty_largek}). We can refine this asymptotic expansion and determine the correction of order $\tO(1)$ to $\hl_\infty$. Inserting this first order determination in the correction terms explicitly of order $1/k$ shown above we get after some manipulations: 
\beq
\hl_\infty(k) = \frac{k}{2} \ln 2 + (\ln k)^2 - \frac{1}{4} (\ln(k w_k))^2 + (1+\gamma) \ln k + \frac{\gamma}{2} + \frac{\gamma^2}{4} - \frac{1}{2 w_k} + \tO(\frac{1}{k}) \ , \qquad \text{with} \ \gamma = \ln\left( \frac{\ln 2}{2} \right) \ .
\eeq
This is still non-explicit because $w_k$ is given as the solution of a self-consistent equation, and its expansion contains (iterated) logarithms. Up to twice iterated logarithms one finally gets
\beq
\hl_{\infty}(k) = k \frac{\ln 2}{2} + \frac{3}{4} (\ln k)^2 - \frac{1}{2} (\ln k) (\ln \ln k) + O((\ln k) (\ln \ln \ln k)) \ .
\eeq

\section{The large $k$ limit of the threshold $l_T(k)$ for $1<T<\infty$}
\label{app_largek_Tarb}

We justify here the formula given in Eq.~(\ref{eq_lT_largek}) about the large $k$ behavior of the threshold $l_T(k)$; the case $T=1$ was already treated in Appendix~\ref{app_largek_T1}, we shall follow the same steps in the derivation of this generalization to $T>1$. Recall that the threshold $l_T(k)$ is defined by the cancellation of the entropy of the tipping point, $s(T,\ttip(T))$; this computation thus amounts to solve the RS equations (\ref{eq_reg_Qt}-\ref{eq_reg_rhot}), complemented by the tipping point condition (\ref{eq_Qinfty}-\ref{eq_Qinfty_deriv}) to fix the value of $\epsilon$, then to compute the corresponding entropy as a function of $k$ and $l$ (with (\ref{eq_phi_T_reg}-\ref{eq_theta_T_reg})), and finally to impose the vanishing of this entropy to deduce $l_T$ as a function of $k$.

To simplify these computations in the large $k$ limit we shall first get rid of exponentially small corrections, using the following ansatz on the RS unknowns:
\bea
Q_1 &=& \frac{1}{2} \left(1 + \frac{1}{l} Q'_1 + \ho \left(\frac{1}{l}\right) \right) \ , \\
Q_t &=& \frac{1}{2} Q''_t + \hO \left(\frac{1}{l}\right) \qquad \text{for} \ t \in[2,T] \ , \\
R_t &=& R'_t + \hO \left(\frac{1}{l}\right)  \qquad \text{for} \ t \in[1,T] \ , \\
\hQ_t &=& \frac{1}{2^{k-1}} \left(\hQ'_t + \hO \left(\frac{1}{l}\right) \right) 
\qquad \text{for} \ t \in[1,T-1] \ , \\
\hQ_\infty &=& \frac{1}{2^{k-1}} \left(1 + \hO \left(\frac{1}{l}\right) \right) \ , \\
\hrho_t &=& 1 + \frac{1}{l} \hrho'_t + \ho \left(\frac{1}{l}\right) \qquad \text{for} \ t \in[1,T]  \ ,
\eea
where the primed and doubly primed quantities are slowly depending on $k$. Inserting this ansatz in (\ref{eq_reg_Qt}-\ref{eq_reg_rhot}) and expanding at the leading exponential order yields equations between the primed quantities; one can then eliminate the $Q''_t$'s and the $R'_t$'s, to obtain $2T-1$ equations on the $2T-1$ unknowns $\hQ'_1,\dots,\hQ'_{T-1}$, $\hrho_1,\dots,\hrho_T$:
\bea
\frac{1}{2}\left( (1-\hQ'_{t-1})^\frac{1}{k-1} - (1-\hQ'_t)^\frac{1}{k-1} \right) &=& 
e^{\hrho'_t} \left(e^{\hl \hQ'_{t-1}} - e^{\hl \hQ'_{t-2}}  \right) \qquad \text{for} \ t \in [1,T-1] \ ,
\label{eq_largek_Tarb_1}\\
\frac{1}{2}(1-\hQ'_{T-1})^\frac{1}{k-1} &=& e^{\hrho'_T} \left(e^{\hl \hQ'_{T-1}} - e^{\hl \hQ'_{T-2}}  \right) + e^{\epsilon+\hrho'_T} \left(e^{\hl} - e^{\hl \hQ'_{T-1}}  \right) \ , \label{eq_largek_Tarb_2} \\
\hrho'_t - \hrho'_{t+1} &=& 2 \hl (k-1) (1-\hQ'_t)^\frac{k-2}{k-1} e^{\hl \hQ'_t} 
(e^{\hrho'_{t+1}}-e^{\hrho'_{t+2}}) \qquad \text{for} \ t \in[1,T-2] \ , \label{eq_largek_Tarb_3} \\
\hrho'_{T-1} - \hrho'_T &=& 2 \hl (k-1) (1-\hQ'_{T-1})^\frac{k-2}{k-1} e^{\hl \hQ'_{T-1}+\hrho'_T} (1-e^\epsilon) \ ,\label{eq_largek_Tarb_4}
\eea
where in the first line one has to interpret by convention: $\hQ'_0=0$, $e^{\hl \hQ'_{-1}}=0$. From the solution of these equations one can deduce $Q'_1$ from:
\beq
\hrho'_1 = (k-1) Q'_1 - \hl (k+1) + 2 \hl (k-1) \left[ e^{\hrho'_2 + \hl \hQ'_1} + 
\sum_{t=3}^T e^{\hrho'_t} (e^{\hl \hQ'_{t-1}} -e^{\hl \hQ'_{t-2}} )
+ e^{\epsilon + \hrho'_T} (e^\hl - e^{\hl \hQ'_{T-1}})
\right] \ .
\eeq
Inserting the ansatz in the expression (\ref{eq_phi_T_reg}-\ref{eq_zv_reg}) for the free-entropy yields, after some simplifications based on the above equations,
\beq
\phi = - \hrho'_1 - \hl - \frac{\hl}{k} + \sum_{t=1}^{T-1} (\hrho'_t - \hrho'_{t+1} ) (1-\hQ'_t)^\frac{1}{k-1} + \hO\left(\frac{1}{l} \right) \ .
\eeq
At this order the fraction of frozen variables is obtained as
\beq
\theta = 2 e^{\epsilon+\hrho'_T} (e^\hl - e^{\hl \hQ'_{T-1}}) \ ,
\eeq
while the tipping condition translates into
\beq
e^{\epsilon+\hrho'_T + \hl} = \frac{1}{2} \left(\frac{1}{\hl}\right)^{\frac{1}{k-1}} w'_k \ ,
\label{eq_largek_Tarb_tipping}
\eeq
where the series $w'_k$ was defined in Eq.~(\ref{eq_ttip_T1_largek}) while treating the large $k$ asymptotics of the $T=1$ case.

This set of equations allows to determine an approximation of $l_T(k)$ where only exponentially small corrections have been neglected. To have a more explicit expression we need to make further simplifications, and to keep now only the leading order in an expansion in powers of $k$ (keeping however all logarithmic corrections for the moment). Recalling that $\hl_1(k)$ was shown to be $\tO(k)$ in Appendix~\ref{app_largek_T1}, and that $l_T$ is growing with $T$, we can safely assume here that $\hl=\tO(k)$; we can also assume self-consistently that $\hrho'_t - \hrho'_{t+1} = \tO(k)$, and that the $\hQ'_t$ are strictly increasing with $t$, and are strictly smaller than $1$. These assumptions lead to a simplified version of (\ref{eq_largek_Tarb_1},\ref{eq_largek_Tarb_3}), which reads:
\bea
\frac{1}{2k} \ln\left(\frac{1-\hQ'_{t-1}}{1-\hQ'_t} \right) &=& e^{\hrho'_t + \hl \hQ'_{t-1}}\qquad \text{for} \ t \in [1,T-1] \ , \\
\hrho'_t - \hrho'_{t+1} &=& 2 \hl k (1-\hQ'_t) e^{\hrho'_{t+1} + \hl \hQ'_t} \qquad \text{for} \ t \in[1,T-2] \ . 
\eea
We thus obtain $\hrho'_t=-\hl \hQ'_{t-1} + \tO(1)$ for $t \in [1,T-1]$, and can close these two equations under a single recursion of the form
\beq
\frac{1-\hQ'_{t-1}}{1-\hQ'_t} = g\left( \frac{1-\hQ'_t}{1-\hQ'_{t+1}} \right) \qquad
\text{with} \ \ g(x)=1+\ln x \ .
\eeq
This allows to express $\hQ'_1,\dots,\hQ'_{T-2}$ as
\beq
\frac{1}{1-\hQ'_t} = \prod_{t'=T-t-1}^{T-2} g^{\circ t'}(x) \qquad \text{for} \ t \in[1,T-2] \ ,
\label{eq_1over1mhqPt}
\eeq
where $g^{\circ t}$ denotes the $t$-th iterate of the function $g$, and $x=\frac{1-\hQ'_{T-2}}{1-\hQ'_{T-1}}$.
The two boundary equations (\ref{eq_largek_Tarb_2},\ref{eq_largek_Tarb_4}) can now be simplified to obtain that $\hrho'_T=-\hl \hQ'_{T-1} + \tO(1)$, that (\ref{eq_1over1mhqPt}) is also valid for $t=T-1$ with the convention $g^{\circ 0}(x)=x$, and that $x=x(k,\hl)$ is the solution of the self-consistent equation
\beq
\alpha_{k,\hl} + 1 = x + \sum_{t=0}^{T-2} \ln ( g^{\circ t}(x)) \ ,
\label{eq_largek_Tarb_x}
\eeq
where the factor
\beq
\alpha_{k,\hl} =  \ln \hl - \ln w_k - \frac{1}{w_k} + \tO\left(\frac{1}{k} \right)
\eeq
comes from the expansion of (\ref{eq_largek_Tarb_tipping}). From the solution of the equation on $x$ all the $\hQ'_t$ and $\hrho'_t$ can be deduced at their leading order in powers of $1/k$. At this point one can express the entropy of the tipping point, which leads after a short computation to
\beq
s(T,\ttip(k,l),k,l=2^{k-1} \, \hl) = \ln 2 - \frac{\hl}{k} \left[1 + \ln\left(\frac{1}{1-\hQ'_1}\right) + \hQ'_{T-1} \right] + \tO \left(\frac{1}{k}\right) \ .
\eeq
Imposing the condition of vanishing of this entropy yields a value of $\hl_T(k)$ that is exact at the order $1/k$. To obtain the asymptotic expansion stated in Eq.~(\ref{eq_lT_largek}) one finally studies the equation (\ref{eq_largek_Tarb_x}): its left hand side is asymptotic to $\ln \hl \sim \ln k$, the dominant term of the right hand side is $x$, hence the solution is $x \sim \ln k$. Moreover the dominant term of the square bracket in the above expression of the entropy is $g^{\circ T}(x) \sim \ln^{\circ T}(k)$, which yields Eq.~(\ref{eq_lT_largek}).

\bibliography{myentries,added_GS}

\end{document}